\documentclass[preprint,11pt,times]{elsarticle}

\usepackage[top=2.5cm,bottom=2.5cm,left=2.0cm,right=2.0cm]{geometry}
\usepackage[symbol]{footmisc}
\usepackage[labelsep=period]{caption}
\usepackage[figurename=Fig.]{caption}
\usepackage{booktabs} 
\usepackage{multirow}
\usepackage{comment}
\usepackage{mathtools}
\usepackage{textcomp}
\usepackage{makecell}
\usepackage{gensymb} 
\usepackage{graphicx}
\usepackage{subcaption}
\usepackage{threeparttable}
\usepackage{amssymb}
\usepackage{enumitem}
\usepackage{algorithm}
\usepackage{algpseudocode}
\usepackage{amsthm}
\newtheorem{lemma}{Lemma} 
\newtheorem{proposition}{Proposition}
\usepackage{microtype}

\usepackage{lineno,hyperref}
\hypersetup{unicode=true}


\DeclareMathAlphabet{\mathcal}{OMS}{cmsy}{m}{n}
\DeclareMathOperator*{\argmin}{arg\,min}

\biboptions{sort&compress}

\begin{document}

\begin{frontmatter}



\title{Balancing the exploration-exploitation trade-off in active learning for \\ surrogate model-based reliability analysis via multi-objective optimization} 



\author[1]{Jonathan A. Moran\corref{cor1}}
\ead{jmoran@uliege.be}
\author[2]{Pablo G. Morato}

\cortext[cor1]{Corresponding author}
\address[1]{Urban and Environmental Engineering, University of Liege, 4000 Liege, Belgium}
\address[2]{Engineering Risk Analysis Group, Technical University of Munich, 80333 Munich, Germany}

\begin{abstract}
Reliability assessment of engineering systems often requires repeated evaluations of limit-state functions that may rely on computationally expensive high-fidelity models, rendering direct sampling-based reliability analysis impractical.
An effective solution is to approximate the limit-state function with a surrogate model that can be iteratively refined through active learning, thereby reducing the number of model evaluations. 
At each iteration, an acquisition strategy selects the next sample for evaluation by balancing two competing objectives: exploration, to reduce global predictive uncertainty, and exploitation, to improve accuracy near the failure boundary.
Conventional strategies such as the U-function, EFF, ERF, REIF, and portfolio-based schemes encode this balance through single pointwise scores, concealing the underlying trade-off. 
In this work, we formulate sample acquisition as a multi-objective optimization (MOO) problem in which exploration and exploitation are explicit competing objectives, yielding a compact Pareto set that provides a quantifiable trade-off representation.
To select samples from the Pareto set, we investigate principled MOO criteria and propose adaptive trade-off rules, including a scheduled exploration-to-exploitation shift and a reliability-aware selection rule.
Across diverse limit-state functions, we evaluate all tested strategies through relative failure-probability error trajectories, sample-efficiency comparisons, and global rankings, showing that the adaptive MOO-based strategies achieve robust overall performance while consistently meeting strict error targets.
\end{abstract}

\end{frontmatter}

\section{Introduction}
\label{sec:intro}

Assessing the reliability of engineering systems is challenging when performance is evaluated through computationally expensive, high-fidelity simulations (e.g., finite element analysis or computational fluid dynamics).
Active learning (AL) for surrogate model-based reliability analysis addresses this challenge by iteratively selecting samples for evaluation according to an acquisition strategy, thereby requiring fewer computationally expensive evaluations than conventional probabilistic analysis approaches. 
The integration of AL with surrogate modeling has supported a broad range of engineering tasks, including reliability analysis, design optimization, and sensitivity analysis, among other engineering applications \cite{sudret2008global,bichon2009reliability,teixeira2021adaptive,TEIXEIRA2021107248,ehre2022sequential,MORATO2022102140, moran2023active}.
Reliability estimation itself can be performed using a range of methods, selected based on the failure probability regime, problem dimensionality, and available computational budget. These include Monte Carlo sampling, first- and second-order reliability methods (i.e., FORM and SORM) \cite{nc1973FORM,breitung1984SORM}, importance sampling \cite{melchers1989importance}, line sampling \cite{koutsourelakis2004LineSampling}, and subset simulation \cite{au2001SubSetSim}. 
Subset simulation relies on conditional sampling schemes, often implemented via Markov chain Monte Carlo (MCMC) \cite{papaioannou2015mcmc}.
More specialized MCMC-based samplers, such as Hamiltonian Monte Carlo and quasi-Newton Hamiltonian Monte Carlo, have also been investigated \cite{papakonstantinou2023QnHMCMC}. 
In surrogate model-based reliability analysis, subset simulation can also be integrated with surrogate models in two complementary roles: (i) to estimate the failure probability via subset simulation on a computationally inexpensive surrogate of the limit-state function; and (ii) to generate conditional samples across intermediate failure levels, which can be used as a candidate pool to guide sample acquisition within an AL framework \cite{MOUSTAPHA2022102174}.  
Related AL formulations cast the failure probability estimation as a Bayesian inference problem, in which a GP prior on the performance function induces a posterior distribution over the failure probability. 
These approaches explicitly aim to reduce the posterior failure probability \cite{dang2022structural,wei2023expected}.

More generally, surrogate model-based reliability analysis relies on approximations of the limit-state function constructed using a range of surrogate model families, including Gaussian processes (GPs), polynomial chaos expansion, polynomial chaos Kriging, and artificial neural networks, enabling computationally efficient reliability analysis, e.g., \cite{zhang2019reif, marelli2018active, Schobi_PCKriging, bao2021adaptive}. Among these, GPs are particularly suitable for AL because they capture uncertainty in model predictions, which can be directly used to guide sample acquisition \cite{chauhan2024active, moustapha2024reliability, dhulipala2022reliability, gaspar2017adaptive}. 
In AL for reliability, sample efficiency critically depends on the trade-off between two competing objectives during sample acquisition: (i) \textit{exploration}, which addresses global model uncertainty, and (ii) \textit{exploitation}, which focuses on refining the surrogate model estimated failure boundaries. 
At each AL iteration, an acquisition strategy must carefully select the next sample for evaluation, weighing exploration against exploitation, a challenge amplified by the large candidate pools typical in reliability analysis.

In this context, acquisition strategies can be broadly classified into look-ahead and point-wise methods, depending on whether they explicitly account for the expected impact of a new sample on the surrogate model. 
Look-ahead methods evaluate candidate samples based on their anticipated effect on the updated surrogate model and associated reliability quantities, such as failure probability uncertainty or global misclassification. This evaluation requires approximating conditional surrogate updates, often over auxiliary integration sets, which can lead to a high computational cost. Representative examples include stepwise uncertainty reduction and expected integrated error reduction criteria (EIER), which assess candidates through anticipated reductions in global uncertainty- or error measures \cite{zhou2024look, bect2012sequential, wei2023expected}. 
In contrast, pointwise methods rank candidates using only local surrogate predictions, avoiding the need for conditional surrogate updates and auxiliary integration, and are thus substantially less computationally demanding than look-ahead approaches while often achieving competitive performance in practice.
Common examples include the expected feasibility function (EFF) \cite{bichon2008} and the U-function \cite{ECHARD2011145}, and more recent strategies, such as the expected risk function (ERF) \cite{erf_yang2015}, the reliability-based expected improvement function (REIF and REIF2) \cite{reif_zhang2019}, and the portfolio-based strategy \cite{portfolio_hong2023}. 
Despite their practical efficiency, these pointwise strategies implicitly condense exploration and exploitation into a single scalar score, making the relative influence of each component opaque and potentially biasing sample selection due to scaling effects or problem-dependent objective magnitudes \cite{IDas1997sums_objectives, marler2010weighted}.

Multi-objective optimization (MOO) provides a principled framework for problems with competing objectives by identifying Pareto-optimal (non-dominated) solutions that represent alternative trade-offs rather than a single outcome \cite{marler2004survey, miettinen1999nonlinear}.
In reliability-based design optimization (RBDO), Pareto-based extensions have been proposed to balance competing design goals under reliability constraints, frequently combining surrogate models and AL to reduce the cost of expensive simulations while optimizing performance within admissible reliability levels \cite{bichon2009reliability, dubourg2011reliability, lv2019surrogate, xiao2020system, shi2024active, chen2025reliability}. Although effective, these approaches remain predominantly optimization-oriented, with AL primarily serving to accelerate the identification of feasible or optimal designs rather than to directly support reliability estimation.
Beyond RBDO, Pareto-based concepts have been widely adopted in Bayesian optimization (BO) frameworks for constrained multi-objective design under expensive evaluations, where GP surrogates and AL are used to identify Pareto-optimal designs while adaptively learning unknown constraints \cite{li2026bayesian, khatamsaz2023bayesian}. More recently, adaptive BO strategies have been proposed in which the acquisition function dynamically balances objective optimization and constraint learning based on evolving uncertainty or regions of interest \cite{zhang2023constrained}.
Closely related work has shown that acquisition strategies can be interpreted as implicitly selecting Pareto-optimal compromises between predictive uncertainty and performance, and has proposed explicit multi-objective or scalarization-based acquisition rules to control this balance during feasible-region identification or constraint learning \cite{nikova2025trade}. 
However, these formulations remain primarily focused on design-space exploration or optimization rather than on adaptive trade-off considerations for reliability analysis.
Related Pareto-guided sampling strategies have also been proposed outside classical BO, including surrogate-assisted optimization methods that prioritize candidates near evolving design Pareto fronts defined in performance space \cite{LIU2025119541} and successive Pareto simulation approaches that preferentially retain samples near failure by contrasting capacity and demand components \cite{DEOLIVEIRA2025103819}.
 
\begin{figure*}[t]
  \graphicspath{ {./figures/} }
  \centering
  \includegraphics{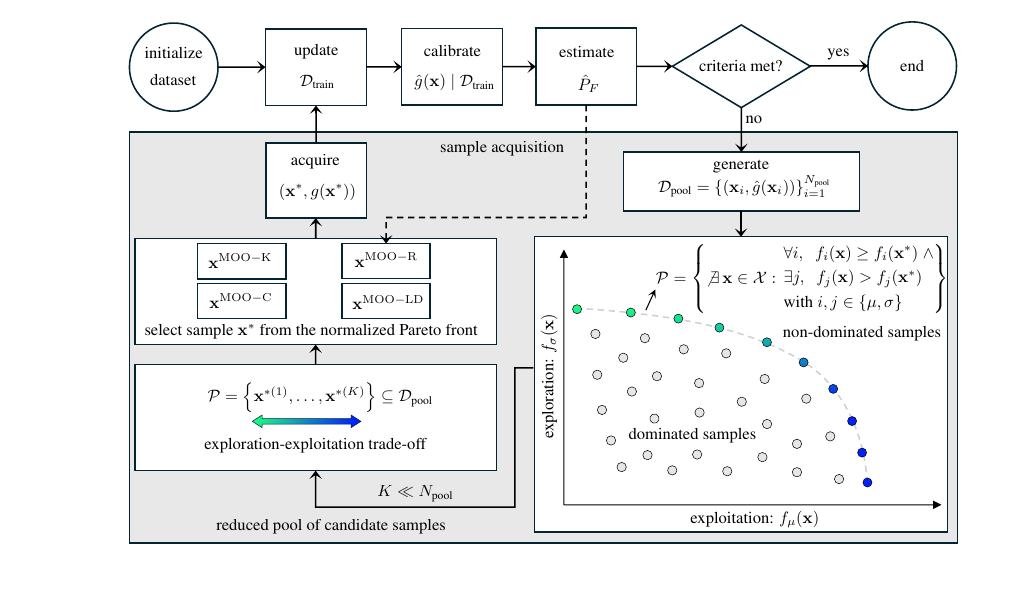}
  \caption{Multi-objective optimization framework for sample acquisition in active learning for reliability analysis. The framework integrates sample generation, Pareto front identification, sample acquisition, surrogate calibration, and failure probability estimation. At each iteration, the candidate pool, $\mathcal{D}_{\text{pool}}$, containing $N_{\text{pool}}$ samples is reduced to a Pareto-optimal set, $\mathcal{P}=\{\mathbf{x}^{*(1)},\dots,\mathbf{x}^{*(K)}\}$, of size $K$. Each non-dominated sample in $\mathcal{P}$ represents a distinct trade-off between exploration (high predictive uncertainty) and exploitation (proximity to the failure boundary). From this reduced set, a sample is selected using either principled Pareto-based strategies (knee point, MOO-K; compromise solution, MOO-C) or the proposed adaptive strategies (MOO-LD; MOO-R). The selected sample is then evaluated and incorporated into the training set, after which the surrogate model is recalibrated, and failure probability is re-estimated.}
  \label{moo_al_framework}
\end{figure*}

Widely used pointwise acquisition strategies in AL for reliability typically combine exploration and exploitation into a single scalar metric, implicitly fixing their trade-off and limiting interpretability, while potentially biasing sampling toward either uncertainty-driven exploration or failure-driven exploitation.
Although Pareto-based approaches have shown promise in related optimization and inference contexts, they have not been explicitly leveraged to formulate sample acquisition as a multi-objective problem in active learning for surrogate-based reliability analysis.
In this paper, we formulate sample acquisition in active learning for reliability as a multi-objective optimization (MOO) problem, explicitly treating exploitation and exploration as two conflicting objectives derived from the surrogate model predictive mean and standard deviation, respectively. 
Exploration promotes sampling in regions of high model uncertainty, while exploitation focuses on refining predictions near the estimated failure boundaries.
By treating surrogate predictions as conflicting objectives, the acquisition step yields a Pareto-optimal set of candidate samples rather than a single scalar optimum. This reduces the original candidate pool to a smaller non-dominated subset, each representing an interpretable trade-off between exploration and exploitation, from which samples can be selected using principled MOO-based decision rules.
In particular, we examine established MOO-based solutions, such as the knee point \cite{bechikh2010searching_knee} and the compromise solution \cite{zeleny1973compromise}, which balance the trade-off between competing objectives. 
Although these approaches were not originally designed for reliability analysis, we show that they can serve as effective strategies for acquiring samples in certain limit-state functions. 
In addition, we introduce adaptive MOO-based acquisition strategies that dynamically adjust objective preference using scheduled weighting and convergence in real-time reliability estimates.
The overarching workflow of our proposed MOO-based sample acquisition approach is illustrated in Figure~\ref{moo_al_framework}, which depicts both the identification of the Pareto-front and the subsequent sample acquisition from the non-dominated set.

In this work, we also connect conventional pointwise acquisition strategies, including EFF, U, ERF, REIF, and portfolio allocation, to the MOO-based framework. 
Under this interpretation, each acquisition strategy selects candidate points that correspond to particular solutions in a bi-objective space defined by the surrogate predictive mean and associated uncertainty, and often aligning with points on the Pareto front. 
Across diverse limit-state functions commonly studied in reliability analysis, we evaluate both MOO-based and state-of-the-art acquisition strategies, including pointwise and look-ahead approaches, in terms of sample efficiency and reliability prediction error. 
We further analyze how each strategy manages the exploration-exploitation balance over active learning iterations, highlighting case-dependent patterns that clarify the strengths and limitations of the examined acquisition strategies. 
Our experiments show that widely used acquisition strategies exhibit case-dependent performance across different limit-state functions, whereas principled MOO-based strategies are generally effective but may underperform in certain settings.
By contrast, adaptive MOO-based strategies that dynamically adjust the exploration–exploitation preference demonstrate robust performance across diverse examples, consistently meeting all imposed prediction error targets within the allocated acquisition budget. Overall, this paper proposes a multi-objective formulation of the acquisition step in active learning for surrogate-based reliability analysis, offering a unifying framework that captures optimal exploration–exploitation trade-offs at every learning iteration, connects classical and Pareto-based acquisition strategies, and enables interpretable, case-specific analysis of their effectiveness for efficient reliability estimation. 
In summary, the primary contributions of this work are:
\begin{itemize}
\itemindent=1pt
\item Introducing a multi-objective optimization (MOO) framework for sample acquisition to explicitly balance exploration-exploitation in active learning for surrogate model-based reliability analysis. 
\item A unifying interpretation that connects conventional acquisition strategies to the proposed MOO framework.
\item The introduction of Pareto-based acquisition strategies for efficient reliability analysis.
\item A comprehensive global numerical assessment across diverse limit-state functions, analyzing sample efficiency, exploration–exploitation dynamics, and patterns that clarify strategy-specific strengths and limitations.
\end{itemize}

The remainder of this paper is organized as follows: Section~\ref{sec:background} introduces the problem statement and relevant background. Section \ref{sec:tradeoff-multiobjective}, formulates sample acquisition in active learning as a multi-objective optimization and establishes connections with widely used acquisition strategies. Focusing on sample selection from the Pareto front. Section \ref{sec:bi-objective_reliability} focuses on Pareto-based sample selection, presenting principled solutions such as the knee point and compromise solution, along with adaptive acquisition strategies driven by schedule objective preference and real-time reliability estimates. The numerical experiments across a diverse set of limit-state functions commonly used in reliability analysis are reported in Section~\ref{sec:experiments}. In Section~\ref{sec:discussion}, we analyze how the investigated strategies manage the exploration–exploitation trade-off and what the observed patterns reveal about their strengths and limitations, whereas Section~\ref{sec:conclusions} summarizes the main findings and outlines directions for future research.

\section{Preliminaries}
\label{sec:background}

\subsection{Problem statement}
Consider a reliability analysis in which the limit-state function is defined as $g:\mathcal{X} \subset \mathbb{R}^M\rightarrow \mathbb{R}, \, \mathbf{x} \mapsto y=g(\mathbf{x})$, where $\mathbf{x} \in \mathcal{X}$ denotes the input parameters (e.g., material properties, geometric dimensions, loading conditions), with dimensionality $M$, and $y \in \mathbb{R}$ stands for the scalar response.
The limit-state function $g(\mathbf{x})$ partitions the input space $\mathcal{X}$ into a safe domain $\mathcal{D}_\text{s} \coloneqq \left\{  \mathbf{x} \in  \mathcal{X} \mid g(\mathbf{x})>0 \right\}$ and a failure domain $\mathcal{D}_\text{f}\coloneqq\left\{  \mathbf{x} \in  \mathcal{X} \mid g(\mathbf{x}) \leq  0 \right\}$. 
The hypersurface $g(\mathbf{x})=0$ thus marks the boundary beyond which the system does not meet the defined performance criteria. 
The probability of failure $P_F$ can then be expressed by the multivariate integral:
\begin{equation}
P_F = \int_{\mathcal{D}_{\text{f}}\coloneqq \left\{  \mathbf{x} \in  \mathcal{X} \mid g(\mathbf{x}) \leq  0 \right\} } f_{\mathbf{X}}  \left( \mathbf{x} \right) d\mathbf{x},
\label{eq:Pf_integral}
\end{equation}
where the inputs are modeled as a random vector, $\mathbf{X}$, with joint probability density function (PDF) $f_{\mathbf{X}}\left( \mathbf{x} \right)$.
In scenarios where the limit-state function is not available in closed form, directly evaluating the integral becomes infeasible. 
In such cases, sampling methods such as Monte Carlo sampling can be applied to estimate $P_F$ by repeatedly sampling from $f_{\mathbf{X}}(\mathbf{x})$ and evaluating $g(\mathbf{x})$. 
However, if estimating the failure probability, $P_F$, via sampling methods (e.g., Monte Carlo sampling) is computationally infeasible due to the high cost of evaluating the corresponding limit-state function, the response can instead be approximated by a surrogate model, $\hat{y}=\hat{g}(\mathbf{x}) \approx g(\mathbf{x})$, iteratively calibrated via active learning.
At each iteration, an acquisition strategy selects a sample from a candidate pool. 
The selected sample is then added to the training dataset, and the surrogate model is subsequently calibrated.
In this context, the goal is to define an acquisition strategy such that the failure probability estimated by the surrogate model achieves a desired approximation error using the minimal number of actively acquired samples.

\subsection{Gaussian process-based surrogate models}
Gaussian process-based surrogate models approximate the response of a limit-state function, $y=g(\mathbf{x})$, as a realization of a Gaussian process, $\mathcal{GP}(\cdot)$.
Without considering the training dataset, a prior is initially placed on the limit-state function:
\begin{equation}
g(\mathbf{x}) \sim \mathcal{GP}\left( m_{\text{GP}}(\mathbf{x}), k_{\text{GP}}(\mathbf{x}, \mathbf{x}^{\prime};\boldsymbol{\theta}) \right),
\label{eq:GP_model}
\end{equation}
where $m_{\text{GP}}(\mathbf{x}) = \mathbb{E}[g(\mathbf{x})]$ is the mean function that captures the underlying trend of the response, which can be modeled as a constant (simple), a known value (ordinary), or a low-order polynomial (universal) trend.
The function \( k_{\text{GP}}(\mathbf{x}, \mathbf{x}^{\prime};\boldsymbol{\theta}) \) is the covariance or kernel function, which represents assumptions about the correlation between responses at different input locations \( \mathbf{x} \) and \( \mathbf{x}' \). It is parameterized by a set of hyperparameters \( \boldsymbol{\theta} = \{\sigma_f^2, \boldsymbol{\ell\}} \in \mathbb{R}_{>0}^{M+1}\), where \(  \sigma_f^2 \) denotes the process variance and \( \boldsymbol{\ell} \) is a vector of characteristic length-scales associated with each input dimension.
Given a training dataset containing $N$ samples, $\mathcal{D}_{\text{train}} = \{ (\mathbf{x}_i, y_i) \}_{i=1}^N$, and assuming a noise-free response $y_i$, the posterior predictive distribution $\hat{g}(\mathbf{x}_*)$ at a new input point $\mathbf{x}_*$ is then expressed as a Gaussian random variable:
\begin{equation}
\hat{g}(\mathbf{x}_*) \mid \mathcal{D}_{\text{train}} \sim \mathcal{N}\left( \mu_{\hat{y}}(\mathbf{x}_*), \sigma_{\hat{y}}^2(\mathbf{x}_*) \right),
\label{eq:GP_predictive_model}
\end{equation}
where $\mu_{\hat{y}}(\mathbf{x}_*)$ and $\sigma_{\hat{y}}^2(\mathbf{x}_*)$ are the posterior mean and predictive variance, respectively, defined as:
\begin{align}
\mu_{\hat{y}}(\mathbf{x}_*) &= m_{\text{GP}}(\mathbf{x}_*) + \mathbf{k}_{\text{GP}}^\top(\mathbf{x}_*) \mathbf{K}_{\text{GP}}^{-1} \left( \mathbf{y} - \mathbf{m}_{\text{GP}} \right), \label{eq:pred_mu} \\
\sigma_{\hat{y}}^2(\mathbf{x}_*) &= k_{\text{GP}}(\mathbf{x}_*, \mathbf{x}_*) - \mathbf{k}_{\text{GP}}^\top(\mathbf{x}_*) \mathbf{K}_{\text{GP}}^{-1} \mathbf{k}_{\text{GP}}(\mathbf{x}_*)\,, \label{eq:pred_sigma}
\end{align}
where $\mathbf{y}$ is the vector of observed limit-state function responses and 
\(\mathbf{m}_{\text{GP}}\)
contains the mean function at training inputs $\{\mathbf{x}_i \}_{i=1}^N$:
\[
  \mathbf{y} = [y_1, y_2, \dots, y_N]^\top,
  \qquad
  \mathbf{m_{\text{GP}}} = [m_{\text{GP}}(\mathbf{x}_1), m_{\text{GP}}(\mathbf{x}_2), \dots, m_{\text{GP}}(\mathbf{x}_N)]^\top.
\]
In this formulation, $\mathbf{K}_{\text{GP}}$ represents an $N \times N$ covariance matrix with $(i,j)$-th entry $[\text{K}_{\text{GP}}]_{ij} = k_{\text{GP}}(\mathbf{x}_i, \mathbf{x}_j)$, and $\mathbf{k}_{\text{GP}}(\mathbf{x}_*)$ is an $N \times 1$ vector with entries $[\text{k}_{\text{GP}}]_i=k_{\text{GP}}(\mathbf{x}_*,\mathbf{x}_i)$. 
A key feature offered by Gaussian process-based surrogate models is that they indicate the model uncertainty associated with the predictions through the predictive variance $\sigma_{\hat{y}}^2(\mathbf{x}_*)$.

\subsection{Multi-objective optimization}
Multi‐objective optimization (MOO) seeks to optimize \(L\) potentially conflicting objectives,
where \(\mathbf{f} : \mathcal{X} \rightarrow \mathbb{R}^L\) is a vector-valued objective function defined over the feasible set \(\mathcal{X} \subseteq \mathbb{R}^M\). 
Unlike single-objective problems, which yield a unique solution, MOO produces a Pareto front of trade‐offs, where improvement in one objective may degrade another. 
Formally, the problem is formulated as:

\begin{equation}
\label{eq:multiobjective_optimization}
\text{maximize}\quad  \mathbf f(\mathbf x) = \left[f_1(\mathbf{x}), f_2(\mathbf{x}), \ldots, f_L(\mathbf{x}) \right] \,. 
\end{equation}
In this context, a solution $\mathbf{x}^* \in \mathcal{X}$ is \emph{Pareto optimal} if there exists no other $\mathbf{x} \in \mathcal{X}$ that improves at least one objective without degrading another. 
The set of non-dominated solutions, known as the \emph{Pareto front}, $\mathcal{P} \subseteq \mathcal{X}$, contains all solutions that are not dominated by any other feasible point in $\mathcal{X}$, and is defined as:

\begin{equation}
\mathcal{P} = \left\{ \mathbf{x}^* \in \mathcal{X} \;\middle|\;
\not\exists\,\mathbf{x} \in \mathcal{X}:
\Bigl[
  \forall i,\; f_i(\mathbf{x}) \ge f_i(\mathbf{x}^*)
\Bigr]
\;\wedge\;
\Bigl[
  \exists j,\; f_j(\mathbf{x}) > f_j(\mathbf{x}^*)
\Bigr]
\right\}
\quad \text{with } i, j \in \{1, \dots, L\}.
\label{eq:pareto_front_def}
\end{equation}
The Pareto front, $\mathcal{P}$, defines the frontier of optimal trade-offs among competing objectives, and thus each point on $\mathcal{P}$ represents a valid compromise solution.
To select a specific solution from $\mathcal{P}$, one must impose additional criteria based on the decision-maker's preferences.

\section{Balancing the exploration–exploitation trade-off in reliability analysis via multi-objective optimization}
\label{sec:tradeoff-multiobjective}
Active learning for Gaussian process–based reliability analysis typically defines acquisition strategies by considering that samples near the estimated failure boundary may improve failure probability predictions, while also accounting for the fact that samples from underexplored regions may reduce global uncertainty. 
Various strategies for sample acquisition have been proposed, seeking a balance between both objectives for efficient surrogate refinement. 
Conventional approaches leverage the surrogate model's mean and standard deviation predictions to rank samples, for example, by using direct ratio-based metrics \cite{ECHARD2011145}, by defining a probabilistic window around the failure boundary to prioritize near-boundary samples \cite{bichon2008}, or by combining these quantities through fixed linear weighting schemes \cite{reif_zhang2019}. 
Although these strategies can be effective in practice, the trade-off is implicitly addressed, making it challenging to interpret or explicitly adjust the balance between these two objectives.

In this work, we formalize sample acquisition in active learning as a multi-objective optimization (MOO) problem. 
Exploiting the knowledge captured by the surrogate model, one objective is to acquire samples near the failure boundary $\hat{g}(\mathbf{x})=0$. 
This can be achieved by acquiring points that minimize the absolute value computed from the predictive mean, $|\mu_{\hat{y}}(\mathbf{x})|$, as they are expected to reduce the discrepancy between the surrogate model and the ground-truth limit-state function near the failure surface.
However, the surrogate model may still be inaccurate, and potential failure regions may remain unknown.
In that case, acquiring samples from underexplored regions may be beneficial. 
Encouraging exploration, the second objective acquires samples that maximize the uncertainty captured by the surrogate model through $\sigma_{\hat{y}}(\mathbf{x})$. 

\begin{figure}[t]
  \graphicspath{ {./figures/} }
  \centering
  \includegraphics{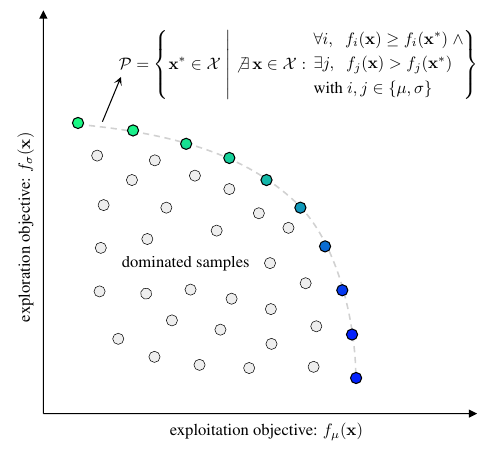}
  \caption{Pareto front in the exploration-exploitation objective space. The bi-objective space is defined by $f_{\mu}(\mathbf{x})$ (exploitation) and $f_{\sigma}(\mathbf{x})$ (exploration). The Pareto front $\mathcal{P}$ (colored points) represents the set of non-dominated solutions for which improving one objective necessarily worsens the other. Grey points denote dominated samples that are excluded from $\mathcal{P}$. This front captures the trade-off between exploration and exploitation in active learning for surrogate model-based reliability analysis.}
  \label{pareto_front}
\end{figure}

\subsection{Formulation}
\label{sec:moo_formulation}
Let $\mathcal{X} \subset \mathbb{R}^M$ be the input design space, as defined previously in Section \ref{sec:background}. 
The predictive mean $\mu_{\hat{y}}(\mathbf{x})$ and standard deviation $\sigma_{\hat{y}}(\mathbf{x})$ can be obtained from a Gaussian process-based surrogate model for each $\mathbf{x} \in \mathcal{X}$. 
We can then formulate the following non-constrained bi-objective optimization problem \footnote{Throughout the paper, we adopt the notation 
\(f_{\mu}(\mathbf{x}) \coloneqq -|\mu_{\hat{y}}(\mathbf{x})|\) (i.e., exploitation objective) and 
\(f_{\sigma}(\mathbf{x}) \coloneqq \sigma_{\hat{y}}(\mathbf{x})\) (i.e., exploration objective) .}:
\begin{equation}
\begin{aligned}
\makebox[2.5cm][r]{maximize:} &\quad \mathbf{f}(\mathbf{x}) \coloneqq
\begin{bmatrix}
f_{\mu}(\mathbf{x}) \\
f_{\sigma}(\mathbf{x})
\end{bmatrix}, \, f_{\mu}(\mathbf{x}) \coloneqq -|\mu_{\hat{y}}(\mathbf{x})|, \, f_{\sigma}(\mathbf{x}) \coloneqq \sigma_{\hat{y}}(\mathbf{x}) , \\[6pt]
\end{aligned}
\label{eq:maximization_problem_ALreliability}
\end{equation}
where the first objective, $f_{\mu}(\mathbf{x})$, \textit{exploits} the surrogate model to refine the failure boundary, whereas the second objective, $f_{\sigma}(\mathbf{x})$, \textit{explores} the design space to globally reduce uncertainty.
Formally, the set of non-dominated solutions defines the \emph{Pareto front} $\mathcal{P} \subset \mathcal{X}$:
\begin{equation}
\mathcal{P} = \left\{ \mathbf{x}^* \in \mathcal{X} \;\middle|\;
\not\exists\,\mathbf{x} \in \mathcal{X}:
\Bigl[
  \forall i,\; f_i(\mathbf{x}) \ge f_i(\mathbf{x}^*)
\Bigr]
\;\wedge\;
\Bigl[
  \exists j,\; f_j(\mathbf{x}) > f_j(\mathbf{x}^*)
\Bigr]
\right\}
\quad \text{with } i, j \in \{\mu, \sigma\}.
\label{eq:pareto_front_AL}
\end{equation}
In Figure~\ref{pareto_front}, the Pareto front $\mathcal{P}$ is represented in the bi-objective space defined by $f_{\mu}(\mathbf{x})$ and $f_{\sigma}(\mathbf{x})$. 
Points on $\mathcal{P}$ correspond to optimal exploration-exploitation trade-off solutions: an improvement over $f_{\mu}(\mathbf{x})$ results in a degradation over  $f_{\sigma}(\mathbf{x})$, and vice versa. 
By formulating the acquisition process as a MOO problem, we reduce the search space from the original domain $\mathcal{X}$ to a much smaller set of Pareto-optimal candidate samples $\mathcal{P} = \{ \mathbf{x}^{*(1)}, \dots, \mathbf{x}^{*(K)} \}$, where $K \ll N_{\text{pool}}$.
The Pareto front $\mathcal{P}$ can be constructed using a deterministic non-dominated sorting procedure applied to the surrogate predictions of the entire pool of candidate samples. Because evaluating the Gaussian process-based objectives $f_{\mu}(\mathbf{x})$ and $f_{\sigma}(\mathbf{x})$ is computationally inexpensive, the non-dominated set can be identified through direct exhaustive comparison.
 
\subsection{Are conventional pointwise acquisition strategies Pareto-optimal solutions?}
\label{sec:u_eff_ParetoSolutions}

In the engineering reliability community, widely used acquisition strategies for active learning are often formulated as pointwise criteria derived from the surrogate predictive mean and associated uncertainty at the current iteration. 
These include the U-function \cite{ECHARD2011145}, the expected feasibility function (EFF) \cite{bichon2008}, the expected risk function (ERF) \cite{erf_yang2015}, the reliability-based expected improvement function (REIF and REIF2) \cite{reif_zhang2019} and the portfolio allocation strategy \cite{portfolio_hong2023}.
Although these strategies were not originally proposed within the context of a multi-objective optimization problem, they implicitly address two conflicting objectives: sampling in regions near the limit-state function boundary (exploitation) versus in regions characterized by high predictive uncertainty (exploration). 
In the following, we interpret how these strategies acquire samples that lie on the Pareto front obtained by applying our MOO formulation, provided they satisfy a monotonicity condition with respect to the predictive mean and standard deviation.

The Pareto set proposed in this framework enables an interpretable selection of samples with optimal contribution in the exploration and exploitation trade-off. 
An additional advantage of this framework is clarifying how other strategies implicitly manage this balance. 
In this context, we can verify that the monotonicity of an acquisition function with respect to the input predictive mean and standard deviation is a sufficient condition to prove that its optimal sample lies on the Pareto front.
Specifically, for a bi-objective space defined by exploitation 
$f_{\mu}(\mathbf{x}) = -|\mu_{\hat{y}}(\mathbf{x})|$
and exploration 
$f_{\sigma}(\mathbf{x}) = \sigma_{\hat{y}}(\mathbf{x})$, a scalar acquisition $a(\mathbf{x})$ function  yields a Pareto-optimal solution if its maximization (or minimization) contradicts the existence of a dominating sample. 
To formalize this concept, we introduce a general proposition that encompasses pointwise acquisition strategies commonly used in reliability analysis.

\begin{proposition}\label{prop:monotonicity_condition}
\noindent\leavevmode\\
Let $a(\mathbf{x})$ be a scalar acquisition function. If $a(\mathbf{x})$ is strictly monotonic with respect to the exploitation objective $f_{\mu}(\mathbf{x}) = -|\mu_{\hat{y}}(\mathbf{x})|$ and the exploration objective $f_{\sigma}(\mathbf{x}) = \sigma_{\hat{y}}(\mathbf{x})$, then any sample $\mathbf{x}^*$  that optimizes $a(\mathbf{x})$ over the design space $\mathcal{X}$ is Pareto optimal with respect to maximizing $\mathbf{f}(\mathbf{x}) = \left[f_{\mu}(\mathbf{x}),\, f_{\sigma}(\mathbf{x}) \right]^\top$.
\end{proposition}
This proposition allows us to identify whether acquisition functions are specific instances of Pareto-optimal selections. 
Formal definitions and mathematical derivations are provided in \ref{sec:Appendix_acquisitions} and \ref{sec:appendix:proofs}.
Table \ref{tab:strategy_categorization} categorizes conventional active learning strategies based on their solution type, the hyperparameters that govern their behavior, and their alignment with the Pareto-optimal set $\mathcal{P}$. Solutions identified as belonging to $\mathcal{P}$ represent a mathematically optimal improvement between reducing global model uncertainty and refining the failure boundary within the predictive space of the surrogate model.
As summarized in Table \ref{tab:strategy_categorization}, strategies like U, EFF, ERF, and REIF strictly follow the monotonicity conditions. Strategies like REIF2 diverge from the Pareto set because they introduce external weighting from the physical input space $f_{\mathbf{X}}(\mathbf{x})$. This non-constant scaling can lead to selecting samples that are dominated in the surrogate's objective space but reside in high-probability zones. In contrast, the Portfolio allocation strategy maintains Pareto optimality as long as its constituent functions are themselves non-dominated.

\begin{table}[t]
\centering
\caption{Categorization of conventional pointwise acquisition strategies within the proposed MOO framework. Each strategy is positioned according to its implicit emphasis on competing objectives, namely exploration of uncertain regions and exploitation of the current surrogate prediction.
This mapping clarifies how widely used acquisition functions can be interpreted as specific scalarizations or projections of a broader MOO-based formulation, thereby providing a unifying perspective on their behavior.}
\label{tab:strategy_categorization}
\begin{threeparttable}
\begin{tabular}{lllll}
\toprule
Strategy & Solution & Parameters & Pareto optimal ($\in \mathcal{P}$)? & Objective treatment \\ 
\midrule
U-function \cite{ECHARD2011145} & $\mathbf{x}^{\rm U}$ & None & Yes  & Nonlinear ratio-based \\
EFF \cite{bichon2008} & $\mathbf{x}^{\rm EFF}$ & $\epsilon(\mathbf{x})$ \tnote{(1)} & Yes & Probabilistic integration \\
ERF \cite{erf_yang2015}& $\mathbf{x}^{\rm ERF}$ & None & Yes & Magnitude-weighted CDF \\
REIF \cite{reif_zhang2019} & $\mathbf{x}^{\rm REIF}$ & $\xi$ \tnote{(2)} & Yes & Static linear scalarization \\
REIF2 \cite{reif_zhang2019} & $\mathbf{x}^{\rm REIF2}$ & $\xi$ \tnote{(3)} & No & Density-weighted improvement \\
Portfolio \cite{portfolio_hong2023} & $\mathbf{x}^{\rm Port}$ & $\rho, \tau$ \tnote{(4)} & Yes & Probabilistic selection \\ \bottomrule
\end{tabular}
\begin{tablenotes}
\item[(1)] The parameter $\epsilon(\mathbf{x})$ defines the width of a symmetric integration window around the predicted limit-state boundary.
\item[(2)] The weight $\xi$ specifies a fixed linear scalarization between exploration and exploitation objectives.
\item[(3)] It incorporates the input probability density $f_{\mathbf{X}}(x)$, prioritizing high-density regions, which may lead to sample acquisition in dominated areas of the bi-objective space.
\item[(4)] $\rho$ represents the memory factor and $\tau$ controls the selection probability across constituent acquisition strategies.
\end{tablenotes}
\end{threeparttable}
\end{table}

\begin{figure}
  \graphicspath{ {./figures/} }
  \centering
  \includegraphics{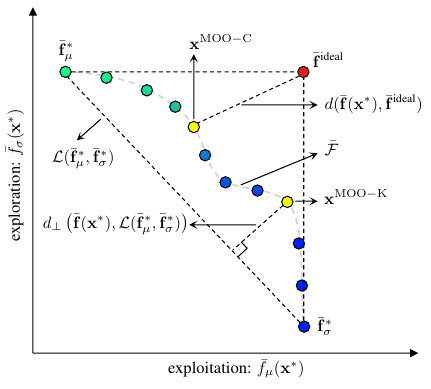}
  \caption{Knee point and compromise solution on the normalized Pareto front. The normalized bi-objective space is defined by exploitation $\bar{f}_{\mu}(\mathbf{x}^*)$ and exploration $\bar{f}_{\sigma}(\mathbf{x}^*)$. The Pareto front $\bar{\mathcal{F}}$ (colored points) represents the set of non-dominated solutions, bounded by the extreme points $\bar{\mathbf{f}}_{\mu}^*$ and $\bar{\mathbf{f}}_{\sigma}^*$. The boundary line $\mathcal{L}$ connects these extreme points. The knee point $\mathbf{x}^{\rm MOO-K}$ is identified as the point on $\bar{\mathcal{F}}$ with the maximum perpendicular distance between $\bar{\mathbf{f}}(\mathbf{x}^*)$ and $\mathcal{L} (\bar{\mathbf{f}}_{\mu}^*, \bar{\mathbf{f}}_{\sigma}^*)$. The compromise solution $\mathbf{x}^{\rm MOO-C}$ is identified as the point on $\bar{\mathcal{F}}$ with the minimum distance  between $\bar{\mathbf{f}}(\mathbf{x}^*)$ and $\bar{\mathbf{f}}^{\text{ideal}}$.}
  \label{pareto_front_points}
\end{figure}

\section{MOO-based acquisition strategies}
\label{sec:bi-objective_reliability}
By solving the multi-objective optimization problem formulated in Equation~\ref{eq:maximization_problem_ALreliability}, the search space is narrowed from the original candidate pool to a reduced set of $K$ Pareto-optimal solutions, denoted by $\mathcal{P} = \{ \mathbf{x}^{*(1)}, \dots, \mathbf{x}^{*(K)} \} \subset \mathcal{X}$. 
A question that needs to be addressed is which sample from the Pareto front should be selected for evaluation. 
In this section, we describe how principled multi-objective approaches, such as the knee-point and compromise solutions, can be straightforwardly employed for selecting samples from the Pareto front. 
Beyond these static selection criteria, we argue that the efficiency of reliability analysis can be significantly improved by prioritizing global exploration in the early stages to detect failure boundaries, subsequently transitioning toward exploitation for refinement.
To address this, we introduce adaptive Pareto-based acquisition strategies that explicitly steer this preference over iterations. Specifically, we propose a linear-decay preference strategy that enforces a structured transition from exploration to exploitation, and an adaptive acquisition strategy that dynamically adjusts this preference based on real-time reliability convergence indicators.

Following the formulation introduced in Section~\ref {sec:tradeoff-multiobjective}, the Pareto front in the objective space can be defined as 
$\mathcal F \coloneqq \{\mathbf f(\mathbf x^*) = (f_\mu(\mathbf x^*), f_\sigma(\mathbf x^*)) \mid \mathbf x^* \in \mathcal P\} \subset \mathbb R^2$. 
To avoid bias when selecting a sample from the Pareto front due to scaling differences between objectives, it is recommended to apply min–max normalization \cite{bechikh2010searching_knee}.
The normalized Pareto front can then be defined as:
\begin{equation}
\bar{\mathcal{F}} \coloneqq \left\{ \bar{\mathbf{f}}(\mathbf{x}^*) = \left( \bar{f}_{\mu}(\mathbf{x}^*), \bar{f}_{\sigma}(\mathbf{x}^*) \right) \;\middle|\; \mathbf{x}^* \in \mathcal{P} \right\} \subset [0,1]^2 \,, 
\label{eq:norm_pareto}
\end{equation}
normalizing each sample in the objective space as:
\begin{equation}
    \bar{f}_{\mu}(\mathbf{x}^*)=\frac{f_{\mu}(\mathbf{x}^*)-f_{\mu}^{\min }}{f_{\mu}^{\max }-f_{\mu}^{\min }}, \quad \bar{f}_{\sigma}(\mathbf{x}^*)=\frac{f_{\sigma}(\mathbf{x}^*)-f_{\sigma}^{\min }}{f_{\sigma}^{\max }-f_{\sigma}^{\min }},
\end{equation}
where $f_{i}^{\min }$ and $ f_{i}^{\max}$ denote the minimum and maximum values respectively of the $i$-th objective observed within the current Pareto set $\mathcal P$. It is important to note that since the surrogate model is updated at each active learning iteration, the non-dominated set $\mathcal P$ evolves dynamically. Consequently, these normalization bounds ($f_{i}^{\min }$ and $ f_{i}^{\max}$) are recomputed at every iteration to ensure that the objective space is consistently mapped to the unit square $[0,1]^2$.

\subsection{Knee point (MOO-K) solution}
The knee point on the Pareto front represents a solution at which a small improvement in one objective results in a significant deterioration in another, thereby capturing the best trade-off in the conflicting MOO problem \cite{branke2004knee, deb2011knee}. 
It corresponds to the location on the Pareto front where the curvature reaches its maximum, indicating the region of greatest sensitivity to trade-offs between the two objectives. 
The knee point, $\mathbf{x}^{\rm MOO-K}$, can be determined using geometric or analytical methods that evaluate the geometry of the Pareto front, such as curvature-based measures or geometric deviation from the boundary line connecting extreme solutions of the Pareto front. 
The method proposed by \cite{bechikh2010searching_knee} leverages the geometry of the normalized Pareto front $\bar{\mathcal{F}}\subset [0, 1]^2$. This method computes the perpendicular distance of each normalized objective vector $\bar{\mathbf{f}}(\mathbf{x}^*) \in \bar{\mathcal{F}}$ from the straight line $\mathcal{L}$ connecting the two extreme normalized points, $\bar{\mathbf{f}}_{\mu}^*$ and $\bar{\mathbf{f}}_{\sigma}^*$, as illustrated in Figure \ref{pareto_front_points}.
These extreme points represent solutions exhibiting the strongest preference for a specific objective and are defined in the objective space as:
\begin{itemize}
\itemindent=1pt
\item $\bar{\mathbf{f}}_{\mu}^* \coloneqq \left( \bar{f}_{\mu}^{\max}, \bar{f}_{\sigma}^{\min} \right)$: the solution with the maximum value of $\bar{f}_{\mu}( \mathbf{x}^*)$ and the minimum value of $\bar{f}_{\sigma}( \mathbf{x}^*)$,
\item $\bar{\mathbf{f}}_{\sigma}^* \coloneqq \left( \bar{f}_{\mu}^{\min}, \bar{f}_{\sigma}^{\max} \right)$: the solution with the minimum value of $\bar{f}_{\mu}(\mathbf{x^*})$ and the maximum value of $\bar{f}_{\sigma}( \mathbf{x^*})$.
\end{itemize}
Let \( \mathbf{u} \coloneqq \bar{\mathbf{f}}_{\sigma}^* - \bar{\mathbf{f}}_{\mu}^* \) be the direction vector of the line connecting the two extreme points. Then, for any point \( \bar{\mathbf{f}}(\mathbf{x}^*) \in \bar{\mathcal{F}} \) in the normalized objective space, its perpendicular distance to $\mathcal{L}\,(\bar{\mathbf{f}}_{\mu}^*, \bar{\mathbf{f}}_{\sigma}^*)$ is the norm of the residual after orthogonal projection:
\begin{equation}
d_{\perp}\left(\bar{\mathbf{f}}(\mathbf{x}^*), \mathcal{L}(\bar{\mathbf{f}}_{\mu}^*, \bar{\mathbf{f}}_{\sigma}^*)\right) = \left\| \bar{\mathbf{f}}(\mathbf{x}^*) - \left( \bar{\mathbf{f}}_{\mu}^* + \frac{(\bar{\mathbf{f}}(\mathbf{x}^*) - \bar{\mathbf{f}}_{\mu}^*) \cdot \mathbf{u}}{\|\mathbf{u}\|^2} \, \mathbf{u} \right) \right\|,
\label{eq:knee_projection_distance}
\end{equation}
where \( (\cdot) \) denotes the dot product and \( \|\cdot\| \) the Euclidean norm.
The knee point is then defined as the solution \( \mathbf{x}^{\rm MOO\text{-}K} \in \mathcal{P} \) whose normalized objective vector \( \bar{\mathbf{f}}(\mathbf{x}^*) \in \bar{\mathcal{F}} \) attains the maximum perpendicular distance from the reference line $\mathcal{L}(\bar{\mathbf{f}}_{\mu}^*, \bar{\mathbf{f}}_{\sigma}^*)$ connecting the two extreme points of the Pareto front:
\begin{equation}
\mathbf{x}^{\rm MOO-K} = \arg\max_{\mathbf{x}^* \in \mathcal{P}} d_{\perp}\left(\bar{\mathbf{f}}(\mathbf{x}^*), \mathcal{L}(\bar{\mathbf{f}}_{\mu}^*, \bar{\mathbf{f}}_{\sigma}^*)\right)\,.
\end{equation}

\subsection{Compromise solution (MOO-C)}
The compromise solution represents the point on the Pareto front that lies closest to the \textit{ideal point} in the objective space, as illustrated in Figure \ref{pareto_front_points}. 
Rooted in classical multi-criteria decision-making theory \cite{zeleny1973compromise}, the compromise solution provides a balanced trade-off between competing objectives by minimizing the distance to the ideal point, $\bar{\mathbf{f}}^{\text{ideal}}$, which is defined in the normalized objective space as:
\begin{equation}
\bar{\mathbf{f}}^{\text{ideal}} = \left(\bar{f}_{\mu}^{\text{ideal}}, \bar{f}_{\sigma}^{\text{ideal}}\right)\coloneqq \left(\bar{f}_{\mu}^{\max}, \bar{f}_{\sigma}^{\max} \right),
\end{equation}
where $\bar{f}_{\mu}^{\max}$ and $\bar{f}_{\sigma}^{\max}$ are the best solutions that can be achieved independently for each individual objective. 
Although $\bar{\mathbf{f}}^{\text{ideal}}$ is a theoretical point that is generally infeasible in practical problems characterized by competing objectives, it serves as a reference from which the proximity of each Pareto-optimal solution to the theoretical best case can be measured. 
To identify the compromise solution, $\mathbf{x}^{\rm MOO-C}\in\mathcal{P}$, we compute the Euclidean distance between each normalized objective vector, \( \bar{\mathbf{f}}(\mathbf{x}^*) \in \bar{\mathcal{F}} \), and the ideal point, $\bar{\mathbf{f}}^{\text{ideal}}$:
\begin{equation}
d(\bar{\mathbf{f}}(\mathbf{x}^*), \bar{\mathbf{f}}^{\text{ideal}})  = \| \bar{\mathbf{f}}^{\text{ideal}} - \bar{\mathbf{f}}(\mathbf{x}^*) \|.
\end{equation}
The compromise solution can then be identified as:
\begin{equation}
\mathbf{x}^{\rm MOO-C} = \arg\min_{\mathbf{x}^* \in \mathcal{P}} d\left(\bar{\mathbf{f}}(\mathbf{x}^*), \bar{\mathbf{f}}^{\text{ideal}}\right).
\end{equation}
By selecting the Pareto-optimal solution closest to the ideal point, the compromise solution offers a systematic and principled approach \cite{zeleny1973compromise} to balance exploration and exploitation. 

\begin{figure*}[t]
  \graphicspath{ {./figures/} }
  \centering
  \includegraphics{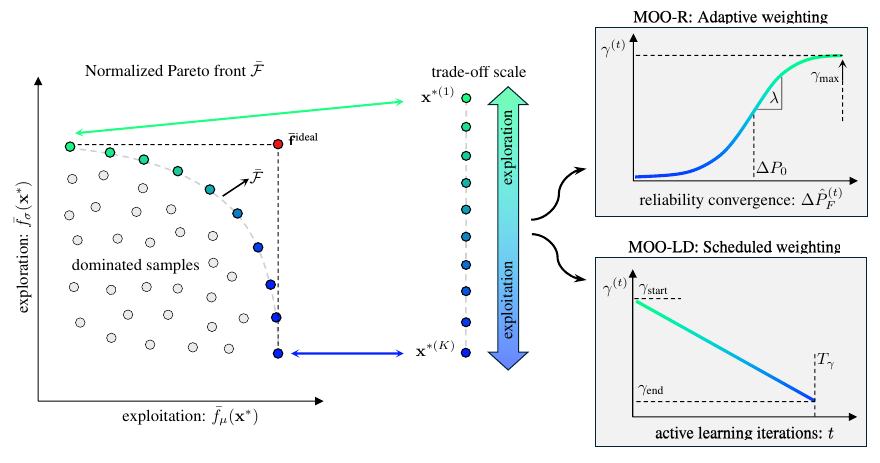}
  \caption{Overview of the proposed MOO-based acquisition strategies for active learning in reliability.
(Left) Normalized Pareto front $\bar{\mathcal{F}}$, spanning from the most exploratory solution 
$\mathbf{x}^{*(1)}$ (green) to the most exploitative solution $\mathbf{x}^{*(K)}$ (blue).
(Center) Preference scale illustrating the selection of Pareto-optimal solutions based on a 
weighted exploration--exploitation trade-off.
(Right) Selection mechanisms for the two proposed strategies:
(Top) The adaptive MOO-R strategy, where the exploration weight $\gamma^{(t)}$ is determined by a 
logistic mapping of the reliability convergence indicator $\Delta \hat{P}_F^{(t)}$, controlled by 
parameters $\lambda$ and $\Delta P_0$.
(Bottom) The scheduled MOO-LD strategy, where the exploration weight $\gamma^{(t)}$ follows a 
linear decay from $\gamma_{\text{start}}$ to $\gamma_{\text{end}}$ over a transition horizon 
$T_\gamma$.
Both strategies select the evaluation sample at each iteration by minimizing a weighted squared 
Euclidean distance to the ideal point $\bar{\mathbf{f}}^{\text{ideal}}$ in the normalized objective space.}
  \label{MOOR_graph}
\end{figure*}

\subsection{Linear-decay preference (MOO-LD)}
While the MOO-C strategy relies on a fixed trade-off between objectives, the effectiveness of AL for reliability depends on how sample acquisition evolves over iterations. 
Early global exploration is needed to identify potentially disjoint failure regions, followed by a gradual shift toward exploitation to refine the limit-state boundaries. 
To operationalize this behavior, we introduce the linear-decay (MOO-LD) acquisition strategy, which replaces the static trade-off with a time-dependent preference.
Following the normalization procedure in Equation~\ref{eq:norm_pareto}, candidate solutions on the Pareto front are ranked according to their distance to the ideal point 
$\bar{\mathbf{f}}^{\text{ideal}}$. 
A time-dependent preference parameter, $\gamma^{(t)}$, controls the exploration-exploitation preference, evolving according to a linearly decaying schedule:
\begin{equation}
\gamma^{(t)} \coloneqq \gamma_{\text{start}} 
+ \left( \gamma_{\text{end}} - \gamma_{\text{start}} \right)
\min \left( 1, \frac{t}{T_\gamma} \right),
\label{eq:gamma_schedule}
\end{equation}
where $t$ denotes the current active learning iteration and $T_\gamma$ defines the decay period.
Figure~\ref{MOOR_graph} illustrates the evolution of the preference parameter, where
$\gamma_{\text{start}} = 1$ promotes exploration in the early iterations and 
$\gamma_{\text{end}} = 0$ progressively shifts the focus toward exploitation.
Samples are selected by minimizing a preference score defined as the 
weighted squared Euclidean distance to the ideal point, $\bar{\mathbf{f}}^{\text{ideal}}$:
\begin{equation}
s\!\left(\bar{\mathbf{f}}(\mathbf{x}^*), \bar{\mathbf{f}}^{\text{ideal}}, \gamma^{(t)}\right)
= \sqrt{(1 - \gamma^{(t)}) \left( \bar{f_{\mu}}^{\text{ideal}} - \bar{f}_{\mu}(\mathbf{x}^*) \right)^2
+ \gamma^{(t)} \left( \bar{f}_{\sigma}^{\text{ideal}} - \bar{f}_{\sigma}(\mathbf{x}^*) \right)^2}\,.
\label{eq:ld_distance}
\end{equation}
At iteration $t$, the sample $\mathbf{x}^{\mathrm{MOO\text{-}LD}}$ is selected as the Pareto-optimal solution that minimizes 
the preference score:
\begin{equation}
\mathbf{x}^{\mathrm{MOO\text{-}LD}}
= \arg\min_{\mathbf{x}^* \in \mathcal{P}}
s\!\left(\bar{\mathbf{f}}(\mathbf{x}^*), \bar{\mathbf{f}}^{\text{ideal}}, \gamma^{(t)}\right).
\label{eq:ld_selection}
\end{equation}
By adopting a linear decay schedule, MOO-LD starts with an exploration phase and gradually shifts toward exploitation.

\subsection{Multi-objective optimization reliability sampling (MOO-R)}
While the previously discussed criteria address the trade-off through static geometric balances or predetermined temporal schedules, active learning for reliability can further benefit from adaptive strategies that dynamically recalibrate the exploration-exploitation balance based on real-time model performance. By monitoring the evolution of reliability estimates, the acquisition logic can react to the actual state of the surrogate rather than following a fixed trajectory.
For instance, large variations in the predicted failure probability $\hat{P}_F$ over successive iterations may indicate that the surrogate model retains substantial epistemic uncertainty in certain regions of the response space, thereby motivating increased exploratory sampling. 
Conversely, minor variations in $\hat{P}_F$ may signal that the surrogate model has globally identified the critical failure regions, suggesting that subsequent sampling should focus on exploitation to refine the approximation of the limit-state function boundary. 
Building on these reliability-based insights, we propose an adaptive acquisition strategy, MOO-R, that selects samples for evaluation based on the average relative variation in $\hat{P}_F$ at active learning iteration, $t$, computed over a sliding window of $N_{\text{it}}$ iterations. 
Let $ \delta \hat{P}_{F}^{(j)}:=\left|\hat{P}_{F}^{(j+1)}-\hat{P}_{F}^{(j)}\right| / \hat{P}_{F}^{(j)}$ denote the relative change between two consecutive iterations $j$ and $j+1$. The average relative variation at iteration $t$ is computed over a sliding window
of length $N_t=\min(t,\,N_{\mathrm{it}})$:
\begin{equation}
    \Delta \hat{P}_{F}^{(t)}=\frac{1}{N_{t}} \sum_{j=t-N_{t}}^{t-1} \delta \hat{P}_{F}^{(j)}, \quad \text { with } \quad N_{t}=\min \left(t, N_{\mathrm{it}}\right)
\end{equation}
where ${N_{t}}$ denotes the number of available past estimations, capped at buffer length $N_{\rm{it}}$. 
The average relative variation $\Delta \hat{P}_F^{(t)}$ serves as an indicator of convergence stability in the predicted failure probability at active learning iteration $t$. 
Since $\Delta \hat{P}_F^{(t)}$ cannot be computed until $N_{\text{it}}$ iterations have elapsed, the acquisition strategy initially defaults to selecting the most exploratory solution to promote global coverage.
MOO-R leverages $\Delta \hat{P}_F^{(t)}$ to adaptively balance exploration and exploitation: (i) when $\Delta \hat{P}_F^{(t)}$ is high, MOO-R prioritizes exploratory sampling to reduce global epistemic uncertainty, and when (ii) $\Delta \hat{P}_F^{(t)}$ is low, MOO-R shifts toward exploitation, refining the surrogate near the limit-state function boundary.
To enable a smooth and adaptive transition between exploration and exploitation, we introduce the weighting parameter, $\gamma^{(t)}$, defined as:
\begin{equation}
\gamma^{(t)} \coloneqq \gamma_{\text{max}} \cdot \frac{1}{1 + \text{e}^{ -\lambda \left( \Delta \hat{P}_F^{(t)} - \Delta P_0 \right)}},
\end{equation}
where $\gamma_{\max}$ denotes the maximum weighting factor assigned to exploration, $\lambda>0$ controls the steepness of the transition, and $\Delta P_0$ is a threshold that centers the sigmoid. 
The dynamics of this adaptive selection are shown in Figure \ref{MOOR_graph}, where the Pareto samples are weighted given the logistic function $\gamma^{(t)}$, with the model-dependent estimate $\hat{P}_F^{(j)}$. If $\Delta \hat{P}_F^{(t)} > \Delta P_0$, then $\gamma^{(t)} \approx \gamma_{\max}$, emphasizing exploration; if $\Delta \hat{P}_F^{(t)} < \Delta P_0$, then $\gamma^{(t)} \approx 0$, emphasizing exploitation. 

To select a sample from the Pareto front, we establish a preference score $s(\bar{\mathbf{f}}(\mathbf{x}^*), \bar{\mathbf{f}}^{\text{ideal}}, \gamma^{(t)})$ based on the weighted squared Euclidean distance to the ideal point $\bar{\mathbf{f}}^{\text{ideal}}$. 
The preference score is defined as: 
\begin{equation} 
s\left(\bar{\mathbf{f}}(\mathbf{x}^*), \bar{\mathbf{f}}^{\text{ideal}}, \gamma^{(t)}\right) = \sqrt{(1 - \gamma^{(t)}) \left( \bar{f}_{\mu}^{\text{ideal}} - \bar{f}_{\mu}(\mathbf{x}^*) \right)^2 + \gamma^{(t)} \left( \bar{f}_{\sigma}^{\text{ideal}} - \bar{f}_{\sigma}(\mathbf{x}^*) \right)^2}\,,
\end{equation} 
where $(1 - \gamma^{(t)})$ and $\gamma^{(t)}$ act as adaptive weights for the exploitative and exploratory components, respectively. 
Following this distance-based scoring, the MOO-R acquisition selects the Pareto-optimal point that minimizes the preference score: 
\begin{equation} 
\mathbf{x}^{\rm MOO-R} = \arg\min_{\mathbf{x^*}  \in \mathcal{P}} s\left(\bar{\mathbf{f}}(\mathbf{x}^*), \bar{\mathbf{f}}^{\text{ideal}}, \gamma^{(t)}\right). 
\end{equation}
The adaptive behavior of MOO-R is governed by the configuration of the logistic function $(\lambda, \Delta P_0)$ and $N_{\rm{it}}$.
 
\section{Numerical experiments}
\label{sec:experiments}
In this section, we evaluate the sampling efficiency of the proposed MOO-based acquisition strategies on well-known limit-state functions with input dimensionalities ranging from 2 to 40. 
We benchmark the MOO-based strategies against conventional pointwise acquisition strategies and examine how each approach trades off exploration and exploitation. 
In addition, we include EIER \cite{wei2023expected} as a representative look-ahead strategy used as a benchmark in the numerical experiments.

\subsection{Experimental protocol}
The numerical experiments follow the protocol described below, which is designed to facilitate a consistent comparison of performance across the investigated acquisition strategies.

\begin{itemize}[leftmargin=*]

\item Acquisition strategies: We evaluate a set of conventional pointwise strategies for reliability analysis (i.e., U-function, EFF, ERF, REIF, REIF2, and a portfolio-based scheme), and MOO-based strategies (i.e., MOO-LD, MOO-R, MOO-K, and MOO-C). 
In addition, we include the EIER look-ahead strategy as a benchmark in the numerical experiments. 
Its implementation follows \cite{wei2023expected}, with details provided in \ref{app:eier}.
The parameter settings for all strategies are summarized in Table~\ref{tab:acq_protocol}, while parameter sensitivity studies for MOO-LD and MOO-R are provided in Figure \ref{moold_hyperp_fb6} and \ref{moor_hyperp_fb6} respectively, in \ref{sec:Appendix_support}.

\begin{table}[t]
\centering
\begin{threeparttable}
\caption{Acquisition strategies considered in the numerical experiments and their configuration parameters. All strategies (i.e., pointwise, MOO-based, and lookahead) are evaluated under a common experimental protocol using the surrogate mean prediction $\mu_{\hat{y}}(x)$ and its associated standard deviation $\sigma_{\hat{y}}(x)$.}
\label{tab:acq_protocol}
\begin{tabular}{ll}
\toprule
Strategy & Parameters \\
\midrule
U-function & None \\
EFF       & $\epsilon(\mathbf{x}) = 2{\sigma}_{\hat{y}}(\mathbf{x})$ \cite{bichon2008} \\
ERF       & None \\
REIF      & $\xi=2.0$ \cite{reif_zhang2019} \\
REIF2     & $\xi=2.0$ \cite{reif_zhang2019} \\
Portfolio & $\rho=2.0$, $\tau=0.7$ \cite{portfolio_hong2023} \\
\midrule
MOO-K     & None \\
MOO-C     & None \\
MOO-LD    & $\gamma_{\text{start}} = 1.0$, $\gamma_{\text{end}} = 0.0$, $T_\gamma=50$ \\
MOO-R     & $N_{\mathrm{it}}=2$, $\Delta P_0 = 0.2$, $k=40$ \\
\midrule
EIER      & $N_{\mathrm{cand, U}}=500$, $N_g=10^3$, $N_{\mathrm{int}}=10^4$ \cite{wei2023expected} \\
\bottomrule
\end{tabular}
\end{threeparttable}
\end{table}

\item Surrogate model: The limit-state function in all experiments is approximated using a noise-free Gaussian process (GP) with a Matérn covariance kernel and a smoothness parameter of $\nu = 3/2$ \cite{rasmussen2005gaussian}. 
The GP adopts a constant mean function, estimated from the training data, to model the global trend of the limit-state function response.
The kernel is parameterized by a set of hyperparameters \( \boldsymbol{\theta} = \{\sigma_f^2, \boldsymbol{\ell}\} \in \mathbb{R}_{>0}^{M+1} \), where \( \sigma_f^2\) denotes the process variance and \( \boldsymbol{\ell} \) represents a distinct characteristic length scale for each input variable.
We recalibrate the surrogate model at every iteration by maximizing the log-marginal likelihood (LML) using the L-BFGS-B optimization algorithm \cite{morales2011_fmin_l_bfgs_b}.
The optimizer is warm-started from the kernel parameters of the previous iteration. 
To reduce convergence to poor local minima, we monitor LML and trigger a global re-optimization when the calibration degrades substantially, indicated by a pronounced drop relative to the previous iteration or an unusually low absolute LML.
This procedure involves nine multi-start initializations from a isotropic starting point (\( \ell_j = 1.0\) for $j = 1,\dots,M$). 

\item Initial dataset:
The training dataset is initialized with $N_0 = 10$ samples generated via Latin Hypercube Sampling (LHS) uniformly distributed in the unit hypercube $[0,1]^M$, where $M$ is the input dimensionality \cite{LHS-McKay}. These samples are mapped to the standard normal space through an isoprobabilistic transformation. This normalization yield inputs with approximately zero mean and unit variance, improving numerical stability during kernel hyperparameter optimization.

\item Reliability estimation: 
At each iteration, the reliability is estimated using Monte Carlo sampling (MCS) with $10^7$ samples independently drawn from a standard multivariate normal distribution.  
We intentionally retain crude MCS as a simple and robust estimator in order to avoid additional assumptions from more specialized reliability methods and to preserve a consistent basis for comparing acquisition strategy performance. 
The failure probability estimate $\hat{P}_F$ is computed as the empirical mean of an indicator function over MCS samples as:
\begin{equation}
\hat{P}_F \;\approx\; \frac{1}{10^7} \sum_{i=1}^{10^7} \mathbf{1}\!\left\{ \hat{\mu}_y(\mathbf{x}_i) \leq 0 \right\},
\end{equation}
where $\mathbf{1}{\cdot}$ denotes the indicator function that equals 1 if the surrogate model posterior mean $\hat{\mu}_y(\mathbf{x}_i)$ indicates failure and 0 otherwise. 

\item Pool of candidate samples: 
Following \cite{ECHARD2011145}, at each active learning iteration, we generate a pool set $\mathcal{D}_{\text{pool}}$, with $N_{\text{pool}} = 10^6$ samples drawn via MCS from a standard multivariate normal distribution, thereby discretizing the continuous input space.

\item Active learning iterations: 
For each experiment, we allocate a fixed budget of training samples to evaluate the acquisition strategies. 
For two-dimensional limit-state functions, the budget is set to 200 samples; for higher-dimensional limit state functions, it is set to 500 samples. In both cases, the process starts from an initial dataset with 10 passive samples and acquire the remaining 190 or 490 samples through the tested acquisition strategies.
Imposing a fixed budget isolates the comparative performance of the analyzed acquisition strategies from potential confounding effects introduced by adaptive stopping criteria \cite{MOUSTAPHA2022102174}.
Throughout the text, we denote each active learning iteration by the index $t$. 
In all experimental settings, 15 independent random seeds are evaluated, each characterized by a different initial training dataset. 

\item Performance metric:
We evaluate each investigated acquisition strategy based on the relative error between the reference failure probability, \(P_F\), and the estimated failure probability, \(\hat{P}_F\). 
At each active learning iteration $t$, we store the failure probability prediction, \(\hat{P}_F\), provided by the surrogate model, and compute the error metric, \(\delta P_{_F}\), as:
\begin{equation} 
\delta_{P_F} = \frac{\left| \hat{P}_F - P_F \right|}{P_F} . 
\end{equation}
We also quantify the number of active learning iterations required by the investigated acquisition strategies to achieve a specified relative $P_F$-error target. 
Because the resulting \(\delta{P_F}\) may fluctuate over consecutive active learning iterations, we consider a target $\delta{P}_\text{F,target}$ to be achieved if the obtained \(\delta P_{_F}\)
remains below the specified threshold for $S$ consecutive iterations:
\begin{equation} 
\frac{\left| \hat{P}_{F,i} - P_F \right|}{P_F} < \delta{P_{F,\text{target}}} \quad \text{for} \quad i = t, t+1, \ldots, t+S-1 . \end{equation}
If the prescribed condition is met, the estimated failure probability, \(\hat{P}_F\), remains within \(\pm \delta P_F\) of the reference reliability, $\hat{P}_F$, over $S$ consecutive iterations. 


\item  Posterior uncertainty in the failure-probability estimate:
At the active learning iteration at which an experiment meets the prescribed relative error in failure probability, we additionally report a posterior uncertainty measure for the GP-based failure-probability estimate \cite{wei2023expected}. 
Using a fixed set of Monte Carlo samples $\{\mathbf{x}^{(j)}\}_{j=1}^{N_{\mathrm{MC}}}$, with $\mathbf{x}^{(j)} \sim f_{\mathbf X}$ of size $N_{\mathrm{MC}}$, we draw $N_g$ realizations from the GP posterior conditioned on the corresponding training set, $\mathcal{D}_{\mathrm{train}}$. The surrogate-based failure-probability estimate is computed as:
\begin{equation}
\hat{P}_F^{(k)}
\approx
\frac{1}{N_{\mathrm{MC}}}
\sum_{j=1}^{N_{\mathrm{MC}}}
\mathbf{1}\!\left\{
\hat{g}^{(k)}\!\left(\mathbf{x}^{(j)}\right)\mid\mathcal{D}_{\mathrm{train}}\le 0
\right\},
\qquad k=1,\dots,N_g,
\label{eq:pf_post_samples}
\end{equation} 
where $\hat{g}^{(k)}(\cdot)\mid\mathcal{D}_{\mathrm{train}}$ denotes the $k$-th GP posterior realization. 
The posterior mean failure-probability estimate is then computed as:
\begin{equation}
\bar{P}_F
\approx
\frac{1}{N_g}
\sum_{k=1}^{N_g}
\hat{P}_F^{(k)},
\label{eq:pf_post_mean}
\end{equation}
and the corresponding posterior coefficient of variation is estimated as:
\begin{equation}
\mathrm{CoV}\!\left[\hat P_{F}\mid \mathcal{D}_{\mathrm{train}}\right]
\approx
\frac{ \sqrt{\frac{1}{N_g-1} \sum_{k=1}^{N_g} \left( \hat{P}_F^{(k)}-\bar{P}_F  \right)^2} }{ \bar{P}_F}.
\label{eq:pf_post_cov}
\end{equation}
In our experiments, we rely on $N_g=250$ posterior realizations and $N_{\mathrm{MC}}=10^7$ Monte Carlo samples. 

\end{itemize}

Each experimental configuration is initialized with a dataset of $N_0 = 10$ samples, generated via LHS. These samples are mapped to the standard normal space and stored in the training dataset, $\mathcal{D}_{\text{train}}$.
For two-dimensional limit-state functions, the active learning loop runs for 190 iterations; for higher-dimensional functions, it runs for 490 iterations.
At each iteration $t$, we calibrate the GP model given the currently available training dataset $\hat{g}(\mathbf{x}) \mid \mathcal{D}_{\text{train}}$, and estimate the failure probability, $\hat{P}_F^{(t)}$, via MCS with $10^7$ samples drawn from a standard normal distribution. We then compute and store the relative error, $\delta{P}_{F}^{(t)}$.
Before acquiring a sample, a pool of $N_{\text{pool}} = 10^6$ candidate samples is regenerated via MCS in the standard normal space, and the surrogate model is used to predict the mean $\hat{\mu}_y(\mathbf{x})$ and standard deviation $\hat{\sigma}_y(\mathbf{x})$ for each candidate $\mathbf{x} \in \mathcal{D}_{\text{pool}}$. 
From the pool of candidate samples, the acquisition strategies select the subsequent training sample $\mathbf{x}$, for evaluation, and the resulting input-output pair $(\mathbf{x}, g(\mathbf{x}))$ is appended to $\mathcal{D}_{\text{train}}$. 
Each setting is evaluated over 15 independent random seeds to support statistical robustness across different dataset initializations. 
The resulting $\delta{P}_F$ metrics are post-processed to quantify the target threshold $\delta{P}_\text{F,target}$ over a specified number of consecutive iterations.

All experiments reported in this paper were implemented in Python and executed on the CPU nodes of the Lyra cluster \cite{ceci_hpc} using SLURM scheduling \cite{slurm_ref}. For a total acquisition budget of 200 samples, an active learning run required on average approximately 25--30 minutes of wall-clock time for the conventional pointwise acquisition strategies, 30--40 minutes for the MOO-based strategies, and 3 hours for the EIER acquisition strategy. The codebase is publicly available at \url{https://github.com/Jonalex7/MOO-AL.git}

\subsection{Analytical limit-state functions}
The investigated acquisition strategies are tested on multiple well-known limit-state functions.
For each limit-state function considered, the reference failure probability, \(P_{F}\), is estimated as the mean of ten independent Monte Carlo simulations, each containing $10^8$ samples. The corresponding reliability index is computed as $\beta = -\Phi^{-1}(P_F)$, where \(\Phi^{-1}(\cdot)\) stands for the inverse standard normal cumulative function. In the following, we define and describe each examined limit-state function.

\subsubsection*{Four-branch function}
The four-branch limit-state function is widely used by the reliability community \cite{waarts2000fbranch, schueremans2005fbranch1, schueremans2005fbranch2}, defined as:
\begin{equation}
g(\boldsymbol {\mathbf{x}} ) =  \min \left\{ 
\begin{array}{l}
         g_1(\mathbf{x})  = 3 + 0.1(x_1-x_2)^2 - \frac{x_1+x_2}{\sqrt{2}}\\ 
         g_2(\mathbf{x})  = 3 + 0.1(x_1-x_2)^2 + \frac{x_1+x_2}{\sqrt{2}}\\
         g_3(\mathbf{x})  = (x_1-x_2) + \frac{k}{\sqrt{2}}\\
         g_4(\mathbf{x})  = (x_2-x_1) + \frac{k}{\sqrt{2}}       
\end{array} \right\},
\end{equation}
where $\mathbf{x} = (x_1, x_2)$ are two independent, standard normal random variables. 
By setting the parameter \(k\) (e.g., $k=6$ or $k=7$), the limit-state function can be adjusted, and the corresponding reference values are:
\begin{equation}
    \begin{aligned} 
    P_{F,k=6} \approx 4.46 \times 10^{-3}, &\quad \beta_{k=6} \approx 2.62,\\ 
    P_{F,k=7} \approx 2.22 \times 10^{-3}, &\quad \beta_{k=7} \approx 2.84 . 
    \end{aligned}
\end{equation}
This nonlinear limit-state function features four distinct failure branches, two linear and two nonlinear with convex–concave curvature, arranged symmetrically in the input space. Initial exploration is required to identify the multiple failure regions, followed by further exploitation to capture the sharp connections between branch boundaries.

\subsubsection*{Himmelblau function}
Adapted from a nonlinear optimization problem \cite{himmelblau2018applied}, the Himmelblau function is a fourth-order polynomial limit-state function resulting in multiple, distinct failure regions. 
We adopt the modified form proposed by \cite{papakonstantinou2023QnHMCMC}:
\begin{equation}
g(\mathbf{x} ) = \left( 
\frac{ (0.75 x_1 - 0.5)^2 }{1.81} + \frac{ 0.75 x_2 - 0.5 }{1.81} - 11
 \right)^2 +
\left( 
\frac{ 0.75 x_1 - 1 }{1.81} + \frac{ (0.75 x _2 - 0.5)^2 }{1.81} - 7
 \right)^2 - 
\kappa \,,
\end{equation}
where $\mathbf{x} = (x_1, x_2)$ are two independent standard normal random variables, and $\xi$ is a constant that shifts the failure boundary. 
In our experiments, we set $\kappa = 95$, which yields a reference failure probability $P_{F} \approx 1.66\times10^{-4}$ and a corresponding reliability index $\beta \approx 3.59$. 
This limit-state function features a continuous circular open failure boundary around the high-density area defined by the corresponding standard normal input distributions. 
An initial exploratory phase allows the identification of the open boundary and can then be followed by the exploitation of the circular failure boundary. 

\subsubsection*{Hat function}
The modified Gayton Hat function \cite{echard2013hat} defines a nonlinear, isolated failure boundary as a function of two independent normal random variables $\mathbf{x} = (x_1, x_2)$:
\begin{equation}
g(\mathbf{x} ) = 20 - \left( \mathrm{x}_1 - \mathrm{x}_2 \right)^2 - 8 \left( \mathrm{x}_1 + \mathrm{x}_2 -4 \right)^3, 
\end{equation}
where $x_1, x_2 \sim \mathcal{N}(\mu=0.25, \text{std}=1)$. 
In this setting, the reference failure probability is $P_{F} \approx 3.87 \times 10^{-4}$, with a corresponding reliability index $\beta \approx 3.36$. 
The isolated, hat-shaped failure boundary lies near one corner of the two-dimensional input domain, making it detectable with limited exploration due to the monotonic decay of the limit-state function response. 

\begin{table}
\centering
\caption{Random variables for the nonlinear oscillator limit-state function. Each input parameter is expressed as a normally distributed random variable, characterized by its mean ($\mu$) and standard deviation (std). 
}
\begin{tabular}{lllll}
\hline
Variable & Description & Distribution & \(\mu\) & std \\
\hline
$c_1$ &  Primary stiffness & Normal       & 1    & 0.1                \\
$c_2$ &  Secondary stiffness & Normal       & 0.1  & 0.01               \\
$m$   & Mass of the system   & Normal       & 1    & 0.05               \\
$r$   & Response amplitude   & Normal       & 0.5  & 0.05               \\
$\tau_1$ & Excitation time  & Normal       & 1    & 0.2                \\
$A_1$   & Force amplitude & Normal       & 1    & 0.2    \\
\hline           
\end{tabular}
\label{nonlinear_oscillator_variables}
\end{table}

\begin{figure}[t]
  \graphicspath{ {./figures/} }
  \centering
  \includegraphics{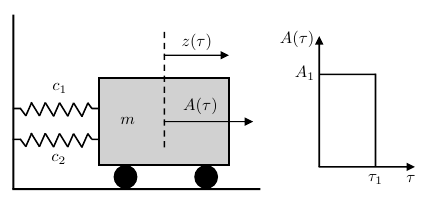}
  \caption{Schematic of the undamped nonlinear oscillator system. The system consists of a single-degree-of-freedom mass $m$ connected to two parallel linear springs with stiffness coefficients $c_1$ and $c_2$. An impulsive load $A(\tau)$ of amplitude $A_1$ is applied at time $\tau_1$. The displacement response $z(\tau)$ is evaluated, and the peak value $z_{\text{max}}$ during excitation is compared against the failure threshold $3r$, where $r$ denotes the allowable displacement limit.}
  \label{nonlinear_oscillator}
\end{figure}

\subsubsection*{Nonlinear oscillator}
\label{sec:Nonlinear_oscillator}
Studied in \cite{ECHARD2011145, nielsen2005benefit, gayton2003cq2rs}, this limit-state function is defined in terms of six independent input random variables $\mathbf{x} = (c_1,\, c_2,\, m,\, r,\, \tau_1,\, A_1)$, listed in Table~\ref{nonlinear_oscillator_variables}, and graphically shown in Figure~\ref{nonlinear_oscillator}. These variables represent physical quantities of the oscillator, including stiffness parameters ($c_1$, $c_2$), mass ($m$), threshold response amplitude ($r$), excitation time ($\tau_1$), and applied force amplitude ($A_1$). Failure is defined as the event in which the maximum displacement $z_{\text{max}}$ exceeds the critical threshold $3r$. 
The governing limit-state function is formulated as:
\begin{equation}
    g(\mathbf{x} ) = 3r - \left| z_{\text{max}} \right| = 3r - \left| \frac{2A_1}{m\omega_0^2} \sin\left( \frac{\omega_0 \tau_1}{2} \right) \right|,
\end{equation}
where $\omega_0 = \sqrt{\frac{c_1 + c_2}{m}}$, is the natural frequency of the system. In this case, the reference failure probability is $P_{F} \approx 2.86 \times 10^{-2}$, corresponding to a reliability index $\beta\approx 1.9$. 

\begin{table}[t]
\centering
\caption{Random variables for the 2-DOF damped oscillator limit-state function. Each input parameter is expressed as a lognormal distributed random variable, characterized by its mean ($\mu$) and standard deviation (std). }
\begin{tabular}{lllll}
\hline
Variable & Description & Distribution & $\mu$ & std \\
\hline
$m_p$    & Primary mass & Lognormal & 1.5 & 0.15 \\
$m_s$    & Secondary mass & Lognormal & 0.01 & 0.001 \\
$k_p$    & Primary stiffness & Lognormal & 1.0 & 0.2 \\
$k_s$    & Secondary stiffness & Lognormal & 0.01 & 0.002 \\
$\zeta_p$ & Primary damping ratio & Lognormal & 0.05 & 0.02 \\
$\zeta_s$ & Secondary damping ratio & Lognormal & 0.02 & 0.01 \\
$F_s$    & Force capacity & Lognormal & 15.0 & 0.0015 \\
$S_0$    & White-noise intensity & Lognormal & 100.0 & 10 \\
\hline
\end{tabular}
\label{2dof_oscillator_variables}
\end{table}

\begin{figure}[t]
\graphicspath{ {./figures/} }
\centering
\includegraphics{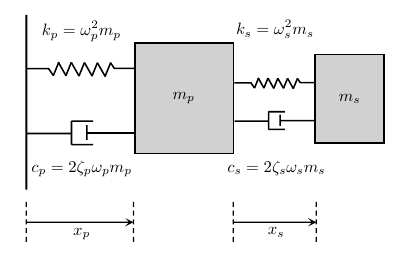}
\caption{Schematic of the two-degree-of-freedom damped oscillator. The primary oscillator is defined by mass $m_p$, stiffness $k_p$, and damping coefficient $c_p$, while the secondary oscillator is defined by $m_s$, $k_s$, and $c_s$. The parameters $x_p$ and $x_s$ denote the relative displacements of the primary and secondary masses, respectively, under stochastic base excitation. Failure occurs when the force induced in the secondary spring exceeds its capacity $F_s$.}
\label{2dof_oscillator_system}
\end{figure}

\subsubsection*{Two-degrees-of-freedom (2-DOF) damped oscillator}
\label{sec:2dof_oscillator}

This limit-state function, originally proposed by \cite{kiureghian1991efficient} and further analyzed in \cite{bourinet2011assessing}, involves a primary-secondary structural system. The system is characterized by eight independent lognormal random variables detailed in Table~\ref{2dof_oscillator_variables}, where the parameters represent the mass, stiffness, and damping ratios of the primary and secondary oscillators, the force capacity of the secondary spring, and the intensity of the white-noise base excitation. The physical configuration of the system is shown in Figure~\ref{2dof_oscillator_system}. The failure criterion is defined by the response of the secondary spring to displacement exceeding its capacity. The limit-state function is expressed as:
\begin{equation} 
g(\mathbf{x}) = 
F_s - 3k_s \sqrt{ \frac{\pi S_0}{4 \zeta_s \omega_s^3} \left[ \frac{\zeta_a \zeta_s}{\zeta_p \zeta_s (4 \zeta_a^2 + \theta^2) + \gamma_m \zeta_a^2} \frac{(\zeta_p \omega_p^3 + \zeta_s \omega_s^3) \omega_p}{4 \zeta_a \omega_a^4} \right] }, 
\end{equation}
where $\omega_p= \sqrt{k_p/m_p}$ and $\omega_s= \sqrt{k_s/m_s}$ are the natural frequencies of the system, $\omega_a = (\omega_p+\omega_s)/2$ and $\zeta_a = (\zeta_p+\zeta_s)/2$ are average mass properties, $\gamma_m=m_s/m_p$ and $\theta = (\omega_p-\omega_s)/\omega_a$ are the mass ratio and the tuning parameter respectively. The mean value of the force capacity $\mu_{F_s}=15.0$ defines a reference failure probability $P_{F} \approx 4.76 \times 10^{-3}$, which corresponds to a reliability index $\beta\approx 2.6$.

\begin{figure*}[t]
  \graphicspath{ {./figures/} }
  \centering
  \includegraphics{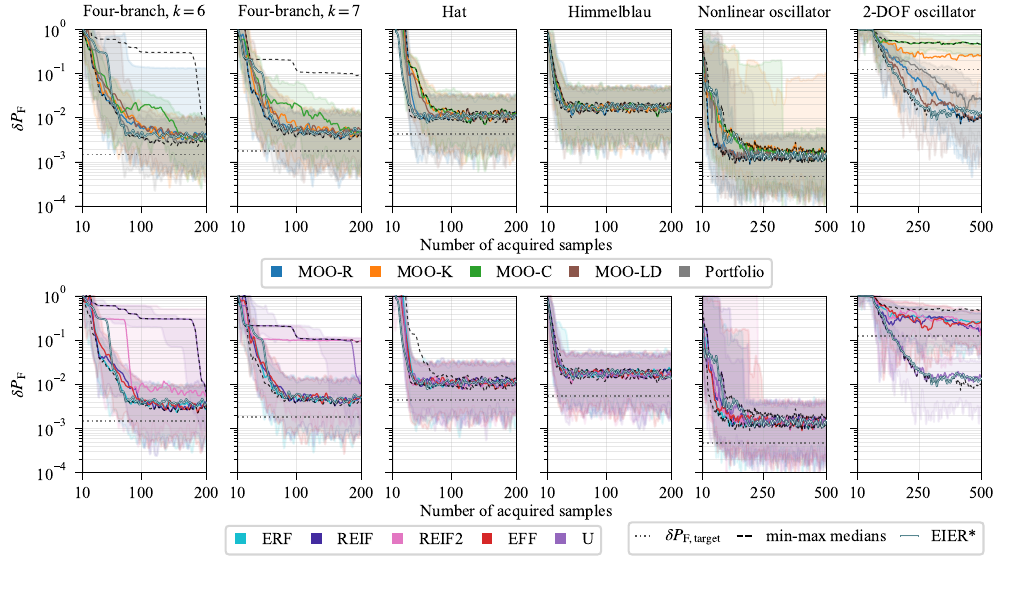}
  \caption{Active learning curves across benchmark limit-state functions. The evolution of the relative error in failure probability, $\delta P_{\mathrm{F}}$, is shown as a function of acquired samples for all investigated limit-state functions. Results are reported for pointwise, MOO-based, and lookahead $(*)$ strategies. Solid lines represent the median across 15 random seeds, while shaded areas indicate the $2.5^\text{th}$--$97.5^\text{th}$ percentiles. Dashed horizontal lines denote case-specific relative error targets $\delta P_{\mathrm{F,target}}$. Dashed black lines indicate the min-max envelope of median performance across all strategies. Lower values of $\delta P_{\mathrm{F}}$ indicate higher accuracy in the estimated failure probability.}
  \label{B_index_evolution}
\end{figure*}

\subsubsection{Results}
The results are organized into five components: (i) active learning curves, (ii) sample efficiency and posterior uncertainty, (iii) global ranking, (iv) high-dimensional limit-state function, and (v) exploration-exploitation balance.

\subsubsection*{Active learning curves}
Figure~\ref{B_index_evolution} illustrates the evolution of the relative error in failure probability, $\delta P_F$, across active learning iterations for the considered limit-state functions. 
The results are organized into two rows: the top row shows MOO-based strategies and the portfolio approach, while the bottom row reports ERF, REIF, REIF2, EFF, and U pointwise acquisition strategies. 
In both rows, the EIER look-ahead strategy is also included as a benchmark for comparison.
Solid lines indicate the median performance across 15 random seeds, while shaded regions represent the $2.5^\text{th}$--$97.5^\text{th}$ percentile range. 
The dashed black lines provide a reference envelope of the minimum and maximum median performance observed across all strategies, while the dotted horizontal line marks the specific performance target, $\delta P_{F,\text{target}}$, used to assess sample efficiency. 
For each limit-state function, the target $\delta P_{F,\text{target}}$ is defined as the minimum median relative error achieved by the 5th best pointwise acquisition strategy.

In the four-branch examples, the Pareto-based strategies (MOO-R, MOO-K, MOO-C, MOO-LD) and EFF exhibit consistent performance, reaching the $\delta P_{F,\text{target}}$ threshold within 100 samples. 
EIER also achieves low relative error in the early iterations, with performance comparable to the best-performing strategies. 
Among the MOO-based strategies, MOO-C is slightly less competitive than MOO-R, MOO-K, and MOO-LD in this regime, while REIF shows comparable performance. 
ERF and Portfolio frequently define the lower envelope of the median relative error. 
In contrast, the U-function defines the upper envelope, often requiring nearly twice as many samples to achieve a comparable relative error. 
A similar trend in relative error reduction is observed for the Hat and Himmelblau functions, where the median error of most strategies falls below the target within 100 samples. 
In both cases, EIER attains a low relative error early in the process and traces the lower envelope of the performance range, particularly within the first 50 samples. 
In Himmelblau, the performance envelope remains narrow, indicating strong agreement across strategies, whereas for the Hat function, MOO-K, MOO-C, and MOO-LD tend toward the upper envelope. 
Portfolio and MOO-R consistently achieve low relative errors in both limit-state functions.

As the dimensionality of the limit-state functions increases (i.e., nonlinear oscillator and 2-DOF oscillator), the sample requirements for achieving a stable relative error extend beyond 200 iterations. 
In the nonlinear oscillator, MOO-R achieves the lowest relative error most rapidly, followed closely by MOO-LD and Portfolio. 
While most strategies reach low relative error levels within 500 samples, the 2-DOF oscillator exhibits a wider spread in performance. In this case, U and EIER initially define the lower envelope of the median relative error up to approximately 300 samples, after which MOO-R and MOO-LD achieve lower relative errors, showing clear improvement in later iterations.

\begin{figure*}
  \graphicspath{ {./figures/} }
  \centering
  \includegraphics{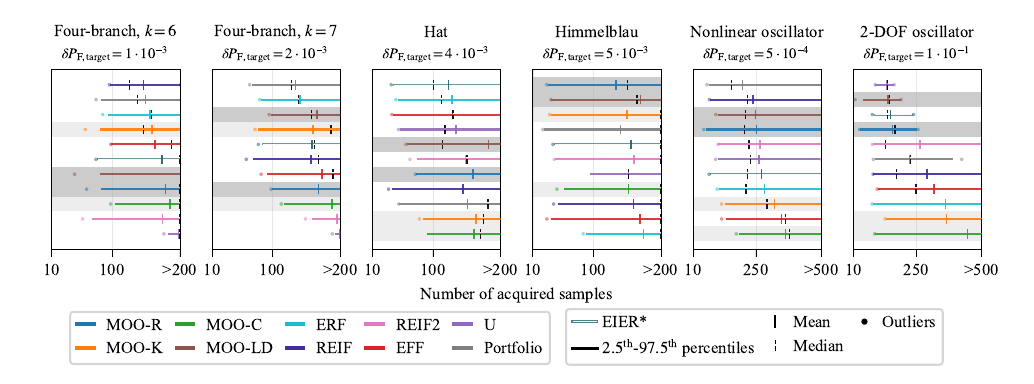}
\caption{Sample efficiency of active learning strategies across benchmark limit-state functions. The number of acquired samples required to achieve the target relative error in failure probability, $\delta P_{F,\text{target}}$ (defined in Figure~\ref{B_index_evolution}), is reported for each strategy. Each column corresponds to a limit-state function, and horizontal lines represent pointwise, MOO-based, and lookahead $(*)$ strategies, ordered locally according to the average per-seed ranking reported in Table~\ref{tab:per_seed_ranking}. Solid horizontal lines represent the $2.5^\text{th}$--$97.5^\text{th}$ percentile range over 15 random seeds, while solid and dashed vertical markers denote the mean and median, respectively. Points indicate outliers. Shaded backgrounds highlight the performance envelope of the MOO-based strategies.}
  \label{boxplots_epsilons_3it}
\end{figure*}

\begin{table*}
\centering
\footnotesize
\setlength{\tabcolsep}{3pt}
\caption{Sample efficiency of active learning strategies across benchmark limit-state functions. The table reports the number of acquired samples required to meet the target relative error in failure probability, $\delta P_{F,\text{target}}$. Values are reported as mean and median sample counts, followed by the ($2.5^\text{th}$--$97.5^\text{th}$) percentile range. Strategies are ordered globally according to the average per-seed ranking reported in Table~\ref{tab:per_seed_ranking}.}
\label{table_sample_efficiency}
\begin{tabular}{lllllll}
\toprule
\multicolumn{1}{l}{Strategy} & 
\multicolumn{1}{c}{Four-branch, {\scriptsize $k=6$}} & 
\multicolumn{1}{c}{Four-branch, {\scriptsize $k=7$}} & 
\multicolumn{1}{c}{Hat} & 
\multicolumn{1}{c}{Himmelblau} & 
\multicolumn{1}{c}{NL-Osc.} & 
\multicolumn{1}{c}{2DOF-Osc.} \\ 
& $\delta P_{F,\text{target}}=1\cdot10^{-3}$ & $\delta P_{F,\text{target}}=2\cdot 10^{-3}$ & $\delta P_{F,\text{target}}=4\cdot 10^{-3}$ & $\delta P_{F,\text{target}}=5\cdot 10^{-3}$ & $\delta P_{F,\text{target}}=5\cdot 10^{-4}$ & $\delta P_{F,\text{target}}=1\cdot10^{-1}$ \\ \midrule

Portfolio & 148.7, 137{\scriptsize(82,201)} & 133.3, 127{\scriptsize(69,201)} & 150.8, 181{\scriptsize(49,201)} & 140.5, 201{\scriptsize(26,201)} & 198.3, 156{\scriptsize(70,501)} & 222.8, 228{\scriptsize(91,390)} \\
MOO-LD    & 177.3, 201{\scriptsize(80,206)} & 153.7, 157{\scriptsize(98,201)} & 132.6, 113{\scriptsize(60,204)} & 146.0, 164{\scriptsize(36,201)} & 247.7, 209{\scriptsize(103,501)} & 140.5, 146{\scriptsize(45,190)} \\
EIER      & 173.7, 201{\scriptsize(78,201)} & 158.1, 162{\scriptsize(85,201)} & 122.5, 100{\scriptsize(40,201)} & 155.7, 201{\scriptsize(42,201)} & 271.3, 218{\scriptsize(78,501)} & 150.6, 139{\scriptsize(93,232)} \\
MOO-R     & 179.0, 201{\scriptsize(83,201)} & 167.7, 201{\scriptsize(98,201)} & 158.5, 201{\scriptsize(68,201)} & 132.9, 150{\scriptsize(31,201)} & 252.5, 207{\scriptsize(57,501)} & 157.9, 168{\scriptsize(33,252)} \\
REIF      & 146.0, 125{\scriptsize(95,201)} & 157.1, 167{\scriptsize(70,201)} & 144.5, 201{\scriptsize(38,201)} & 159.3, 201{\scriptsize(46,201)} & 238.1, 218{\scriptsize(73,501)} & 290.1, 173{\scriptsize(85,501)} \\
ERF       & 154.5, 158{\scriptsize(93,201)} & 140.3, 138{\scriptsize(82,201)} & 128.1, 112{\scriptsize(47,201)} & 174.9, 201{\scriptsize(89,201)} & 281.4, 211{\scriptsize(106,501)} & 361.9, 501{\scriptsize(84,501)} \\
EFF       & 163.0, 187{\scriptsize(98,201)} & 173.5, 189{\scriptsize(91,201)} & 128.9, 129{\scriptsize(39,201)} & 162.0, 201{\scriptsize(36,201)} & 340.3, 363{\scriptsize(135,501)} & 317.7, 248{\scriptsize(103,501)} \\
MOO-K     & 158.9, 146{\scriptsize(80,201)} & 158.7, 184{\scriptsize(77,201)} & 163.3, 174{\scriptsize(85,201)} & 142.8, 142{\scriptsize(37,201)} & 321.1, 292{\scriptsize(131,501)} & 365.7, 501{\scriptsize(134,501)} \\
REIF2     & 174.4, 201{\scriptsize(69,201)} & 195.9, 201{\scriptsize(158,201)} & 148.3, 150{\scriptsize(76,201)} & 160.1, 201{\scriptsize(42,201)} & 265.4, 223{\scriptsize(105,501)} & 263.6, 132{\scriptsize(83,501)} \\
U         & 198.8, 201{\scriptsize(182,201)} & 200.1, 201{\scriptsize(193,201)} & 133.8, 117{\scriptsize(49,201)} & 152.2, 152{\scriptsize(94,201)} & 261.0, 229{\scriptsize(103,496)} & 138.0, 139{\scriptsize(93,165)} \\
MOO-C     & 185.7, 201{\scriptsize(104,201)} & 188.5, 201{\scriptsize(116,201)} & 160.8, 170{\scriptsize(90,201)} & 146.6, 166{\scriptsize(55,201)} & 362.1, 378{\scriptsize(184,501)} & 446.6, 501{\scriptsize(92,501)} \\

\bottomrule
\end{tabular}
\label{table_epsilon}
\end{table*}

\begin{figure*}
  \graphicspath{ {./figures/} }
  \centering
  \includegraphics{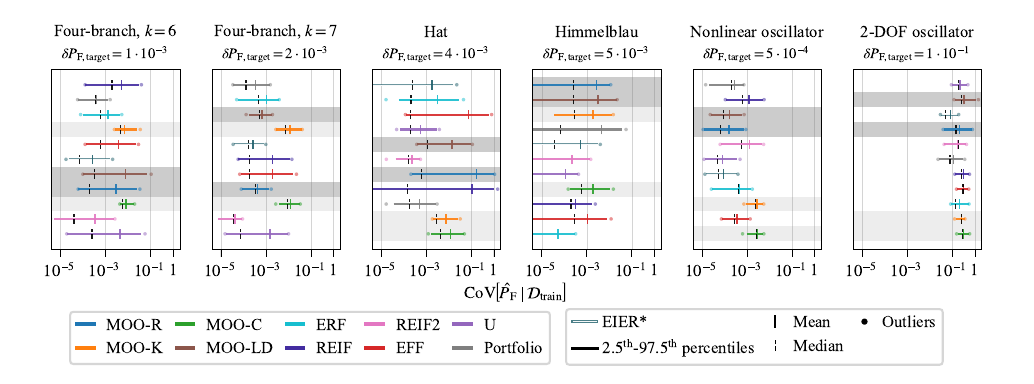}
\caption{Posterior uncertainty in failure probability estimates across benchmark limit-state functions. Each column corresponds to a limit-state function, and horizontal lines represent pointwise, MOO-based, and look-ahead ($*$) strategies, ordered locally according to the average per-seed ranking reported in Table~\ref{tab:per_seed_ranking}. For each strategy, the coefficient of variation of the posterior failure probability estimate, $\mathrm{CoV}\,[\hat P_{\mathrm F}\mid \mathcal{D}_{\mathrm{train}}]$, is reported across 15 random seeds at the iteration where the relative error $\delta P_{\mathrm{F}}$ meets the prescribed target $\delta P_{\mathrm{F,target}}$ (Figure~\ref{B_index_evolution}). Solid horizontal lines denote the $2.5^\text{th}$--$97.5^\text{th}$ percentile range over 15 random seeds; solid and dashed vertical markers indicate the mean and median, respectively; and points denote outliers. Shaded backgrounds highlight MOO-based strategies.}
  \label{fig:pf_post_cov_threshold}
\end{figure*}

\subsubsection*{Sample efficiency}

Figure~\ref{boxplots_epsilons_3it} illustrates the distribution of sample efficiency for each strategy using percentile intervals, together with mean and median indicators. 
Each horizontal line represents the $2.5^\text{th}$ to $97.5^\text{th}$ percentile range of the number of samples required to meet the prescribed relative error target over 15 random seeds. 
Experiments that do not meet the target within the allocated budget (i.e., 200 samples for 2D limit-state functions, and 500 for the remaining limit-state functions) are right-censored and displayed at the upper bound of the horizontal axis.
In addition, Table~\ref{table_epsilon} reports the corresponding statistics, including the mean, median, and the $2.5^\text{th}$--$97.5^\text{th}$ percentile interval for each strategy and limit-state function.

In the four-branch ($k=\{6, 7 \}$) limit-state functions, REIF and Portfolio exhibit the lowest median sample counts. REIF performs the best for $k=6$ (125 samples), whereas Portfolio achieves the lowest median for $k=7$ (127 samples). 
While MOO-K and MOO-R generally outperform the U strategy, Portfolio is characterized by a narrower percentile interval. 
In contrast, U frequently approaches the maximum budget and often fails to reach the target within 200 samples. 
The hat limit-state function requires rapid exploitation of an isolated failure zone. 
In this setting, EIER attains the lowest median sample count (100 samples), while ERF, MOO-LD, and U remain competitive, with MOO-LD and ERF also exhibiting relatively low dispersion. 
In the Himmelblau function, MOO-R is the most sample efficient (150 samples), followed by U and MOO-LD. 
These strategies clearly outperform their counterparts, which fail to meet the target with the entire sample budget.

In the nonlinear oscillator, Portfolio achieves the lowest median count (156 samples), followed by MOO-R and MOO-LD with median values of 207 and 209 samples, respectively. 
In contrast, MOO-C and EFF require substantially more samples, with their mean and median indicators often exceeding 350 samples. 
For the 2-DOF oscillator, the percentile range indicates increased variability across strategies in this higher-dimensional setting. 
REIF2 attains the lowest median sample count (132 samples), followed by U and EIER (139 samples), while MOO-LD remains competitive at 146 samples. 
MOO-R follows with a median of 168 samples. 
In contrast, ERF, MOO-K, and MOO-C fail to reach the target with the entire sample budget.

To complement the sample-efficiency analysis, Figure~\ref{fig:pf_post_cov_threshold} reports the posterior coefficient of variation of the GP-based failure probability estimator, $\mathrm{CoV}\,[\hat P_{\mathrm F}\mid \mathcal{D}_{\mathrm{train}}]$, for all strategies and limit-state functions, where the GP is conditioned on the training set available at the iteration at which the relative error target is met. 
Across the considered limit-state functions, U, $\mathrm{EIER}$, REIF2, and Portfolio consistently exhibit lower posterior CoV values, with relatively narrow percentile intervals. 
In contrast, MOO-K and MOO-C tend to produce higher CoV values and wider dispersion, indicating greater variability across seeds. 
The ERF, REIF, MOO-LD, and EFF strategies show intermediate behavior, with performance depending on the specific limit-state function.
Differences across limit-state functions are also evident. 
The four-branch and Himmelblau functions generally lead to lower CoV values across most strategies, whereas the oscillator problems exhibit larger CoV levels and greater variability. 
In several cases, strategies that do not meet the relative error target within the prescribed budget still attain comparatively low CoV values, particularly in the four-branch and Himmelblau functions.
Overall, the posterior dispersion of $\hat P_{\mathrm F}$ depends both on the selected training set and on the underlying problem characteristics, and may not fully align with the relative error $\delta P_F$ evaluated at the same iteration. 
This pattern is illustrated by the U strategy, which in the four-branch function with $k=6$ exhibits low posterior CoV despite relatively large errors in $\delta P_F$, whereas for the 2-DOF oscillator it achieves low relative error with comparatively higher CoV.

\begin{table}
\centering
\footnotesize
\setlength{\tabcolsep}{5pt}
\caption{Per-seed ranking across benchmark limit-state functions. For each limit-state function and each of the 15 random seeds, strategies are ranked according to the number of acquired samples required to reach the target relative error in failure probability, $\delta P_{\mathrm{F,target}}$. Reported local ranking values correspond to the mean rank across seeds for each limit-state function. The last column reports the mean of these local ranks across all limit-state functions, and strategies are sorted in ascending order of this global ranking (lower is better).}
\label{table_global_rankings}
\begin{tabular}{llccccccc}
\toprule
Rank & Strategy & Four-branch, {\scriptsize $k=6$ } & Four-branch, {\scriptsize $k=7$ } & Hat & Himmelblau & NL-Osc. & 2DOF-Osc. & \textbf{Avg. seed rank} $\downarrow$\\ \midrule
1 & Portfolio    & 51.27  & 41.00  & 82.33 & 72.27 & 49.40  & 71.13  & 61.23 \\
2 & MOO-LD       & 80.13  & 55.60  & 67.93 & 62.00 & 66.47  & 47.13  & 63.21  \\
3 & EIER         & 77.26  & 64.20  & 52.00 & 73.13 & 73.60  & 50.13  & 65.05 \\
4 & MOO-R        & 82.40  & 72.87  & 77.93 & 57.80 & 67.13  & 52.53  & 68.44 \\
5 & REIF         & 47.27  & 65.93  & 82.00 & 83.00 & 63.73  & 72.60  & 69.09  \\
6 & ERF          & 56.93  & 44.93  & 59.40 & 97.33 & 75.87  & 92.07  & 71.09 \\
7 & EFF          & 63.87  & 72.40  & 64.80 & 83.80 & 96.73  & 90.80  & 78.73  \\
8 & MOO-K        & 62.27  & 63.00  & 87.80 & 69.60 & 92.27  & 100.33 & 79.21  \\
9 & REIF2        & 94.47  & 117.53 & 77.00 & 74.07 & 67.93  & 65.40  & 82.73 \\
10 & U           & 124.73 & 128.87 & 67.87 & 76.60 & 73.00  & 44.27  & 85.89  \\
11 & MOO-C       & 91.67  & 92.87  & 87.93 & 78.53 & 102.47 & 118.73 & 95.37  \\ \bottomrule
\end{tabular}
\label{tab:per_seed_ranking}
\end{table}

\begin{figure*}
  \graphicspath{ {./figures/} }
  \centering
  \includegraphics{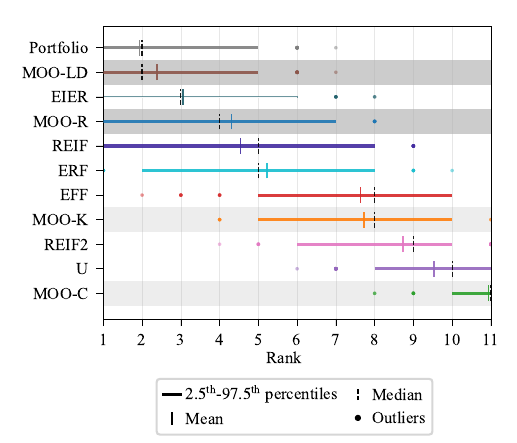}
    \caption{Bootstrap distribution of global rank positions across benchmark limit-state functions. For each bootstrap replicate, the global ranking is recomputed by resampling with replacement the sample-efficiency results for each limit-state function/strategy pair. Each horizontal bar represents the distribution of rank positions for each strategy using $2.5^\text{th}$--$97.5^\text{th}$ percentile intervals, mean and median markers, and outlier points. Shaded backgrounds highlight MOO-based strategies.}
  \label{fig:bootstrap_rank_positions}
\end{figure*}

\subsubsection*{Global ranking}

To report a global performance measure across the considered limit-state functions, we implement a hierarchical ranking procedure based on sample efficiency. 
For each seed, we rely on the number of samples required to meet the prescribed target $\delta P_{F,\text{target}}$. 
All 165 experiments (11 strategies $\times$ 15 seeds), are ranked within each limit-state function according to this sample count. 
The resulting ranks are first averaged over the 15 seeds for each strategy and limit-state function, and subsequently averaged across all limit-state functions to obtain the global ranking reported in Table~\ref{tab:per_seed_ranking}.
The resulting per-seed ranking indicates a structured ordering among the evaluated acquisition strategies. 
Portfolio and MOO-LD occupy the first two positions. 
EIER follows in third position, closely followed by MOO-R. 
REIF and ERF form the next group, while EFF and MOO-K occupy intermediate-to-lower positions. 
REIF2 and U rank near the bottom overall, despite showing competitive behavior in specific cases such as the 2DOF oscillator. 
MOO-C consistently ranks last, indicating that a fixed trade-off between exploration and exploitation over iterations is less effective than the adaptive strategies adopted by MOO-LD and MOO-R.

To assess the sensitivity of the global ranking to the finite number of random seeds, Figure~\ref{fig:bootstrap_rank_positions} reports the distribution obtained via stratified bootstrap resampling. 
For each limit-state function and strategy, the results from the 15 available runs are resampled with replacement, and the full ranking procedure is repeated. 
In each bootstrap replicate, the sample-efficiency rankings are recomputed within each limit-state function and subsequently aggregated into a global ranking, expressed as ordinal positions from $1$ to $11$. 
The resulting distributions indicate a clear separation between a leading group, an intermediate group, and a lower-ranked tail. 
Portfolio and MOO-LD exhibit the most concentrated distributions, with most of their mass in the top three positions and a low probability of falling beyond rank five. 
$\mathrm{EIER}$ also remains within the leading group, with a distribution centered around rank three and confined to the top six positions, indicating competitive performance together with relatively stable ranking. 
MOO-R, REIF, and ERF display broader distributions, indicating that while they can attain leading positions, their rankings are more sensitive to bootstrap resampling. 
By contrast, EFF and MOO-K are primarily concentrated in the middle-to-lower range of the ranking, whereas REIF2 and U are predominantly assigned to lower positions. 
MOO-C is almost entirely confined to the last two ranks, indicating a consistently low ranking.

\begin{figure*}
  \graphicspath{ {./figures/} }
  \centering
  \includegraphics{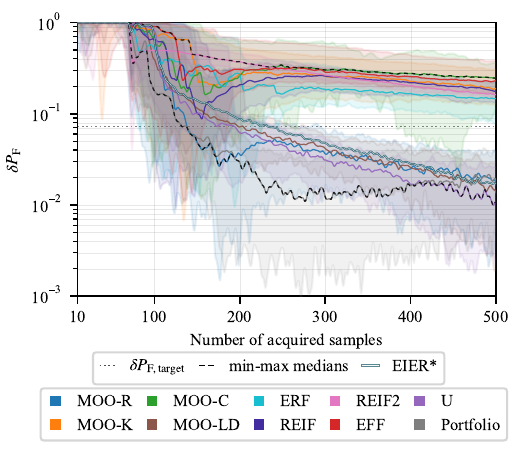}
    \caption{Active learning curves for the high-dimensional limit-state function. The evolution of the relative error in failure probability, $\delta P_{\mathrm{F}}$, is represented as a function of acquired samples for all investigated limit-state functions. 
    Results are reported for pointwise, MOO-based, and lookahead $(*)$ strategies. Solid lines represent the median across 15 random seeds, while shaded areas indicate the $2.5^\text{th}$--$97.5^\text{th}$ percentiles. The horizontal dashed line denotes the target threshold $\delta P_{\mathrm{F,target}}=7 \cdot 10^{-2}$. Dashed black lines indicate the min-max envelope of median performance across all strategies. Lower values of $\delta P_{\mathrm{F}}$ indicate higher accuracy in the estimated failure probability.}
  \label{pf_evolution_highdim}
\end{figure*}

\begin{table}
\centering
\caption{Per-seed ranking and sample efficiency for the high-dimensional limit-state function. For each random seed, strategies are ranked according to the number of acquired samples required to reach the target relative error in failure probability, $\delta P_{\mathrm{F,target}}$. Reported ranking values correspond to the mean rank across seeds, ordered in ascending order (lower is better).
The last column reports the sample-efficiency mean and median sample counts, followed by the ($2.5^\text{th}$--$97.5^\text{th}$) percentile range.}
\label{tab:high_dim_rankings}
\small
\begin{tabular}{lcl}
\toprule
Strategy & \textbf{Avg. seed rank} $\downarrow$ & Sample efficiency \\ 
& & $\delta P_{F,\text{target}} = 7 \cdot 10^{-2}$ \\
\midrule
MOO-R     & 49.20  & 132.1, 133 {\scriptsize(88,185)} \\
REIF      & 55.93  & 138.5, 143 {\scriptsize(103,177)} \\
Portfolio & 57.00  & 141.6, 131 {\scriptsize(96,193)} \\
MOO-C     & 64.53  & 200.1, 140 {\scriptsize(98,501)} \\
EFF       & 69.87  & 228.5, 135 {\scriptsize(85,501)} \\
MOO-K     & 73.67  & 227.1, 153 {\scriptsize(60,501)} \\
ERF       & 83.27  & 225.7, 181 {\scriptsize(91,501)} \\
U         & 87.13  & 172.5, 170 {\scriptsize(121,227)} \\
MOO-LD    & 95.47  & 191.8, 201 {\scriptsize(121,261)} \\
EIER      & 117.07 & 233.8, 238 {\scriptsize(175,292)} \\
REIF2     & 118.93 & 444.1, 501 {\scriptsize(75,501)} \\ 
\bottomrule
\end{tabular}
\end{table}

\begin{figure*}
  \graphicspath{ {./figures/} }
  \centering
  \includegraphics{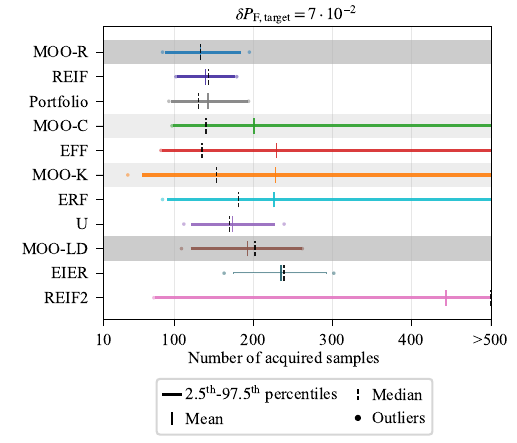}
    \caption{Sample efficiency for the high-dimensional limit-state function.
    The number of acquired samples required to achieve the target relative error in failure probability, $\delta P_{F,\text{target}}$, is reported for each strategy. Horizontal lines represent pointwise, MOO-based, and lookahead $(*)$ strategies, ordered according to the average per-seed ranking reported in Table~\ref{tab:high_dim_rankings}. Solid horizontal lines represent the $2.5^\text{th}$--$97.5^\text{th}$ percentile range over 15 random seeds, while solid and dashed vertical markers denote the mean and median, respectively. Points indicate outliers. Shaded backgrounds highlight the performance envelope of the MOO-based strategies.}
  \label{highdim_boxplots_epsilons_3it}
\end{figure*}

\begin{figure*}
  \graphicspath{ {./figures/} }
  \centering
  \includegraphics{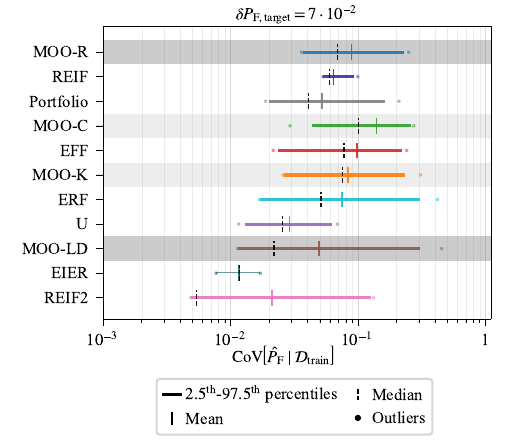}
    \caption{Posterior uncertainty in failure probability estimates for the high-dimensional limit-state function. 
    Horizontal lines represent pointwise, MOO-based, and look-ahead ($*$) strategies, ordered according to the average per-seed ranking reported in Table~\ref{tab:high_dim_rankings}. For each strategy, the coefficient of variation of the posterior failure probability estimate, $\mathrm{CoV}\,[\hat P_{\mathrm F}\mid \mathcal{D}_{\mathrm{train}}]$, is reported across 15 random seeds at the iteration where the relative error $\delta P_{\mathrm{F}}$ meets the prescribed target $\delta P_{\mathrm{F,target}}$. Solid horizontal lines denote the $2.5^\text{th}$--$97.5^\text{th}$ percentile range over 15 random seeds; solid and dashed vertical markers indicate the mean and median, respectively; and points denote outliers. Shaded backgrounds highlight MOO-based strategies.}
  \label{pf_post_cov_highdim}
\end{figure*}

\subsubsection*{High-dimensional limit-state function}

This high-dimensional limit-state function was originally introduced by \cite{rackwitz2001reliability}. 
Its formulation allows the function to be defined based on a specified number of random variables $M$, making it well-suited for testing the scalability of active learning strategies. 
Specifically, the limit-state function is defined as:
\begin{equation} 
g(\mathbf{x}) = \left( M + 3\,\text{std} \cdot \sqrt{M} \right) - \sum_{i=1}^{M} x_i \,,
\end{equation}
where $\mathbf{x} = (x_1, x_2, \ldots, x_M)$ are $M$ independent lognormal random variables with unit mean and standard deviation $\text{std}=0.2$. 
In our experiments, we set $M=40$, which results in a reference failure probability $P_{F} \approx 1.98 \times 10^{-3}$, corresponding to a reliability index $\beta \approx 2.88$. 
Despite its higher dimensionality relative to the other examined functions, this case follows the same experimental protocol, i.e., an initial dataset is generated with $N_0=10$ samples drawn via LHS, followed by 490 samples acquired throughout the active learning loop. Following the evaluation framework established in the previous section, we analyze the performance of the acquisition strategies.

Figure \ref{pf_evolution_highdim} illustrates the evolution of the relative error in failure probability over active learning iterations. 
All evaluated strategies maintain a relative error, $\delta P_{\mathrm{F}} \approx 1$, during the first 75 iterations, indicating that the initial design does not yet capture the failure region.
Between approximately 75 and 150 iterations, a sharp decrease in relative error is observed across most strategies, corresponding to the first identification of the limit-state boundary.
Portfolio and REIF2 exhibit early reductions in error, around 75 iterations, while the MOO-based strategies and U follow shortly thereafter. 
By 150 iterations, most strategies transition from global exploration to local refinement of the failure boundary.
Beyond 150 iterations, the reduction in relative error slows for most strategies. 
MOO-R, MOO-LD, Portfolio, EIER, and U reach relatively stable error levels, with $\delta P_{\mathrm{F}}$ typically ranging between $1 \cdot 10^{-2}$ and $5 \cdot 10^{-2}$. 
In this regime, intermittent increases in relative error are observed for several strategies, indicating variability in the refinement of the failure boundary.

Non-monotonic error evolution is observed for certain strategies (e.g., MOO-C and REIF), where the relative error increases after an initial reduction. 
This behavior is primarily attributed to the interaction between high-dimensional geometry and GP hyperparameter optimization. 
In high-dimensional settings, the failure boundary spans a large region, and the GP hyperparameters become sensitive to local sampling density.
When a strategy becomes predominantly exploitative (e.g., MOO-R or U), the L-BFGS-B optimizer adjusts the length-scales ($\ell_j$) to fit locally dense regions near the boundary. 
While this refinement can improve local accuracy, it can lead to excessively small length-scales, reducing the GP model’s ability to generalize across sparsely sampled regions of the failure domain. 
As a result, the estimated failure probability may shift, leading to increases in relative error.
This effect is amplified by the location of the failure region in the low-density tails of the input distribution, where the GP is weakly informed. 
Consequently, variations in hyperparameters during the multi-start re-optimization can impact the estimated tail probabilities, resulting in the observed oscillations. 

We additionally rank the strategies for this high-dimensional limit-state function following the per-seed ranking procedure described in the previous subsection.
Table \ref{tab:high_dim_rankings} shows that MOO-R, REIF, and Portfolio occupy the leading positions. 
While Portfolio often traces the lower envelope of the median evolution, MOO-R attains a higher ranking due to more consistent achievement of the relative error target across seeds. 
This highlights the distinction between aggregate median behavior and the seed-wise ranking. 
Figure \ref{highdim_boxplots_epsilons_3it} also indicates competitive median performance for U, MOO-LD, MOO-R, and EIER. 
EFF and MOO-K exhibit competitive medians but substantially higher average sample counts, with their $97.5^\text{th}$ percentiles reaching the budget limit.
This indicates that a significant fraction of runs does not meet the target within the allocated budget.
MOO-R achieves the best average rank, reflecting its consistency, whereas Portfolio attains the lowest median sample count. 
EIER ranks below U and MOO-LD as fewer seeds meet the target early despite comparable median sample count behavior. 
REIF2 ranks last, with an average rank of 118.93 and an average sample count of 444.1. 

The posterior $\mathrm{CoV}\!\left[\hat P_{\mathrm F}\mid \mathcal{D}_{\mathrm{train}}\right]$ distributions in Figure \ref{pf_post_cov_highdim} provide a complementary perspective to the ranking reported in Table~\ref{tab:high_dim_rankings}. 
Highly ranked strategies (e.g., MOO-R, REIF, and Portfolio) do not exhibit the lowest CoV values. 
By contrast, $\mathrm{EIER}$ shows one of the smallest CoVs, with a very narrow percentile range. 
REIF2 attains an even smaller median CoV, but with a substantially broader upper tail, indicating greater variability across seeds. 
A similar, though less pronounced, pattern is observed for U, which also achieves relatively low CoV while remaining mid-ranked. 
Overall, these results indicate that, for this high-dimensional limit-state function, achieved relative errors in failure probability and posterior confidence are not necessarily aligned.

\begin{figure*}[t]
  \graphicspath{ {./figures/} }
  \centering
  \includegraphics{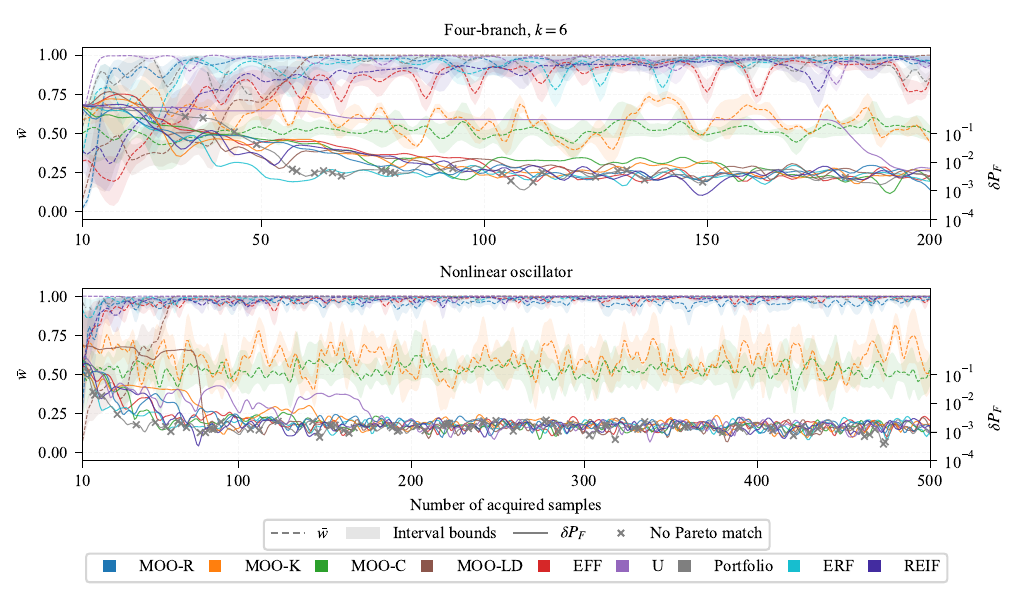}
    \caption{Exploration-exploitation preference and relative $\delta P_F$-error across iterations. 
    The derived exploration-exploitation weight intervals $[w_{\min},w_{\max}]$ (shaded regions, left axis) together with their mean values $\bar w$ (dashed lines, left axis) are shown alongside the corresponding relative error, $\delta P_F$ (solid lines, right axis), over active learning iterations. All curves are smoothed with a Gaussian filter ($\text{std}=1.5$). 
    Results are presented for the four‐branch $k=6$ (top) and the nonlinear oscillator (bottom). 
    Pareto-based strategies (MOO-R, MOO-C, MOO-K, and MOO-LD) are compared with EFF, U, ERF, REIF, and Portfolio. When the Portfolio strategy selects REIF2, the acquired samples may not lie on the Pareto front; in these cases, the weight interval is not shown, and only the performance error is reported using `$\times$' markers. By construction, $w=0$ corresponds to pure exploration and $w=1$ to pure exploitation, providing a quantitative interpretation of the balance maintained by each strategy over active learning iterations.}
  \label{invertals_evolution_comparison}
\end{figure*}

\subsubsection*{Exploration--exploitation balance}
To further investigate how each pointwise acquisition strategy balances exploitation and exploration, we compute the weight interval, $\bigl[w_{\min},\,w_{\max}\bigr]$, that would yield the selected Pareto sample under a weighted-preference scalarization.
Following a scheme based on an Euclidean-compromise scalarization, we determine the sample, $\mathbf{x}^*_{(w)}$, given a weight, $w$, as:
\begin{equation} 
\mathbf{x}^*_{(w)} = \argmin_{\mathbf{x}^* \in \mathcal{P}} \;\sqrt{\,w\,(1 - \bar{f}_{\mu}(\mathbf{x}^*))^2 \;+\;(1-w)\,(1 - \bar{f}_{\sigma}(\mathbf{x}^*))^2\,}
\quad\text{for}\quad w\in[0,1],
\label{Eucledian_compromise}
\end{equation}
where \((1 - \bar{f}_{\mu}(\mathbf{x}^*))\) and \((1 - \bar{f}_{\sigma}(\mathbf{x}^*))\) stand for the component-wise distance between a sample on the normalized Pareto and the ideal point \(\bar{\mathbf{f}}^{\mathrm{ideal}}\). 
By back-computing $\mathbf{x}^*_{(w)}$ from a grid of weights, we identify the contiguous interval 
\(\bigl[w_{\min},\,w_{\max}\bigr] \;\subset\;[0,1]\) that defines the minimum and maximum weights for which $\mathbf{x}^*_{(w)}$ would be selected. 
This interval captures the trade‐off preferences between pure exploration (\(i.e., w=0\)) and pure exploitation (\(i.e., w=1\)), associated with a given sample on the Pareto front. 
To assess the exploration-exploitation balance achieved by an acquisition strategy through a single-number metric, we summarize the weight interval by its mean value: 
\begin{equation} 
\bar w \;=\;\frac{w_{\min} + w_{\max}}{2}.
\label{mean_weight_interval}
\end{equation}
In this analysis, we focus on pointwise acquisition strategies and assess their performance on the four-branch and nonlinear oscillator limit-state functions. 
We study these cases because of the contrasting performance requirements: the four-branch function demands significant initial exploration to identify multiple failure modes, while the nonlinear oscillator is primarily exploitative. 
In Figure \ref{invertals_evolution_comparison}, we present the mean preference weight $\bar w$ and the relative $\delta P_F$-error corresponding to the sample $\mathbf{x}^*$ selected in each active learning iteration. 

Inspecting Figure \ref{invertals_evolution_comparison}, we observe that EFF initially exhibits an equally balanced exploration–exploitation behavior and smoothly transitions to the exploitation regime (i.e, $\bar w \approx 0.8 - 0.9$) after approximately 25 iterations. In contrast, U fully prioritizes exploitation (i.e., $\bar w \approx 1.0$) throughout the entire active learning loop. 
Although the described behavior remains consistent across the two analyzed limit-state functions, the resulting relative $\delta P_F$-error achieved by the strategies differs. 
By smoothly transitioning from exploration to exploitation, EFF requires only a few iterations to achieve a low relative $\delta P_F$-error on the four-branch function. 
However, this behavior demands a high number of iterations to reach $\delta P_\text{F,target}=1\cdot10^{-3}$ on the nonlinear oscillator as reported in Figure~\ref{boxplots_epsilons_3it}. 
By instead focusing on exploitation from the first iterations, U achieves a low relative $\delta P_\text{F,target}$-error on the nonlinear oscillator but straggles to reach $\delta P_\text{F,target}=1\cdot10^{-3}$ on the four-branch limit-state function within 200 active learning iterations.

ERF and REIF exhibit comparable behaviors in Figure \ref{invertals_evolution_comparison}. In the four-branch function, REIF maintains a preference around $\bar w \approx 0.8$ with a smooth, prolonged transition that certainly does not reach full exploitation. In the same limit-state function, ERF achieves quicker reliability convergence by arriving at the exploitation preference faster, despite proposing occasional exploration peaks (e.g., $\bar w \approx 0.7$). For the nonlinear oscillator, both strategies behave similarly, transitioning smoothly from exploration to almost full exploitation throughout the iteration process.
The portfolio strategy, assembled from conventional acquisition functions, provides a combined learning strategy that can be evaluated through our Pareto-based framework. Since it incorporates REIF2, some selected samples do not lie strictly on the Pareto front; these instances are marked with dots in the $\delta P_F$ evolution of Figure \ref{invertals_evolution_comparison}, and no weight interval is calculated for them. Despite these events, we observe a smooth transition toward exploitation in the portfolio strategy. This provides quick convergence in the nonlinear oscillator, while in the four-branch limit-state, the transition is fairly effective as exploitation is prioritized early, potentially at the expense of identifying all failure boundaries.

In the MOO-based approaches (i.e., MOO-C, MOO-K, MOO-LD, and MOO-R), we observe a consistent interpretable balance between objectives.
We observe that MOO-C, as expected, maintains an equally balanced behavior (i.e., $\bar w \approx 0.5$) throughout the active learning loop for both limit-state functions. 
Although MOO-K also exhibits an equally balanced behavior, it occasionally leans toward $\bar w \approx 0.75$, favoring exploitation more strongly than MOO-C. As observed in Figures~\ref{B_index_evolution} and~\ref{boxplots_epsilons_3it}, this behavior benefits MOO-K over MOO-C in all reliability scenarios. On the four-branch function, MOO-K achieve the target relative $\beta$-error with fewer iterations than EFF, whereas their equally balanced behavior causes them to underperform relative to the U strategy on the nonlinear oscillator.
The MOO-LD strategy transitions from exploration to exploitation according to its linear decay scheduling within 50 iterations. In the nonlinear oscillator, this transition clearly marks the point where the reliability estimate improves significantly; immediately after shifting toward exploitation, $\delta P_F$ drops.
Adapting the exploration-exploitation trade-off over iterations, MOO-R initially explores (i.e., $\bar w <0.25 $) and shifts toward exploitation as reliability estimates stabilize. 
In both limit-state functions, MOO-R consistently prioritizes exploitation after the initial exploration phase, with fewer fluctuations than the U strategy on the four-branch function. 
Finally, we observe that strategies consistently maintaining an equally balanced trade-off (i.e., EFF, MOO-K, and MOO-C) yield wider weight-interval bounds than strategies that focus on exploitation, such as U and MOO-R.

\begin{figure}[t]
  \graphicspath{ {./figures/} }
  \centering
  \includegraphics{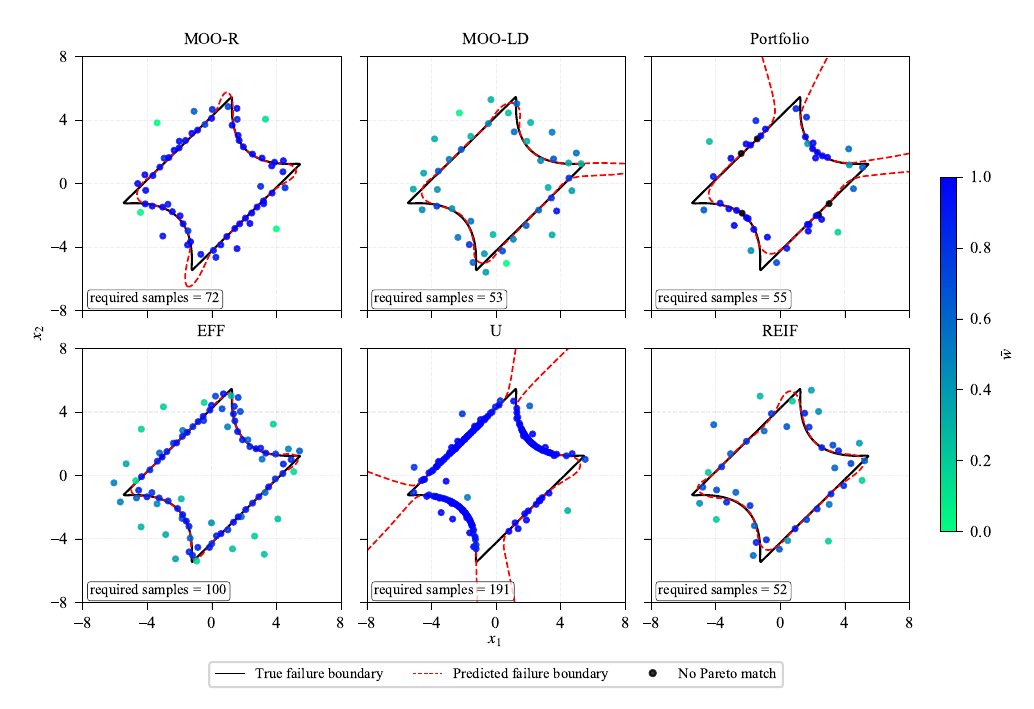}
    \caption{Acquired samples in the input space for the four-branch ($k=6$) limit-state function. Results are shown for the MOO-R, MOO-LD, Portfolio, EFF, U, and REIF strategies. In each subplot, the failure boundary $g(\mathbf{x})=0$ is plotted with a solid black contour, and the GP-predicted boundary is shown with a dashed red line.
    For visualization purposes, active samples are represented up to the point where the relative error in failure probability remains below $\delta P_{F,\text{target}} = 1 \times 10^{-2}$ for three consecutive iterations. Active samples are color-coded by the mean exploration–exploitation weight $\bar{w}$, ranging from green ($\bar w=0$, pure exploration) to blue ($\bar w=1$, pure exploitation) according to their mean exploration–exploitation weight $\bar w$. Samples with no assigned preference are labeled as ``No Pareto match" and shown in black. The required number of samples for each case is reported in the lower-left corner of each panel, while the colorbar on the right maps $\bar w$ over $[0,1]$.}
  \label{four_branch_6_activesamples_evolution}
\end{figure}

\begin{figure}
  \graphicspath{ {./figures/} }
  \centering
  \includegraphics{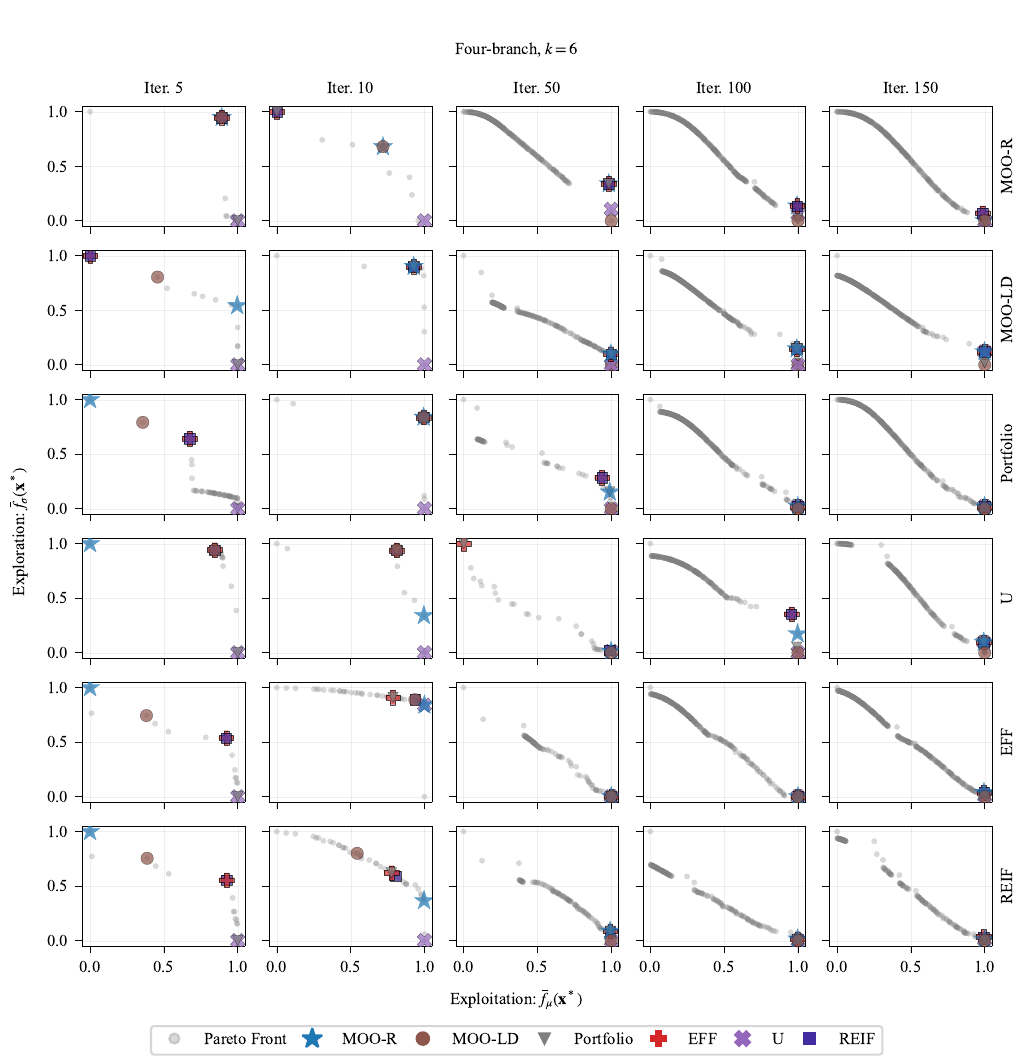}
    \caption{Evolution of normalized Pareto fronts and strategy-specific acquisition snapshots for the four-branch ($k=6$) limit-state function. Each row illustrates a sequence of active learning iterations driven by a given acquisition strategy (indicated on the right), while columns represent snapshots at fixed iteration intervals. Background markers (gray circles) depict the Pareto front in the normalized bi-objective space, where the horizontal axis represents the exploitation objective $\bar{f}_{\mu}(\mathbf{x}^*)$, and the vertical axis represents the exploration objective $\bar{f}_{\sigma}(\mathbf{x}^*)$. In each subplot, the sample acquired by the strategy indicated on the right edge of the figure is marked, together with the samples that other strategies would have acquired at the same iteration.}
  \label{fig:four_branch_pf_w_comparison}
\end{figure}

\begin{figure}
  \graphicspath{ {./figures/} }
  \centering
  \includegraphics{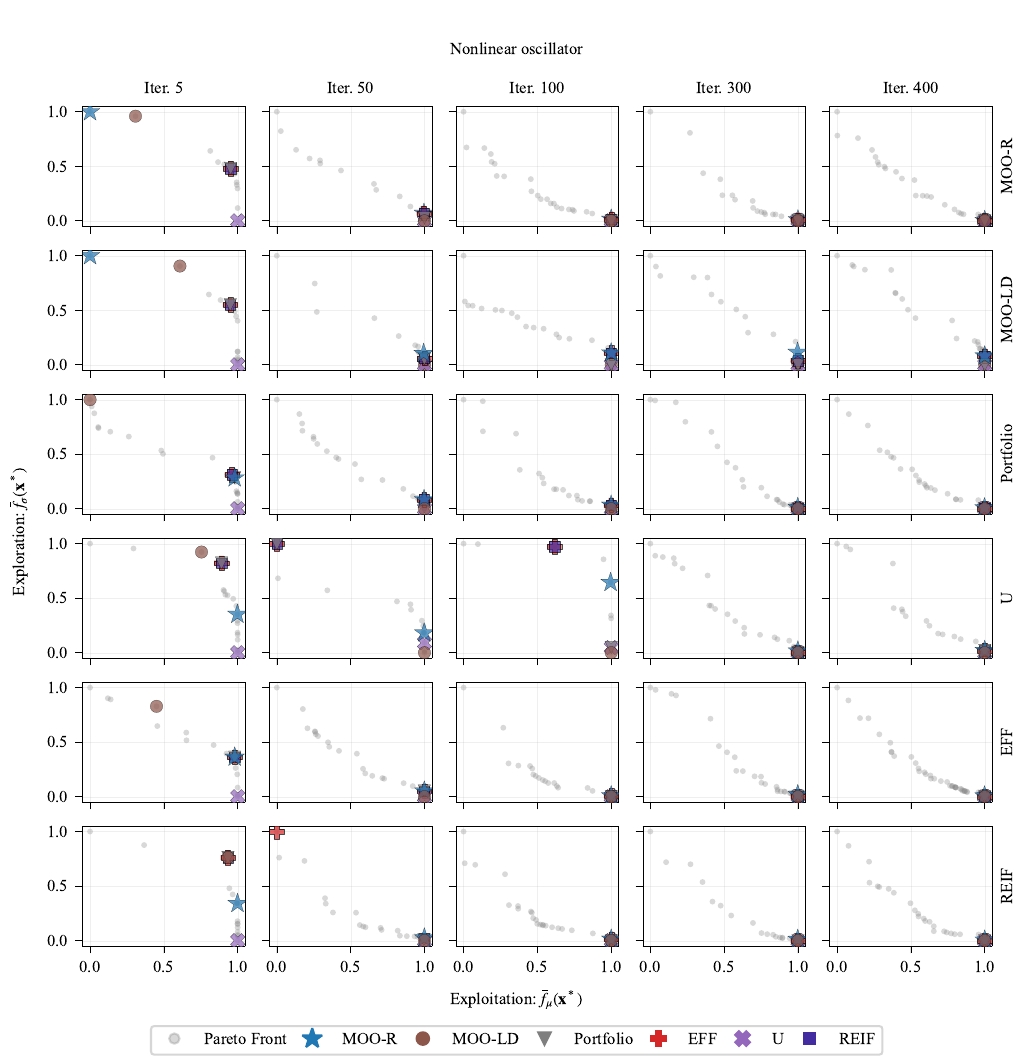}
    \caption{Evolution of normalized Pareto fronts and strategy-specific acquisition snapshots for the nonlinear oscillator limit-state function. 
    Each row illustrates a sequence of active learning iterations driven by a given acquisition strategy (indicated on the right), while columns represent snapshots at fixed iteration intervals. Background markers (gray circles) depict the Pareto front in the normalized bi-objective space, where the horizontal axis represents the exploitation objective $\bar{f}_{\mu}(\mathbf{x}^*)$, and the vertical axis represents the exploration objective $\bar{f}_{\sigma}(\mathbf{x}^*)$. In each subplot, the sample acquired by the strategy indicated on the right edge of the figure is marked, together with the samples that other strategies would have acquired at the same iteration.}
  \label{fig:nonlinear_oscillator_pf_w_comparison}
\end{figure}

\begin{figure}[ht]
  \graphicspath{ {./figures/} }
  \centering
  \includegraphics{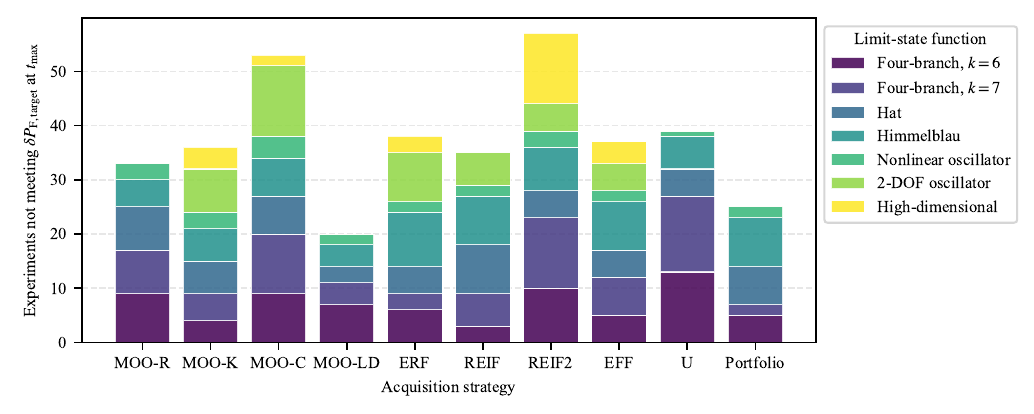}
  \caption{Experiments not meeting the relative error target within the acquisition budget. For each acquisition strategy, stacked bars represent the total number of experiments (random seeds) that failed to meet the $\delta P_{F,\text{target}}$ within the allocated acquisition budget (i.e., 200 and 500 samples for 2-dimensional and higher-dimensional limit-state functions, respectively). Colors within each bar denote the corresponding limit-state functions. Lower bar heights indicate a smaller number of runs failing to reach $\delta P_{F,\text{target}}$ within the acquisition budget, aggregated across all limit-state functions.}
 \vspace{-0.1cm}
  \label{aggregated_N_failed_exp}
\end{figure}

\section{Discussion}
\label{sec:discussion}
In this section, we analyze the behavior of the investigated acquisition strategies.
Figures \ref{four_branch_6_activesamples_evolution} and \ref{activesamples_evolution_nonlinear_oscillator} summarize the spatial distribution of acquired samples across a representative subset of six acquisition strategies, ranging from the proposed MOO-based variants (MOO-R, MOO-LD) to conventional strategies (U, EFF, REIF) and the portfolio approach. Given the observed outcomes,
our results indicate that strategies with a balanced exploration–exploitation trade-off (e.g., MOO-K, MOO-C) are not always effective and may underperform in settings like the nonlinear oscillator, where adaptively shifting toward exploitation improves performance. 
While REIF2 demonstrated efficiency in specific instances, such as the 2-DOF oscillator, its overall performance across seeds was inconsistent, and its selection logic does not strictly align with Pareto optimality. This section focuses on the strategies that show the high performance in the global rankings, omitting REIF2 and ERF in favor of the more representative acquisition strategies.

Figures \ref{four_branch_6_activesamples_evolution} and \ref{activesamples_evolution_nonlinear_oscillator} show that U exhibits focused exploitation enabling fast convergence, EFF promotes diversity that facilitates failure boundary discovery but may require additional samples to reach convergence.
In contrast, the designed MOO-based strategies offer a clearer exploration to exploitation transition. MOO-R strikes a balance by employing moderate initial exploration ($\bar w \approx 0.0$), as evidenced by the localized samples in Figure \ref{four_branch_6_activesamples_evolution}. MOO-LD further refines this by explicitly transitioning from global exploration to local refinement. This is reflected in a more spread distribution of samples that softly shifts from exploration to exploitation reaching convergence efficiently without the risk of premature convergence in a purely exploitative state. 
These findings supported by the previous global analysis point out the notion that exploration is most useful in the early active learning stages to detect failure boundaries, whereas exploitation becomes critical later to refine predictions near them.
While REIF show competitive sample counts with a fair sample distribution across all failure branches, the portfolio approach, despite reaching convergence, produces a failure boundary that requires further refinement in the corners of the failure boundary in Figure \ref{four_branch_6_activesamples_evolution}. 

Modulating the transition from exploration to exploitation is particularly important for limit-state functions such as the four-branch function, which require both identification and refinement of boundary regions. 
A consistent exploitation preference, exemplified by the U strategy, may lead to fast convergence in scenarios with well-defined failure boundaries, yet risks missing failure regions in problems with multiple or disconnected boundaries. 
On the other hand, a more pronounced exploration preference throughout the active learning loop, as exhibited by the EFF strategy, may be effective in lower-dimensional limit-state functions but can become sample-inefficient in higher-dimensional settings, where sampling far from the failure boundary is not beneficial.

From a sample-selection perspective, our proposed MOO-based frameworks mitigate the combinatorial search complexity by reducing the pool of candidate samples to the set of non-dominated solutions, as shown in Figure~\ref{Pareto_size_evolution_moo} (\ref{sec:Appendix_support}).
Although the Pareto front varies across active learning iterations, 
the number of non-dominated samples remains a small fraction of the original candidate pool, often below 0.2\%, thereby shrinking the search space by several orders of magnitude compared to conventional acquisition strategies without imposing any low-discrepancy or truncated filtering criteria. 
This is also relevant when compared with EIER in Table \ref{tab:per_seed_ranking}, which ranks candidates from a U-filtered shortlist rather than from the full candidate pool, since a typical one-step look-ahead score requires an expensive double expectation for each evaluated candidate. While this enables competitive behavior in some cases, it also restricts the evaluation to candidates preselected by U, whose seed-wise ranking is itself not always robust, e.g., with himmelblau and nonlinear oscillator where both strategies get similar ranks. Conversely, the competitive performance reached by MOO-R and MOO-LD in such limit-state functions and the 2-DOF oscillator is obtained at a markedly lower computational cost while controlling exploration--exploitation trade-off.

Our work highlights the effectiveness of adaptive acquisition strategies that shift the exploration–exploitation balance over active learning iterations, as illustrated in Figures \ref{fig:four_branch_pf_w_comparison} and \ref{fig:nonlinear_oscillator_pf_w_comparison}. 
For limit-state functions with complex or disconnected failure boundaries, maintaining exploratory tendencies may be beneficial, whereas in cases where the failure boundary is more readily identified, a faster transition toward exploration can improve sample efficiency. 
With the proposed MOO-based active learning formulation, we provide a principled and interpretable mechanism to modulate the exploration-exploitation preference across iterations.
Although the introduced adaptive MOO-based strategies explicitly adjust this trade-off, they may in some cases become overly exploitative.

In addition, the effectiveness of the investigated acquisition strategies can also be assessed by their overall success in meeting the relative error target within the allocated budget, as shown in Figure \ref{aggregated_N_failed_exp}.
MOO-LD and Portfolio exhibit the lowest number of unsuccessful runs (20 and 25, respectively). Notably, MOO-LD is among the few strategies that meet the targets across all runs for both the 2-DOF oscillator and high-dimensional cases. In contrast, REIF2 and MOO-C show the weakest performance in this analysis.
A breakdown by limit-state function reveals that the four-branch functions and the 2-DOF oscillator constitute the most challenging settings for conventional strategies such as U and REIF2. 
For instance, while U is effective in the nonlinear oscillator, it fails to meet the target in nearly all runs for the four-branch function, suggesting that its predominantly exploitative behavior is prone to being trapped in certain failure regions. 
In contrast, Portfolio leverages its ensemble structure to mitigate this effect by incorporating exploratory behavior via random strategy selection, meeting the target in all but two runs in the same setting.


\section{Concluding remarks}
\label{sec:conclusions}

In this paper, we frame sample acquisition in active learning for surrogate model-based reliability analysis as a multi-objective optimization (MOO) problem, with one objective representing the uncertainty in the limit-state function response prediction and the other representing the distance from the prediction to the failure boundary.
By computing the Pareto front of the formulated bi-objective optimization problem, we explicitly identify samples that balance the exploration-exploitation trade-off, substantially reducing the candidate pool to a smaller, representative set of samples.
Although conventional pointwise acquisition strategies normally correspond to Pareto-optimal samples under our proposed MOO-based framework, they often select samples through an implicit preference encoded in their acquisition logic, and their effectiveness depends on the specific limit-state function.
Principled MOO-based strategies, such as the knee point or compromise solutions, generally achieve competitive performance but may underperform these traditional approaches for certain limit-state functions. 
In contrast, adaptive MOO-based strategies that steer the exploration-exploitation preference over active learning iterations, either through explicit schedules (e.g., linear decay) or via adaptive indicators (e.g., convergence of failure probability estimates), can outperform both conventional strategies and standard MOO-based methods across multiple limit-state functions, offering a robust approach for active learning in reliability.
In summary, the key findings of this study are as follows:

\begin{itemize}
\item A multi-objective optimization view of active learning for reliability yields a compact Pareto set that makes exploration-exploitation preferences explicit and traceable over iterations, enabling more sample-efficient and interpretable acquisition strategies.
\item Pareto-front strategies grounded in geometric criteria (e.g., knee-point and compromise solutions) are viable, often competitive alternatives to classical single-objective acquisition strategies, though approaches that enforce a fixed balance can become less efficient as dimensionality grows.
\item Casting conventional pointwise acquisition strategies within our Pareto framework reveals that they often correspond to Pareto-optimal trade-offs with an implicit preference, clarifying their tendencies (e.g., U favoring exploitation and EFF promoting a more balanced trade-off), and providing principled guidance on when each is likely to be effective given the problem structure.
\item Incorporating adaptive multi-objective acquisition strategies that progressively shift the exploration-exploitation preference (e.g., MOO-LD) or adapt it based on failure probability indicators (e.g., MOO-R) generally improves sample efficiency across varying dimensionalities and failure-boundary geometries, relative to strategies that focus purely on exploitation or a constant balanced trade-off.
\end{itemize}
Although the proposed adaptive MOO-based adaptive strategies are sample-efficient across diverse limit-state functions, further investigation of the adaptive mechanisms governing the exploration-exploitation trade-off remains an open direction for future research. 
Moreover, to alleviate the challenges associated with Gaussian process surrogate models, whose scalability and hyperparameter estimation become increasingly difficult in high-dimensional input spaces, future work may explore alternative surrogate modeling approaches integrated within the proposed MOO-based framework.

\section*{CRediT authorship contribution statement}
\textbf{Jonathan A. Moran}: Conceptualization, Methodology, Coding, Validation, Formal analysis, Investigation, Writing - Original Draft, Writing - Review \& Editing, Visualization. 
\textbf{Pablo G. Morato}: Conceptualization, Methodology, Validation, Formal analysis, Writing - Review \& Editing, Supervision.

\section*{Acknowledgements}
Mr. Moran gratefully acknowledges the financial support of the Fonds de la Recherche Scientifique de Belgique (F.R.S.-FNRS). The authors further acknowledge the computational resources provided by the Consortium des Équipements de Calcul Intensif (CÉCI), funded by the F.R.S.-FNRS under Grant No. 2.5020.11 and by the Walloon Region.

\newpage

\bibliographystyle{elsarticle-num} 
\bibliography{bib}

@article{papaioannou2015mcmc,
  title={{MCMC algorithms for subset simulation}},
  author={Papaioannou, Iason and Betz, Wolfgang and Zwirglmaier, Kilian and Straub, Daniel},
  journal={Probabilistic Engineering Mechanics},
  volume={41},
  pages={89--103},
  year={2015},
  publisher={Elsevier}
}

@inproceedings{moran2023active,
  title     = {{Active learning for structural reliability analysis with multiple limit‐state functions through variance‐enhanced PC-Kriging surrogate models}},
  author    = {Moran A., Jonathan and Morato, Pablo Gabriel and Rigo, Philippe},
  booktitle = {Proceedings of the 14th International Conference on Applications of Statistics and Probability in Civil Engineering (ICASP14)},
  location  = {Dublin, Ireland},
  date      = {July 9--13, 2023},
  year      = {2023},
  pages      = {1--8}
}

@article{sudret2008global,
  title={Global sensitivity analysis using polynomial chaos expansions},
  author={Sudret, Bruno},
  journal={Reliability Engineering \& System Safety},
  volume={93},
  number={7},
  pages={964--979},
  year={2008},
  publisher={Elsevier}
}

@article{ehre2022sequential,
  title={Sequential active learning of low-dimensional model representations for reliability analysis},
  author={Ehre, Max and Papaioannou, Iason and Sudret, Bruno and Straub, Daniel},
  journal={SIAM Journal on Scientific Computing},
  volume={44},
  number={3},
  pages={B558--B584},
  year={2022},
  publisher={SIAM}
}

@article{MORATO2022102140,
title = {{Optimal inspection and maintenance planning for deteriorating structural components through dynamic Bayesian networks and Markov decision processes}}, 
journal = {Structural Safety},
volume = {94},
pages = {102140},
year = {2022},
issn = {0167-4730},
author = {P. G. Morato and K. G. Papakonstantinou and C. P. Andriotis and J. S. Nielsen and P. Rigo}
}

@article{chauhan2024active,
  title={{On active learning for Gaussian process-based global sensitivity analysis}},
  author={Chauhan, Mohit S and Ojeda-Tuz, Mariel and Catarelli, Ryan A and Gurley, Kurtis R and Tsapetis, Dimitrios and Shields, Michael D},
  journal={Reliability Engineering \& System Safety},
  volume={245},
  pages={109945},
  year={2024},
  publisher={Elsevier}
}

@article{moustapha2024reliability,
  title={Reliability analysis of arbitrary systems based on active learning and global sensitivity analysis},
  author={Moustapha, Maliki and Parisi, Pietro and Marelli, Stefano and Sudret, Bruno},
  journal={Reliability Engineering \& System Safety},
  volume={248},
  pages={110150},
  year={2024},
  publisher={Elsevier}
}

@article{dhulipala2022reliability,
  title={Reliability estimation of an advanced nuclear fuel using coupled active learning, multifidelity modeling, and subset simulation},
  author={Dhulipala, Somayajulu LN and Shields, Michael D and Chakroborty, Promit and Jiang, Wen and Spencer, Benjamin W and Hales, Jason D and Laboure, Vincent M and Prince, Zachary M and Bolisetti, Chandrakanth and Che, Yifeng},
  journal={Reliability Engineering \& System Safety},
  volume={226},
  pages={108693},
  year={2022},
  publisher={Elsevier}
}

@article{gaspar2017adaptive,
  title={{Adaptive surrogate model with active refinement combining Kriging and a trust region method}},
  author={Gaspar, Bruno and Teixeira, Angelo P and Soares, C Guedes},
  journal={Reliability Engineering \& System Safety},
  volume={165},
  pages={277--291},
  year={2017},
  publisher={Elsevier}
}

@article{teixeira2021adaptive,
  title={Adaptive approaches in metamodel-based reliability analysis: A review},
  author={Teixeira, Rui and Nogal, Maria and O’Connor, Alan},
  journal={Structural Safety},
  volume={89},
  pages={102019},
  year={2021},
  publisher={Elsevier}
}

@article{TEIXEIRA2021107248,
title = {Reliability analysis using a multi-metamodel complement-basis approach},
journal = {Reliability Engineering \& System Safety},
volume = {205},
pages = {107248},
year = {2021},
issn = {0951-8320},
author = {Rui Teixeira and Beatriz Martinez-Pastor and Maria Nogal and Alan O’Connor}
}

@article{MOUSTAPHA2022102174,
title = {Active learning for structural reliability: Survey, general framework and benchmark},
journal = {Structural Safety},
volume = {96},
pages = {102174},
year = {2022},
issn = {0167-4730},
author = {Maliki Moustapha and Stefano Marelli and Bruno Sudret}
}

@article{zhang2019reif,
  title={{REIF: A novel active-learning function toward adaptive Kriging surrogate models for structural reliability analysis}},
  author={Zhang, Xufang and Wang, Lei and S{\o}rensen, John Dalsgaard},
  journal={Reliability Engineering \& System Safety},
  volume={185},
  pages={440--454},
  year={2019},
  publisher={Elsevier}
}

@article{marelli2018active,
  title={An active-learning algorithm that combines sparse polynomial chaos expansions and bootstrap for structural reliability analysis},
  author={Marelli, Stefano and Sudret, Bruno},
  journal={Structural Safety},
  volume={75},
  pages={67--74},
  year={2018},
  publisher={Elsevier}
}

@article{Schobi_PCKriging,
  title={{Polynomial-chaos-based Kriging}},
  author={Schobi, Roland and Sudret, Bruno and Wiart, Joe},
  journal={International Journal for Uncertainty Quantification},
  volume={5},
  number={2},
  year={2015},
  publisher={Begel House Inc.}
}

@article{bao2021adaptive,
  title={Adaptive subset searching-based deep neural network method for structural reliability analysis},
  author={Bao, Yuequan and Xiang, Zhengliang and Li, Hui},
  journal={Reliability Engineering \& System Safety},
  volume={213},
  pages={107778},
  year={2021},
  publisher={Elsevier}
}

@article{bichon2008,
  title={Efficient global reliability analysis for nonlinear implicit performance functions},
  author={Bichon, Barron J and Eldred, Michael S and Swiler, Laura Painton and Mahadevan, Sandaran and McFarland, John M},
  journal={AIAA Journal},
  volume={46},
  number={10},
  pages={2459--2468},
  year={2008}
}

@article{ECHARD2011145,
title = {{AK-MCS: An active learning reliability method combining Kriging and Monte Carlo Simulation}},
journal = {Structural Safety},
volume = {33},
number = {2},
pages = {145-154},
year = {2011},
issn = {0167-4730},
author = {B. Echard and N. Gayton and M. Lemaire},
}

@article{IDas1997sums_objectives,
	author = {Das, I. and Dennis, J. E.},
	date = {1997/08/01},
	id = {Das1997},
	isbn = {1615-1488},
	journal = {Structural Optimization},
	number = {1},
	pages = {63--69},
	title = {{A closer look at drawbacks of minimizing weighted sums of objectives for Pareto set generation in multicriteria optimization problems}},
	volume = {14},
	year = {1997}
}

@article{marler2010weighted,
  title={The weighted sum method for multi-objective optimization: new insights},
  author={Marler, R Timothy and Arora, Jasbir S},
  journal={Structural and Multidisciplinary Optimization},
  volume={41},
  pages={853--862},
  year={2010},
  publisher={Springer}
}

@article{marler2004survey,
  title={Survey of multi-objective optimization methods for engineering},
  author={Marler, R Timothy and Arora, Jasbir S},
  journal={Structural and Multidisciplinary Optimization},
  volume={26},
  pages={369--395},
  year={2004},
  publisher={Springer}
}

@book{miettinen1999nonlinear,
  title={Nonlinear multiobjective optimization},
  author={Miettinen, Kaisa},
  volume={12},
  year={1999},
  publisher={Springer Science \& Business Media}
}

@inproceedings{bichon2009reliability,
  title={Reliability-based design optimization using efficient global reliability analysis},
  author={Bichon, Barron and Mahadevan, Sankaran and Eldred, Michael},
  booktitle={50th AIAA/ASME/ASCE/AHS/ASC Structures, Structural Dynamics, and Materials Conference 17th AIAA/ASME/AHS Adaptive Structures Conference 11th AIAA No},
  pages={2261},
  year={2009}
}

@article{shi2024active,
  title={{Active learning Kriging-based multi-objective modeling and optimization for system reliability-based robust design}},
  author={Shi, Yuwei and Lin, Chenglong and Ma, Yizhong and Shen, Jingyuan},
  journal={Reliability Engineering \& System Safety},
  volume={245},
  pages={110007},
  year={2024},
  publisher={Elsevier}
}

@article{lv2019surrogate,
  title={{Surrogate-assisted particle swarm optimization algorithm with Pareto active learning for expensive multi-objective optimization}},
  author={Lv, Zhiming and Wang, Linqing and Han, Zhongyang and Zhao, Jun and Wang, Wei},
  journal={IEEE/CAA Journal of Automatica Sinica},
  volume={6},
  number={3},
  pages={838--849},
  year={2019},
  publisher={IEEE}
}

@article{xiao2020system,
  title={{A system active learning Kriging method for system reliability-based design optimization with a multiple response model}},
  author={Xiao, Mi and Zhang, Jinhao and Gao, Liang},
  journal={Reliability Engineering \& System Safety},
  volume={199},
  pages={106935},
  year={2020},
  publisher={Elsevier}
}

@article{chen2025reliability,
  title={{A reliability-based design optimization strategy using quantile surrogates by improved PC-Kriging}},
  author={Chen, Junhua and Chen, Zhiqun and Jiang, Wei and Guo, Hun and Chen, Longmiao},
  journal={Reliability Engineering \& System Safety},
  volume={253},
  pages={110491},
  year={2025},
  publisher={Elsevier}
}

@article{dubourg2011reliability,
  title={{Reliability-based design optimization using Kriging surrogates and subset simulation}},
  author={Dubourg, Vincent and Sudret, Bruno and Bourinet, Jean-Marc},
  journal={Structural and Multidisciplinary Optimization},
  volume={44},
  pages={673--690},
  year={2011},
  publisher={Springer}
}

@inproceedings{bechikh2010searching_knee,
  title={Searching for knee regions in multi-objective optimization using mobile reference points},
  author={Bechikh, Slim and Ben Said, Lamjed and Gh{\'e}dira, Khaled},
  booktitle={Proceedings of the 2010 ACM symposium on applied computing},
  pages={1118--1125},
  year={2010}
}

@article{zeleny1973compromise,
  title={Compromise programming},
  author={Zeleny, Milan},
  journal={Multiple Criteria Decision Making},
  year={1973}
}

@inproceedings{cox1992statistical,
  title={A statistical method for global optimization},
  author={Cox, Dennis D and John, Susan},
  booktitle={IEEE international conference on systems, man, and cybernetics},
  pages={1241--1246},
  year={1992},
  organization={IEEE}
}

@inproceedings{branke2004knee,
  title={Finding knees in multi-objective optimization},
  author={Branke, J{\"u}rgen and Deb, Kalyanmoy and Dierolf, Henning and Osswald, Matthias},
  booktitle={Parallel Problem Solving from Nature-PPSN VIII: 8th International Conference, Birmingham, UK, September 18-22, 2004. Proceedings 8},
  pages={722--731},
  year={2004},
  organization={Springer}
}

@book{rasmussen2005gaussian,
  title={{Gaussian processes for machine learning}},
  author={Rasmussen, C.E. and Williams, C.K.I.},
  isbn={9780262182539},
  lccn={2005053433},
  series={Adaptive Computation and Machine Learning series},
  year={2005},
  publisher={MIT Press}
}

@article{morales2011_fmin_l_bfgs_b,
  title={{Remark on “Algorithm 778: L-BFGS-B: Fortran subroutines for large-scale bound constrained optimization”}},
  author={Morales, Jos{\'e} Luis and Nocedal, Jorge},
  journal={ACM Transactions on Mathematical Software (TOMS)},
  volume={38},
  number={1},
  pages={1--4},
  year={2011},
  publisher={ACM New York, NY, USA}
}

@article{LHS-McKay ,
 ISSN = {00401706},
 author = {M. D. McKay and R. J. Beckman and W. J. Conover},
 journal = {Technometrics},
 number = {2},
 pages = {239--245},
 publisher = {[Taylor & Francis, Ltd., American Statistical Association, American Society for Quality]},
 title = {A Comparison of Three Methods for Selecting Values of Input Variables in the Analysis of Output from a Computer Code},
 urldate = {2024-10-27},
 volume = {21},
 year = {1979}
}

@phdthesis{waarts2000fbranch,
  title={{Structural reliability using finite element methods: an appraisal of directional adaptive response surface sampling (DARS)}},
  author={Waarts, PH},
  year={2000},
  school={Technische Universiteit Delft}
}

@article{schueremans2005fbranch1,
  title={Benefit of splines and neural networks in simulation based structural reliability analysis},
  author={Schueremans, Luc and Van Gemert, Dionys},
  journal={Structural Safety},
  volume={27},
  number={3},
  pages={246--261},
  year={2005},
  publisher={Elsevier}
}

@inproceedings{schueremans2005fbranch2,
  title={{Use of Kriging as meta-model in simulation procedures for structural reliability}},
  author={Schueremans, Luc and Van Gemert, Dionys},
  booktitle={9th International Conference on Structural Safety and Reliability},
  pages={2483--2490},
  year={2005}
}

@book{himmelblau2018applied,
  title={Applied nonlinear programming},
  author={Himmelblau, David M and others},
  year={2018},
  publisher={McGraw-Hill}
}

@article{papakonstantinou2023QnHMCMC,
  title={{Hamiltonian MCMC methods for estimating rare events probabilities in high-dimensional problems}},
  author={Papakonstantinou, Konstantinos G and Nikbakht, Hamed and Eshra, Elsayed},
  journal={Probabilistic Engineering Mechanics},
  volume={74},
  pages={103485},
  year={2023},
  publisher={Elsevier}
}

@article{echard2013hat,
  title={{A combined importance sampling and Kriging reliability method for small failure probabilities with time-demanding numerical models}},
  author={Echard, Benjamin and Gayton, Nicolas and Lemaire, Maurice and Relun, Nicolas},
  journal={Reliability Engineering \& System Safety},
  volume={111},
  pages={232--240},
  year={2013},
  publisher={Elsevier}
}

@article{nielsen2005benefit,
  title={Benefit of splines and neural networks in simulation based structural reliability analysis},
  author={Nielsen, Michael Havbro Faber},
  journal={Structural Safety},
  volume={27},
  pages={403--404},
  year={2005},
  publisher={Elsevier}
}

@article{gayton2003cq2rs,
  title={{CQ2RS: A new statistical approach to the response surface method for reliability analysis}},
  author={Gayton, Nicolas and Bourinet, Jean Marc and Lemaire, Maurice},
  journal={Structural Safety},
  volume={25},
  number={1},
  pages={99--121},
  year={2003},
  publisher={Elsevier}
}

@article{rackwitz2001reliability,
  title={{Reliability analysis — A review and some perspectives}},
  author={Rackwitz, R{\"u}diger},
  journal={Structural Safety},
  volume={23},
  number={4},
  pages={365--395},
  year={2001},
  publisher={Elsevier}
}

@article{nc1973FORM,
  title={An exact and invarient first order reliability format},
  author={A. M. Hasofer and M. C. Lind},
  journal={Journal of Engineering Mechanics},
  volume={100},
  number={1},
  pages={111--121},
  year={1974}
}

@article{breitung1984SORM,
  title={Asymptotic approximations for multinormal integrals},
  author={Breitung, Karl},
  journal={Journal of Engineering Mechanics},
  volume={110},
  number={3},
  pages={357--366},
  year={1984},
  publisher={American Society of Civil Engineers}
}

@article{koutsourelakis2004LineSampling,
  title={{Reliability of structures in high dimensions, part I: algorithms and applications}},
  author={Koutsourelakis, Phadeon-Stelios and Pradlwarter, Helmuth J and Schueller, Gerhart Iwo},
  journal={Probabilistic Engineering Mechanics},
  volume={19},
  number={4},
  pages={409--417},
  year={2004},
  publisher={Elsevier}
}

@article{melchers1989importance,
  title={Importance sampling in structural systems},
  author={Melchers, RE},
  journal={Structural Safety},
  volume={6},
  number={1},
  pages={3--10},
  year={1989},
  publisher={Elsevier}
}

@article{au2001SubSetSim,
  title={Estimation of small failure probabilities in high dimensions by subset simulation},
  author={Au, Siu-Kui and Beck, James L},
  journal={Probabilistic Engineering Mechanics},
  volume={16},
  number={4},
  pages={263--277},
  year={2001},
  publisher={Elsevier}
}

@article{deb2011knee,
  title={Understanding knee points in bicriteria problems and their implications as preferred solution principles},
  author={Deb, Kalyanmoy and Gupta, Shivam},
  journal={Engineering optimization},
  volume={43},
  number={11},
  pages={1175--1204},
  year={2011},
  publisher={Taylor \& Francis}
}

@article{ranjan2008sequential,
  title={Sequential experiment design for contour estimation from complex computer codes},
  author={Ranjan, Pritam and Bingham, Derek and Michailidis, George},
  journal={Technometrics},
  volume={50},
  number={4},
  pages={527--541},
  year={2008},
  publisher={Taylor \& Francis}
}

@article{LIU2025119541,
title = {Pareto-guided active learning for accelerating surrogate-assisted multi-objective optimization of arch dam shape},
journal = {Engineering Structures},
volume = {326},
pages = {119541},
year = {2025},
issn = {0141-0296},
author = {Rui Liu and Gang Ma and Fanhui Kong and Zhitao Ai and Kun Xiong and Wei Zhou and Xiaomao Wang and Xiaolin Chang},
}

@article{DEOLIVEIRA2025103819,
title = {{Successive Pareto simulation method for efficient structural reliability analysis}},
journal = {Probabilistic Engineering Mechanics},
volume = {81},
pages = {103819},
year = {2025},
issn = {0266-8920},
author = {Rodrigo S. {de Oliveira} and Mariella F. {de L.O. Santos} and Silvana M.B. Afonso and Renato {de S. Motta}},
}

@InProceedings{slurm_ref,
author={Yoo, Andy B.
and Jette, Morris A.
and Grondona, Mark},
editor={Feitelson, Dror
and Rudolph, Larry
and Schwiegelshohn, Uwe},
title={{SLURM: Simple Linux Utility for Resource Management}},
booktitle={Job Scheduling Strategies for Parallel Processing},
year={2003},
publisher={Springer Berlin Heidelberg},
address={Berlin, Heidelberg},
pages={44--60},
}

@misc{ceci_hpc,
  author       = {{Consortium des Équipements de Calcul Intensif (CÉCI)}},
  title        = {{CÉCI High-performance computing clusters}},
  howpublished = {\url{https://www.ceci-hpc.be/clusters.html}},
  year         = {2025},
  note         = {Accessed: August 21, 2025}
}

@article{erf_yang2015,
  title={Probability and convex set hybrid reliability analysis based on active learning Kriging model},
  author={Yang, Xufeng and Liu, Yongshou and Zhang, Yishang and Yue, Zhufeng},
  journal={Applied Mathematical Modelling},
  volume={39},
  number={14},
  pages={3954--3971},
  year={2015},
  publisher={Elsevier}
}

@article{reif_zhang2019,
  title={REIF: a novel active-learning function toward adaptive Kriging surrogate models for structural reliability analysis},
  author={Zhang, Xufang and Wang, Lei and S{\o}rensen, John Dalsgaard},
  journal={Reliability Engineering \& System Safety},
  volume={185},
  pages={440--454},
  year={2019},
  publisher={Elsevier}
}

@article{portfolio_hong2023,
  title={Portfolio allocation strategy for active learning Kriging-based structural reliability analysis},
  author={Hong, Linxiong and Shang, Bin and Li, Shizheng and Li, Huacong and Cheng, Jiaming},
  journal={Computer Methods in Applied Mechanics and Engineering},
  volume={412},
  pages={116066},
  year={2023},
  publisher={Elsevier}
}

@inproceedings{auer1995gambling,
  title={Gambling in a rigged casino: The adversarial multi-armed bandit problem},
  author={Auer, Peter and Cesa-Bianchi, Nicolo and Freund, Yoav and Schapire, Robert E},
  booktitle={Proceedings of IEEE 36th annual foundations of computer science},
  pages={322--331},
  year={1995},
  organization={IEEE}
}

@article{kiureghian1991efficient,
  title={Efficient algorithm for second-order reliability analysis},
  author={Kiureghian, Armen Der and Stefano, Mario De},
  journal={Journal of engineering mechanics},
  volume={117},
  number={12},
  pages={2904--2923},
  year={1991},
  publisher={American Society of Civil Engineers}
}

@article{bourinet2011assessing,
  title={Assessing small failure probabilities by combined subset simulation and support vector machines},
  author={Bourinet, J-M and Deheeger, Fran{\c{c}}ois and Lemaire, Maurice},
  journal={Structural Safety},
  volume={33},
  number={6},
  pages={343--353},
  year={2011},
  publisher={Elsevier}
}

@article{zhou2024look,
  title={Look-ahead active learning reliability analysis based on stepwise margin reduction},
  author={Zhou, Tong and Guo, Tong and Dong, You and Yang, Fan and Frangopol, Dan M},
  journal={Reliability Engineering \& System Safety},
  volume={243},
  pages={109830},
  year={2024},
  publisher={Elsevier}
}

@article{bect2012sequential,
  title={Sequential design of computer experiments for the estimation of a probability of failure},
  author={Bect, Julien and Ginsbourger, David and Li, Ling and Picheny, Victor and Vazquez, Emmanuel},
  journal={Statistics and Computing},
  volume={22},
  number={3},
  pages={773--793},
  year={2012},
  publisher={Springer}
}

@article{khatamsaz2023bayesian,
  title={Bayesian optimization with active learning of design constraints using an entropy-based approach},
  author={Khatamsaz, Danial and Vela, Brent and Singh, Prashant and Johnson, Duane D and Allaire, Douglas and Arr{\'o}yave, Raymundo},
  journal={npj Computational Materials},
  volume={9},
  number={1},
  pages={49},
  year={2023},
  publisher={Nature Publishing Group UK London}
}

@article{nikova2025trade,
  title={Trade-offs in Bayesian active learning for feasible region identification},
  author={Nikova, Ioana and Rojas Gonzalez, Sebastian and Dhaene, Tom and Couckuyt, Ivo},
  journal={Journal of Intelligent Manufacturing},
  pages={1--24},
  year={2025},
  publisher={Springer}
}

@article{li2026bayesian,
  title={Bayesian optimization with active constraint learning for advanced manufacturing process design},
  author={Li, Guoyan and Wang, Yujia and Kar, Swastik and Jin, Xiaoning},
  journal={IISE Transactions},
  volume={58},
  number={3},
  pages={257--271},
  year={2026},
  publisher={Taylor \& Francis}
}

@article{zhang2023constrained,
  title={Constrained Bayesian optimization with adaptive active learning of unknown constraints},
  author={Zhang, Fengxue and Zhu, Zejie and Chen, Yuxin},
  journal={arXiv preprint arXiv:2310.08751},
  year={2023}
}

@article{wei2023expected,
  title={{An expected integrated error reduction function for accelerating Bayesian active learning of failure probability}},
  author={Wei, Pengfei and Zheng, Yu and Fu, Jiangfeng and Xu, Yuannan and Gao, Weikai},
  journal={Reliability Engineering \& System Safety},
  volume={231},
  pages={108971},
  year={2023},
  publisher={Elsevier}
}

@article{dang2022structural,
  title={{Structural reliability analysis: A Bayesian perspective}},
  author={Dang, Chao and Valdebenito, Marcos A and Faes, Matthias GR and Wei, Pengfei and Beer, Michael},
  journal={Structural Safety},
  volume={99},
  pages={102259},
  year={2022},
  publisher={Elsevier}
}

\clearpage

\appendix

\section{Acquisition functions: definitions and background}
\label{sec:Appendix_acquisitions}

\subsection{U active learning strategy}
\label{app:U-function}
Inspired by previous work on global optimization \cite{cox1992statistical}, the U strategy is formally defined as \cite{ECHARD2011145}:  
\begin{equation}
\mathbf{x}^{\mathrm{U}} = \arg\min_{\mathbf{x}\in\mathcal{X}} U(\mathbf{x}),
\qquad
\text{with } U(\mathbf{x}) \coloneqq \frac{|\mu_{\hat{y}}(\mathbf{x})|}{\sigma_{\hat{y}}(\mathbf{x})} \, ,
\end{equation}
where the numerator corresponds to the negative of the exploitation objective $-f_{\mu}(\mathbf{x})=\bigl|\mu_{\hat{y}}(\mathbf{x})\bigr|$ and the denominator to the exploration objective $f_{\sigma}(\mathbf{x})=\sigma_{\hat{y}}(\mathbf{x})$, as defined in our MOO formulation in Equation \ref{eq:maximization_problem_ALreliability}.  
By combining the two objectives into a single scalar quantity, this strategy reduces the multi-objective problem to a one-dimensional optimization problem, retrieving a single solution $\mathbf{x}^{\mathrm{U}} \in \mathcal{X}$ that satisfies the necessary conditions to belong to the set of non-dominated solutions defined by the Pareto front $\mathcal{P}$ in Equation \ref{eq:pareto_front_AL}.

The minimizer $\mathbf{x}^{\mathrm{U}}$ is guaranteed to be Pareto optimal as it strictly satisfies the monotonicity conditions, as described in \ref{subsec:proof_U_pareto}.
Although $\mathbf{x}^{\mathrm{U}}$ satisfies the conditions for Pareto optimality, the scalar ratio used for sample acquisition does not provide explicit control over the prioritization of the exploration and exploitation objectives. 
In particular, the trade-off is implicitly captured by a scalar, and relative differences in the scaling of the objectives may introduce bias in sample acquisition.
In addition, samples associated with substantially different exploration-exploitation trade-offs may yield identical U-scores, making their selection arbitrary.
As an example, consider the following two candidate samples:
\begin{itemize}[leftmargin=1.5em]
    \item Sample A: \( |\mu_{\hat{y}}(\mathbf{x}_A)| = 0.02 \), \( \sigma_{\hat{y}}(\mathbf{x}_A) = 0.01 \) \(\Rightarrow\) \( \text{U}(\mathbf{x}_A) = 2.0 \)
    \item Sample B: \( |\mu_{\hat{y}}(\mathbf{x}_B)| = 0.2 \), \( \sigma_{\hat{y}}(\mathbf{x}_B) = 0.1 \) \(\Rightarrow\) \( \text{U}(\mathbf{x}_B) = 2.0 \)
\end{itemize}
Both samples yield the same U-score but represent different trade-offs: Sample A lies near the failure boundary in a low-uncertainty region (favoring exploitation), whereas Sample B lies farther from the boundary in a high-uncertainty region (favoring exploration). 
U assigns the same score to both of them despite their contrasting exploration-exploitation trade-offs.

\subsection{Expected feasibility function}
The Expected Feasibility Function (EFF), introduced by \cite{bichon2008}, is inspired by earlier work on contour estimation \cite{ranjan2008sequential} and extends the \textit{Expected Improvement} (EI) framework originally used for Bayesian global optimization. 
Although EI quantifies the expected improvement that a sample provides over the current best solution, based on the predictive mean and variance provided by a Gaussian process \cite{bichon2008}, EI is fundamentally designed for optimizing a response. 
EFF adapts this idea to reliability, where the objective is to identify regions of the space where the response equals a threshold value.
EFF achieves this by quantifying the expected feasibility, i.e., the probability that the true response satisfies an imposed equality constraint (e.g., $g(\mathbf{x})=0$).
Mathematically, the EFF acquisition strategy is defined as:
\begin{equation}
\mathbf{x}^{\mathrm{EFF}} = \arg\max_{\mathbf{x}\in\mathcal{X}}
 \mathrm{EFF}(\mathbf{x}),
\qquad
\text{with }\mathrm{EFF}(\mathbf{x})
\coloneqq
\int_{-\epsilon(\mathbf{x})}^{\epsilon(\mathbf{x})}
\bigl(\epsilon(\mathbf{x})-|y_g|\bigr)\,
\varphi\!\left(y_g;\mu_{\hat{y}}(\mathbf{x}),\sigma_{\hat{y}}(\mathbf{x})\right)\,dy_g,
\label{eq:eff_def_app}
\end{equation}
where $y_g$ denotes an integration variable corresponding to the GP‐predicted response $\hat{g}(\mathbf{x})$, and $\varphi\bigl(y_g; \mu_{\hat{y}}(\mathbf{x}), \sigma_{\hat{y}}(\mathbf{x}))$ is the probability density function of a Gaussian random variable representing $\hat{g}(\mathbf{x})$ with mean $\mu_{\hat{y}}(\mathbf{x})$ and standard deviation $\sigma_{\hat{y}}(\mathbf{x})$, as provided by the Gaussian process-based surrogate model. 
The parameter $\epsilon(\mathbf{x})=c\, \sigma_{\hat{y}}(\mathbf{x})$ defines a symmetric integration interval around the limit-state function surface $\hat{g}(\mathbf{x}) = 0$, where $c$ is a constant, with a recommended value of 2 in the original work \cite{bichon2008}.



\noindent Within our framework, the parameter $\epsilon(\mathbf{x})$ can be interpreted as a proxy for statically adjusting the trade-off between exploration and exploitation, as it sets the width of the integration window around the predicted limit-state function response and remains typically fixed across active learning iterations. 
Smaller values of $\epsilon$ concentrate on predictions near the failure boundary, emphasizing exploitation (i.e., minimizing $|\mu_{\hat{y}}(\mathbf{x})|$). 
In contrast, larger values of $\epsilon$ expand the integration window, indicating a stronger preference for regions characterized by higher model uncertainty and thus favoring exploration (i.e., higher $\sigma_{\hat{y}}(\mathbf{x})$). 

\subsection{Expected Risk Function}
The Expected Risk Function (ERF), proposed by Yang et al. \cite{erf_yang2015}, is an acquisition strategy designed to identify the degree of risk that a surrogate model incorrectly predicts the sign of the limit-state function at a given point. Originally proposed for reliability analysis in scenarios where convex variables exist in the performance function, ERF provides a mathematically distinct approach to judging misclassification probability compared to the U-function.
Mathematically expressed in terms of the posterior mean and standard deviation of the surrogate model, the ERF acquisition strategy can be expressed as: 
\begin{equation}
\mathbf{x}^{\mathrm{ERF}} = \arg\max_{\mathbf{x}\in\mathcal{X}} \mathrm{ ERF}(\mathbf{x}),
\label{eq:erf_argmax_app}
\end{equation}
with:
\begin{equation}
\mathrm{ERF}(\mathbf{x})
\coloneqq
-\operatorname{sign} \, \bigl(\mu_{\hat{y}}(\mathbf{x})\bigr)\,\mu_{\hat{y}}(\mathbf{x})\,
\Phi\!\left(
-\operatorname{sign}\,\bigl(\mu_{\hat{y}}(\mathbf{x})\bigr)\,\frac{\mu_{\hat{y}}(\mathbf{x})}{\sigma_{\hat{y}}(\mathbf{x})}
\right)
+
\sigma_{\hat{y}}(\mathbf{x})\,
\phi\!\left(\frac{\mu_{\hat{y}}(\mathbf{x})}{\sigma_{\hat{y}}(\mathbf{x})}\right),
\label{eq:erf_def_app}
\end{equation}
where $\operatorname{sign}(\cdot)$ is the sign function, which returns $+1$ if the predicted mean is positive and $-1$ if it is negative. $\phi$ and $\Phi$ represent the probability density function (PDF) and the cumulative distribution function (CDF) of the standard normal distribution, respectively.

\subsection{Reliability-based Expected Improvement Function}
The Reliability-based Expected Improvement Function (REIF), introduced by Zhang et al. \cite{reif_zhang2019} is an acquisition strategy designed to balance global uncertainty reduction with local refinement of the failure boundary. REIF evaluates potential training candidates by considering the response of the limit-state function as a random variable following a folded normal distribution. By utilizing the objectives defined in our MOO framework, the REIF acquisition strategy can be formally expressed as:
\begin{equation}
\mathbf{x}^{\mathrm{REIF}} = \arg\max_{\mathbf{x}\in\mathcal{X}}\ \mathrm{REIF}(\mathbf{x}),
\qquad \text{with }
\mathrm{REIF}(\mathbf{x})
\coloneqq
\xi\, f_{\sigma}(\mathbf{x}) + f_{\mu}(\mathbf{x}),
\label{eq:reif_def_app}
\end{equation}
where $\xi$ is a weighting parameter that adjusts the relative importance between exploration and exploitation. 
An extension of this strategy, denoted as REIF2 \cite{reif_zhang2019}, incorporates the input joint probability density function $f_{\mathbf{X}}\left( \mathbf{x} \right)$ to account for the physical space of the variables:
\begin{equation}
\mathbf{x}^{\mathrm{REIF2}}=\arg\max_{\mathbf{x}\in\mathcal{X}} \mathrm{REIF2}(\mathbf{x}),
\qquad \text{with }
\mathrm{ REIF2}(\mathbf{x}) \coloneqq \mathrm{ REIF}(\mathbf{x})\, f_{\mathbf X}(\mathbf{x}).
\label{eq:reif2_def_app}
\end{equation}

\subsection{Portfolio allocation strategy}
\label{app:portfolio}
The Portfolio allocation strategy for reliability analysis \cite{portfolio_hong2023} is inspired by the multi-armed bandit problem \cite{auer1995gambling}.
Instead of committing to a single learning function, the algorithm maintains a portfolio of different acquisition strategies (e.g., U, EFF, ERF, REIF, REIF2) denoted as the set $\mathcal{A} = \{a_1(\mathbf{x}), a_2(\mathbf{x}), \dots, a_n(\mathbf{x})\}$. The goal is to maximize the cumulative reward by selecting the strategy that best refines the surrogate model at any given iteration $t$, defined as:
\begin{equation} 
r_i(t) = f_{\mu}(\mathbf{x}_i^{*(t)}) = -\lvert \mu_{\hat{y}}(\mathbf{x}_i^{*}(t)) \rvert 
\end{equation}
where $x_i^*(t)$ is the optimal sample identified by the $i$-th strategy at that step:

\begin{equation} \mathbf{x}_i^*(t) = 
\begin{cases} 
\mathbf{x}^{\rm U}, & i = 1 \\
\mathbf{x}^{\rm EFF}, & i = 2 \\ 
\mathbf{x}^{\rm ERF}, & i = 3 \\ 
\mathbf{x}^{\rm REIF}, & i = 4 \\ 
\mathbf{x}^{\rm REIF2}, & i = 5 
\end{cases}. 
\end{equation}
The accumulated reward $G_i(t)$ represents the historical success of the $i$-th strategy. 
\begin{equation} 
G_i(t) = \tau G_i(t - 1) + r_i(t). 
\end{equation}
This is governed by a hyperparameter $\tau \in [0, 1]$, which acts as a forgetting factor to prioritize recent performance. It is important to clarify that $\tau$ controls the exploration in the space of the selected acquisition strategies themselves, allowing the algorithm to periodically favor strategies that have not recently been selected, rather than the predictive space of the surrogate model $f_{\sigma}(\mathbf{x})$.
To select the active strategy for the current iteration, we compute a normalized reward $q_i(t-1)$ to ensure all strategies are compared on a consistent scale $[0, 1]$:
\begin{equation} 
q_i(t-1) = \frac{G_i(t-1) - q_{\min}(t-1)}{q_{\max}(t-1) - q_{\min}(t-1)}, 
\end{equation}
where $q_{\max}$ and $q_{\min}$ are the maximum and minimum accumulated rewards across the portfolio at that time. Finally, the selection probability $p_i(t)$ for each strategy is determined using a Softmax distribution:
\begin{equation} 
p_i(t) = \frac{\exp\left(\rho q_i(t - 1)\right)}{\sum_{j=1}^{k} \exp\left(\rho q_j(t - 1)\right)}. 
\end{equation}
The hyperparameter $\rho$ controls the exploitation intensity of the portfolio in the space of the acquisition strategies; higher values force the selection toward the strategy with the highest historical reward. 

\subsection{Expected integrated error reduction function}
\label{app:eier}
The expected integrated error reduction (EIER) function is a look-ahead acquisition strategy for active learning in reliability analysis \cite{wei2023expected}. 
Pointwise strategies score candidate samples using only the current surrogate model and rely on local predictive quantities.
In contrast, EIER evaluates each candidate by approximating its expected effect on the surrogate after conditioning on that sample, and measures the resulting reduction in a global misclassification criterion.
For a Gaussian process surrogate model conditioned on the training set $\mathcal{D}_{\mathrm{train}}$, the probability of misclassifying a point $\mathbf{x}$ is given by:
\begin{equation}
p_{\mathrm{mis}}(\mathbf{x})
=
\Phi\!\left(
-\frac{\left|\mu_{\hat{y}}(\mathbf{x})\right|}{\sigma_{\hat{y}}(\mathbf{x})}
\right),
\label{eq:pmis_eier}
\end{equation}
where $\mu_{\hat{y}}(\mathbf{x})$ and $\sigma_{\hat{y}}(\mathbf{x})$ denote the surrogate model mean prediction and its associated standard deviation, and $\Phi(\cdot)$ is the standard normal cumulative distribution function. 
Let $\mathbf{x}^+$ denote a candidate sample and 
\(
\hat{Y}^+ \sim \mathcal{N}\left(\mu_{\hat{y}}(\mathbf{x}^+), \sigma^2_{\hat{y}}(\mathbf{x}^+)\right)
\)
a hypothetical response drawn from the GP posterior predictive distribution at $\mathbf{x}^+$. 
Conditioning on the augmented dataset $\mathcal{D}_{\mathrm{train}} \cup \{(\mathbf{x}^+, \hat{y}^+)\}$ yields an updated GP model with predictive mean and standard deviation denoted by
\(
\mu_{\hat{y}^{+}}(\mathbf{x})
\)
and
\(
\sigma_{\hat{y}^{+}}(\mathbf{x})
\), respectively.
The local error reduction at point $\mathbf{x}$ induced by the candidate $\mathbf{x}^+$ is defined as:
\begin{equation}
\mathrm{ER}(\mathbf{x} \mid \mathbf{x}^+, \hat{y}^+)
=
\max\!\,\left(
p_{\mathrm{mis}}(\mathbf{x})
-
\Phi\!\left(
-\frac{\left|\mu_{\hat{y}^{+}}(\mathbf{x})\right|}
{\sigma_{\hat{y}^{+}}(\mathbf{x})}
\right),
\,0
\right).
\label{eq:er_eier}
\end{equation}
Integrating $\mathrm{ER}(\mathbf{x}\mid \mathbf{x}^+, \hat{y}^+)$ over the input space yields the Integrated Error Reduction (IER) associated with $\mathbf{x}^+$:
\begin{equation}
\mathrm{IER(\mathbf{x}^+,\hat{y}^+)}
=
\int_{\mathcal{X}}
\mathrm{ER}(\mathbf{x}\mid \mathbf{x}^+, \hat{y}^+)\,
f_{\mathbf{X}}(\mathbf{x})\,d\mathbf{x},
\label{eq:ier_eier}
\end{equation}
where $f_{\mathbf{X}}(\mathbf{x})$ is the input joint probability density function. Since the future response $\hat{Y}^+$ at $\mathbf{x}^+$ is unknown, EIER evaluates its expectation the GP posterior:
\begin{equation}
\mathcal{L}_{\mathrm{EIER}}(\mathbf{x}^+)
=
\mathbb{E}_{\hat{Y}^+ \mid \mathbf{x}^+,\,\mathcal{D}_\mathrm{train}}
\!\left[
\mathrm{IER}(\mathbf{x}^+, \hat{y}^+)
\right]
=
\int
\left[\int_{\mathcal{X}}
\mathrm{ER}(\mathbf{x}\mid \mathbf{x}^+, \hat{y}^+)\,
f_{\mathbf{X}}(\mathbf{x})\,
d\mathbf{x}\right]
f_{\hat{Y}^+ \mid \mathbf{x}^+,\,\mathcal{D}_\mathrm{train}}(\hat{y}^+)\,d\hat{y}^+ .
\label{eq:eier_def_app}
\end{equation}
Accordingly, samples are selected as:
\begin{equation}
\mathbf{x}^{\mathrm{EIER}}
=
\arg\max_{\mathbf{x}^+\in\mathcal{X}}
\mathcal{L}_{\mathrm{EIER}}(\mathbf{x}^+).
\label{eq:eier_argmax_app}
\end{equation}
Equation~\eqref{eq:eier_def_app} involves two nested integrations. 
The inner integral quantifies the reduction in misclassification probability over the input domain induced by augmenting the training set with the candidate-response pair $(\mathbf{x}^+, y^+)$. 
The outer integral then averages this quantity over the posterior predictive distribution of the hypothetical response at $\mathbf{x}^+$. 
In this sense, EIER is a one-step look-ahead strategy that scores candidate samples according to their expected global effect on the surrogate-based classification of the failure domain. 
In practice, the double integral in Eq.~\eqref{eq:eier_def_app} is approximated by Monte Carlo integration. 
Let $\{\mathbf{x}^{(j)}\}_{j=1}^{N_{\mathrm{MC}}}$ be samples drawn from the input density $f_{\mathbf X}(\mathbf{x})$, and let $\{y^{+(k)}\}_{k=1}^{N_g}$ be samples drawn from the posterior predictive distribution $f_{Y^+ \mid \mathbf{x}^+, \mathcal{D}_{\mathrm{train}}}(y^+)$. Then, EIER is approximated as:
\begin{equation}
\mathcal{L}_{\mathrm{EIER}}(\mathbf{x}^+)
\approx
\frac{1}{N_g\,N_{\mathrm{MC}}}
\sum_{k=1}^{N_g}
\sum_{j=1}^{N_{\mathrm{MC}}}
\mathrm{ER}\!\left(
\mathbf{x}^{(j)}
\mid
\mathbf{x}^+,
\hat{y}^{+(k)}
\right).
\label{eq:eier_mc_app}
\end{equation}

\section{Proofs}
\label{sec:appendix:proofs}

\subsection{General monotonicity condition for Pareto-optimality}
\label{subsec:proof_prop_monotonicity}

Let $\mathcal{X}$ denote the design space and consider the bi-objective \emph{maximization} problem:
\begin{equation}
\max_{\mathbf{x}\in\mathcal{X}}\ \mathbf{f}(\mathbf{x})
\qquad\text{with}\qquad
\mathbf{f}(\mathbf{x}) \coloneqq 
\begin{bmatrix}
f_{\mu}(\mathbf{x})\\
f_{\sigma}(\mathbf{x})
\end{bmatrix}
=
\begin{bmatrix}
-|\mu_{\hat{y}}(\mathbf{x})|\\
\sigma_{\hat{y}}(\mathbf{x})
\end{bmatrix}.
\end{equation}

\paragraph{Weak dominance and strict dominance}
For $(\mathbf{x},\mathbf{x}')\in\mathcal{X}$:
\begin{align}
\mathbf{x}\ \succeq\ \mathbf{x}'
\ &\Longleftrightarrow\ 
f_{\mu}(\mathbf{x}) \ge f_{\mu}(\mathbf{x}')
\ \wedge\
f_{\sigma}(\mathbf{x}) \ge f_{\sigma}(\mathbf{x}'),
\\
\mathbf{x}\ \succ\ \mathbf{x}'
\ &\Longleftrightarrow\ 
\mathbf{x}\succeq \mathbf{x}'
\ \wedge\
\bigl(f_{\mu}(\mathbf{x}) > f_{\mu}(\mathbf{x}')
\ \vee\
f_{\sigma}(\mathbf{x}) > f_{\sigma}(\mathbf{x}')\bigr).
\end{align}
If $\mathbf{x}\succeq \mathbf{x}'$, $\mathbf{x}$ \emph{weakly dominates} $\mathbf{x}'$.
If $\mathbf{x}\succ \mathbf{x}'$, $\mathbf{x}$ \emph{strictly dominates} $\mathbf{x}'$.

\paragraph{Pareto optimality and non-dominance}
A point $\mathbf{x}^\star\in\mathcal{X}$ is \emph{Pareto optimal} (equivalently, \emph{non-dominated})
if there exists no $\mathbf{x}\in\mathcal{X}$ such that $\mathbf{x}\succ \mathbf{x}^\star$.
A point $\mathbf{x}^\star$ is \emph{dominated} if there exists $\mathbf{x}\in\mathcal{X}$ with
$\mathbf{x}\succ \mathbf{x}^\star$.

\paragraph{Coordinate-wise monotonicity}
For a scalarization $\phi_a:\mathbb{R}^2\to\mathbb{R}$, we say that $\phi$ is \emph{strictly increasing
in each coordinate} if for all $(u_1,u_2)\in\mathbb{R}^2$:
\begin{align}
u_1>u_1' \ \Rightarrow\ \phi_a(u_1,u_2)>\phi_a(u_1',u_2), \label{eq:strict_mono_coord1}\\
u_2>u_2' \ \Rightarrow\ \phi_a(u_1,u_2)>\phi_a(u_1,u_2'). \label{eq:strict_mono_coord2}
\end{align}


\begin{proposition}[Monotonicity condition for Pareto-optimality]
\label{prop:monotonicity_condition_app}
Let $a(\mathbf{x})$ be a scalar acquisition of the form:
\begin{equation}
a(\mathbf{x}) = \phi_a\!\left(f_{\mu}(\mathbf{x}),\, f_{\sigma}(\mathbf{x})\right),
\end{equation}
where $\phi_a$ is strictly increasing in each coordinate (cf.\ \eqref{eq:strict_mono_coord1}--\eqref{eq:strict_mono_coord2}).
Then any maximizer $\mathbf{x}^\star=\arg\max_{\mathbf{x}\in\mathcal{X}} a(\mathbf{x})$ is Pareto optimal
for maximizing $\mathbf{f}(\mathbf{x})=[f_\mu(\mathbf{x}),\,f_\sigma(\mathbf{x})]^\top$.
\end{proposition}

\begin{proof}
Let $\mathbf{x}^\star=\arg\max_{\mathbf{x}\in\mathcal{X}} a(\mathbf{x})$. Assume for contradiction that
$\mathbf{x}^\star$ is \emph{not} Pareto optimal. Then there exists $\mathbf{x}\in\mathcal{X}$ such that
$\mathbf{x}\succ \mathbf{x}^\star$:
\begin{equation}
f_{\mu}(\mathbf{x}) \ge f_{\mu}(\mathbf{x}^\star),
\qquad
f_{\sigma}(\mathbf{x}) \ge f_{\sigma}(\mathbf{x}^\star),
\qquad
\text{and at least one inequality is strict.}
\label{eq:dominating_point_exists}
\end{equation}
We now derive $a(\mathbf{x})>a(\mathbf{x}^\star)$ by considering each coordinate explicitly.

\paragraph{Step 1 (monotonicity in $f_{\mu}$)}
From \eqref{eq:dominating_point_exists}, we have $f_{\mu}(\mathbf{x}) \ge f_{\mu}(\mathbf{x}^\star)$.
Since $\phi_a$ is (strictly) increasing in its first coordinate, this implies:
\begin{equation}
\phi_a\!\left(f_{\mu}(\mathbf{x}),\, f_{\sigma}(\mathbf{x})\right)
\ \ge\
\phi_a\!\left(f_{\mu}(\mathbf{x}^\star),\, f_{\sigma}(\mathbf{x})\right),
\label{eq:mono_first_coord}
\end{equation}
and the inequality in \eqref{eq:mono_first_coord} is strict if $f_{\mu}(\mathbf{x})>f_{\mu}(\mathbf{x}^\star)$.

\paragraph{Step 2 (monotonicity in $f_{\sigma}$)}
From \eqref{eq:dominating_point_exists}, we also have $f_{\sigma}(\mathbf{x}) \ge f_{\sigma}(\mathbf{x}^\star)$.
Since $\phi_a$ is (strictly) increasing in its second coordinate, this implies:
\begin{equation}
\phi_a\!\left(f_{\mu}(\mathbf{x}^\star),\, f_{\sigma}(\mathbf{x})\right)
\ \ge\
\phi_a\!\left(f_{\mu}(\mathbf{x}^\star),\, f_{\sigma}(\mathbf{x}^\star)\right),
\label{eq:mono_second_coord}
\end{equation}
and the inequality in \eqref{eq:mono_second_coord} is strict if $f_{\sigma}(\mathbf{x})>f_{\sigma}(\mathbf{x}^\star)$.

\paragraph{Step 3 (combine the coordinate-wise bounds)}
Chaining \eqref{eq:mono_first_coord} and \eqref{eq:mono_second_coord}:
\begin{equation}
a(\mathbf{x})
=
\phi_a\!\left(f_{\mu}(\mathbf{x}),\, f_{\sigma}(\mathbf{x})\right)
\ \ge\
\phi_a\!\left(f_{\mu}(\mathbf{x}^\star),\, f_{\sigma}(\mathbf{x})\right)
\ \ge\
\phi_a\!\left(f_{\mu}(\mathbf{x}^\star),\, f_{\sigma}(\mathbf{x}^\star)\right)
=
a(\mathbf{x}^\star).
\end{equation}
Moreover, because at least one of the dominance inequalities in \eqref{eq:dominating_point_exists} is strict,
at least one of the inequalities in \eqref{eq:mono_first_coord}--\eqref{eq:mono_second_coord} is strict, which implies:
\begin{equation}
a(\mathbf{x}) > a(\mathbf{x}^\star).
\end{equation}
This contradicts the assumption that $\mathbf{x}^\star$ maximizes $a(\mathbf{x})$ over $\mathcal{X}$.
Therefore, no $\mathbf{x}\in\mathcal{X}$ can strictly dominate $\mathbf{x}^\star$, and $\mathbf{x}^\star$ is Pareto optimal.
\end{proof}

\subsection{Pareto-optimality of the U acquisition strategy}
\label{subsec:proof_U_pareto}

\begin{lemma}[Ratio monotonicity]
\label{lem:ratio_mono}
Let $a,b\ge 0$ and $c,d>0$. If $a\le b$ and $c\ge d$, then:
\begin{equation}
\frac{a}{c} \le \frac{b}{d},
\end{equation}
and the inequality is strict if either $a<b$ or $c>d$.
\end{lemma}

\begin{proof}
Since $c\ge d>0$, we have $\frac{1}{c}\le \frac{1}{d}$. Then:
\begin{equation}
\frac{a}{c} \le \frac{b}{c} \le \frac{b}{d},
\end{equation}
where the first inequality is strict if $a<b$, and the second is strict if $c>d$.
\end{proof}

\begin{proposition}[Minimizing $U$ yields a Pareto-optimal point]
\label{prop:U_pareto}
Any minimizer $\mathbf{x}^{\mathrm{U}}=\arg\min_{\mathbf{x}\in\mathcal{X}} U(\mathbf{x})$ is Pareto optimal
for maximizing $\mathbf{f}(\mathbf{x})=\bigl[-|\mu_{\hat{y}}(\mathbf{x})|,\ \sigma_{\hat{y}}(\mathbf{x})\bigr]^\top$.
\end{proposition}

\begin{proof}
Assume for contradiction that $\mathbf{x}^{\mathrm{U}}$ minimizes $U$ but is not Pareto optimal.
Then there exists $\mathbf{x}\in\mathcal{X}$ such that $\mathbf{x}\succ \mathbf{x}^{\mathrm{U}}$:
\begin{equation}
-|\mu_{\hat{y}}(\mathbf{x})| \ge -|\mu_{\hat{y}}(\mathbf{x}^{\mathrm{U}})|,
\qquad
\sigma_{\hat{y}}(\mathbf{x}) \ge \sigma_{\hat{y}}(\mathbf{x}^{\mathrm{U}}),
\qquad
\text{and at least one inequality is strict.}
\label{eq:U_domination_assumption}
\end{equation}
The first inequality in \eqref{eq:U_domination_assumption} is equivalent to:
\begin{equation}
|\mu_{\hat{y}}(\mathbf{x})| \le |\mu_{\hat{y}}(\mathbf{x}^{\mathrm{U}})|.
\label{eq:mu_abs_ineq}
\end{equation}
Apply Lemma~\ref{lem:ratio_mono} with:
\[
a=|\mu_{\hat{y}}(\mathbf{x})|,\quad
b=|\mu_{\hat{y}}(\mathbf{x}^{\mathrm{U}})|,\quad
c=\sigma_{\hat{y}}(\mathbf{x}),\quad
d=\sigma_{\hat{y}}(\mathbf{x}^{\mathrm{U}}),
\]
and use \eqref{eq:mu_abs_ineq} together with the second inequality in \eqref{eq:U_domination_assumption} to obtain:
\begin{equation}
U(\mathbf{x})
=
\frac{|\mu_{\hat{y}}(\mathbf{x})|}{\sigma_{\hat{y}}(\mathbf{x})}
<
\frac{|\mu_{\hat{y}}(\mathbf{x}^{\mathrm{U}})|}{\sigma_{\hat{y}}(\mathbf{x}^{\mathrm{U}})}
=
U(\mathbf{x}^{\mathrm{U}}),
\end{equation}
where the inequality is strict because at least one of the dominance inequalities in \eqref{eq:U_domination_assumption} is strict.
This contradicts the minimality of $\mathbf{x}^{\mathrm{U}}$. Hence, no point strictly dominates $\mathbf{x}^{\mathrm{U}}$,
and $\mathbf{x}^{\mathrm{U}}$ is Pareto optimal.
\end{proof}

\paragraph{Notes}
\begin{enumerate}[label=(\roman*)]
\item The condition $\sigma_{\hat{y}}(\mathbf{x})>0$ is required for $U(\mathbf{x})$ to be well-defined and for Lemma~\ref{lem:ratio_mono}.
\item Minimizing $U(\mathbf{x})$ is sufficient for Pareto optimality, but not necessary. Other Pareto-optimal points need not minimize $U$.
\end{enumerate}

\subsection{Pareto-optimality of the EFF acquisition strategy}
\label{subsec:proof_EFF_pareto}

\paragraph{EFF as a scaled function of standardized distance}
Let:
\begin{equation}
\hat{\mu} \coloneqq \mu_{\hat{y}}(\mathbf{x}),\qquad
\hat{\sigma} \coloneqq \sigma_{\hat{y}}(\mathbf{x}),\qquad
\alpha(\mathbf{x}) \coloneqq \frac{\hat{\mu}}{\hat{\sigma}},
\end{equation}
and define $Z\sim\mathcal{N}(0,1)$ and $(u)^+\coloneqq\max\{u,0\}$. Using the change of variables
$y_g = \hat{\sigma} z + \hat{\mu}$ and $\epsilon(\mathbf{x})=c\hat{\sigma}$, one obtains:
\begin{equation}
\mathrm{EFF}(\mathbf{x})
=
\hat{\sigma}\,\Psi\!\left(\alpha(\mathbf{x})\right),
\qquad
\Psi(\alpha)
\coloneqq
\mathbb{E}\Bigl[\bigl(c-|Z+\alpha|\bigr)^+\Bigr].
\label{eq:eff_sigma_Psi_app}
\end{equation}

\paragraph{Key monotonicity property}
The function $\Psi(\alpha)$ is even and strictly decreasing on $[0,\infty)$.\footnote{This is a standard
property; it follows from the symmetry of $Z$ and the fact that shifting $Z$ away from $0$ stochastically decreases the mass of $|Z+\alpha|$ within the band $[-c,c]$ (strict decrease for $\alpha>0$).}
Consequently, $\Psi(|\alpha|)$ is strictly decreasing in $|\alpha|$, and \eqref{eq:eff_sigma_Psi_app} yields the
following coordinate-wise monotonicities:

\begin{align}
\text{(i) Exploitation monotonicity:}\quad
&\hat{\sigma}_1=\hat{\sigma}_2,\ \ |\hat{\mu}_1|<|\hat{\mu}_2|
\ \Longrightarrow\
|\alpha_1|<|\alpha_2|
\ \Longrightarrow\
\Psi(\alpha_1)>\Psi(\alpha_2) \\
&\hspace{32.5mm} \Longrightarrow\
\mathrm{EFF}(\mathbf{x}_1)>\mathrm{EFF}(\mathbf{x}_2),
\label{eq:EFF_mono_mu_app}
\\
\text{(ii) Exploration monotonicity:}\quad
&|\hat{\mu}_1|=|\hat{\mu}_2|,\ \ \hat{\sigma}_1>\hat{\sigma}_2
\ \Longrightarrow\
|\alpha_1|=\frac{|\hat{\mu}_1|}{\hat{\sigma}_1}<\frac{|\hat{\mu}_2|}{\hat{\sigma}_2}=|\alpha_2|
\ \Longrightarrow\
\Psi(\alpha_1)>\Psi(\alpha_2)
\nonumber\\[-2pt]
&\hspace{32.5mm}\Longrightarrow\
\hat{\sigma}_1\,\Psi(\alpha_1)>\hat{\sigma}_2\,\Psi(\alpha_2)
\ \Longrightarrow\
\mathrm{EFF}(\mathbf{x}_1)>\mathrm{EFF}(\mathbf{x}_2).
\label{eq:EFF_mono_sigma_app}
\end{align}

\begin{proposition}[Maximizing EFF yields a Pareto-optimal point]
\label{prop:EFF_pareto}
Assume $\sigma_{\hat{y}}(\mathbf{x})>0$ for all $\mathbf{x}\in\mathcal{X}$ and $\epsilon(\mathbf{x})=c\,\sigma_{\hat{y}}(\mathbf{x})$ with $c>0$.
Then any maximizer $\mathbf{x}^{\rm EFF}=\arg\max_{\mathbf{x}\in\mathcal{X}} \mathrm{EFF}(\mathbf{x})$ is Pareto optimal
for maximizing $\mathbf f(\mathbf x)=[-|\mu_{\hat y}(\mathbf x)|,\ \sigma_{\hat y}(\mathbf x)]^\top$.
\end{proposition}

\begin{proof}
Let $\mathbf{x}^{\rm EFF}=\arg\max_{\mathbf{x}\in\mathcal{X}} \mathrm{EFF}(\mathbf{x})$. Suppose, for contradiction, that
$\mathbf{x}^{\rm EFF}$ is dominated. Then there exists $\mathbf{x}\in\mathcal{X}$ such that $\mathbf{x}\succ \mathbf{x}^{\rm EFF}$:
\begin{equation}
-|\mu_{\hat{y}}(\mathbf{x})| \ge -|\mu_{\hat{y}}(\mathbf{x}^{\rm EFF})|
\quad\wedge\quad
\sigma_{\hat{y}}(\mathbf{x}) \ge \sigma_{\hat{y}}(\mathbf{x}^{\rm EFF}),
\quad\text{with at least one strict.}
\label{eq:EFF_dom_assumption_app}
\end{equation}
Equivalently,
\begin{equation}
|\mu_{\hat{y}}(\mathbf{x})| \le |\mu_{\hat{y}}(\mathbf{x}^{\rm EFF})|
\quad\wedge\quad
\sigma_{\hat{y}}(\mathbf{x}) \ge \sigma_{\hat{y}}(\mathbf{x}^{\rm EFF}),
\quad\text{with at least one strict.}
\label{eq:EFF_dom_assumption_app2}
\end{equation}
By the coordinate-wise monotonicities in \eqref{eq:EFF_mono_mu_app}--\eqref{eq:EFF_mono_sigma_app},
simultaneously decreasing $|\mu_{\hat{y}}|$ (improving exploitation) and/or increasing $\sigma_{\hat{y}}$
(improving exploration), with at least one strict improvement, implies a strict increase in $\mathrm{EFF}$:
\begin{equation}
\mathrm{EFF}(\mathbf{x})>\mathrm{EFF}(\mathbf{x}^{\rm EFF}).
\end{equation}
This contradicts the maximality of $\mathbf{x}^{\rm EFF}$. Therefore no $\mathbf{x}\in\mathcal{X}$ can strictly dominate
$\mathbf{x}^{\rm EFF}$, and $\mathbf{x}^{\rm EFF}$ is Pareto optimal.
\end{proof}

\paragraph{Notes}
\begin{enumerate}[label=(\roman*)]
\item The argument relies on $\sigma_{\hat y}(\mathbf x)>0$ and on the EFF convention $\epsilon(\mathbf x)=c\,\sigma_{\hat y}(\mathbf x)$.
\end{enumerate}

\subsection{Pareto-optimality of the ERF acquisition strategy}
\label{subsec:proof_ERF_pareto}

\paragraph{Simplified ERF form}
Let $\hat\mu\coloneqq \mu_{\hat y}(\mathbf x)$, $\hat\sigma\coloneqq \sigma_{\hat y}(\mathbf x)$, and
$\alpha(\mathbf x)\coloneqq \hat\mu/\hat\sigma$. Using
\(
-\operatorname{sign}(\hat\mu)\hat\mu = -|\hat\mu|
\)
and
\(
-\operatorname{sign}(\hat\mu)\alpha = -|\alpha|,
\)
\eqref{eq:erf_def_app} can be rewritten as:
\begin{equation}
\mathrm{ERF}(\mathbf x)
=
-|\hat\mu|\,\Phi\!\left(-|\alpha(\mathbf x)|\right)
+
\hat\sigma\,\phi\!\left(\alpha(\mathbf x)\right)
=
\hat\sigma\Bigl(-|\alpha(\mathbf x)|\,\Phi(-|\alpha(\mathbf x)|)+\phi(\alpha(\mathbf x))\Bigr).
\label{eq:erf_simplified_app}
\end{equation}
Define the scalar function:
\begin{equation}
\Psi_{\rm ERF}(\alpha)
\coloneqq
-|\alpha|\,\Phi(-|\alpha|)+\phi(\alpha),
\qquad\text{so that}\qquad
\mathrm{ERF}(\mathbf x)=\sigma_{\hat y}(\mathbf x)\,\Psi_{\rm ERF}\!\left(\alpha(\mathbf x)\right).
\label{eq:PsiERF_app}
\end{equation}

\paragraph{Coordinate-wise monotonicity}
The map $\Psi_{\rm ERF}$ is even and strictly decreasing on $[0,\infty)$ with respect to $\alpha$.\footnote{For $\alpha\ge 0$,
$\Psi_{\rm ERF}(\alpha)= -\alpha\,\Phi(-\alpha)+\phi(\alpha)$. Using $\frac{d}{d\alpha}\phi(\alpha)=-\alpha\phi(\alpha)$ and
$\frac{d}{d\alpha}\Phi(-\alpha)=-\phi(\alpha)$, we obtain
\(
\frac{d}{d\alpha}\Psi_{\rm ERF}(\alpha)= -\Phi(-\alpha)<0
\)
for $\alpha>0$. Evenness follows from $\phi$ being even and the dependence on $|\alpha|$.}
Hence, $\Psi_{\rm ERF}(|\alpha|)$ decreases strictly with $|\alpha|$, and \eqref{eq:PsiERF_app} implies:

\begin{align}
\text{(i) Exploitation monotonicity:}\quad
&\hat\sigma_1=\hat\sigma_2,\ \ |\hat\mu_1|<|\hat\mu_2|
\ \Longrightarrow\
|\alpha_1|<|\alpha_2|
\ \Longrightarrow\
\Psi_{\rm ERF}(\alpha_1)>\Psi_{\rm ERF}(\alpha_2) \\
&\hspace{32.5mm} \Longrightarrow\
\mathrm{ERF}(\mathbf x_1)>\mathrm{ERF}(\mathbf x_2),
\label{eq:ERF_mono_mu_app}
\\
\text{(ii) Exploration monotonicity:}\quad
&|\hat\mu_1|=|\hat\mu_2|,\ \ \hat\sigma_1>\hat\sigma_2
\ \Longrightarrow\
|\alpha_1|=\frac{|\hat\mu_1|}{\hat\sigma_1}<\frac{|\hat\mu_2|}{\hat\sigma_2}=|\alpha_2|
\ \Longrightarrow\
\Psi_{\rm ERF}(\alpha_1)>\Psi_{\rm ERF}(\alpha_2)
\nonumber\\[-2pt]
&\hspace{32.5mm}\Longrightarrow\
\hat\sigma_1\,\Psi_{\rm ERF}(\alpha_1)>\hat\sigma_2\,\Psi_{\rm ERF}(\alpha_2)
\ \Longrightarrow\
\mathrm{ERF}(\mathbf x_1)>\mathrm{ERF}(\mathbf x_2).
\label{eq:ERF_mono_sigma_app}
\end{align}

\begin{proposition}[Maximizing ERF yields a Pareto-optimal point]
\label{prop:ERF_pareto}
Assume $\sigma_{\hat y}(\mathbf x)>0$ for all $\mathbf x\in\mathcal X$. Then, any maximizer
$\mathbf x^{\rm ERF}=\arg\max_{\mathbf x\in\mathcal X}\mathrm{ERF}(\mathbf x)$ is Pareto optimal for maximizing
$\mathbf f(\mathbf x)=[-|\mu_{\hat y}(\mathbf x)|,\ \sigma_{\hat y}(\mathbf x)]^\top$.
\end{proposition}

\begin{proof}
Let $\mathbf x^{\rm ERF}$ maximize $\mathrm{ERF}$. 
Suppose it is dominated, then there exists $\mathbf x\in\mathcal X$ with
$\mathbf x\succ \mathbf x^{\rm ERF}$:
\begin{equation}
|\mu_{\hat y}(\mathbf x)| \le |\mu_{\hat y}(\mathbf x^{\rm ERF})|,
\qquad
\sigma_{\hat y}(\mathbf x) \ge \sigma_{\hat y}(\mathbf x^{\rm ERF}),
\qquad
\text{and at least one strict.}
\label{eq:ERF_dom_assumption_app}
\end{equation}
By the coordinate-wise monotonicities \eqref{eq:ERF_mono_mu_app}--\eqref{eq:ERF_mono_sigma_app},
any strict improvement in $f_\mu=-|\mu_{\hat y}|$ and/or $f_\sigma=\sigma_{\hat y}$ implies a strict increase in $\mathrm{ERF}$:
$\mathrm{ERF}(\mathbf x)>\mathrm{ERF}(\mathbf x^{\rm ERF})$, contradicting maximality. Hence, $\mathbf x^{\rm ERF}$ is non-dominated,
i.e., Pareto optimal.
\end{proof}

\subsection{Pareto-optimality of the REIF acquisition strategy}
\label{subsec:proof_REIF_pareto}

\begin{proposition}[Maximizing REIF yields a Pareto-optimal point]
\label{prop:REIF_pareto}
Assume $\xi>0$. Then any maximizer $\mathbf{x}^{\rm REIF}=\arg\max_{\mathbf{x}\in\mathcal{X}}\mathrm{REIF}(\mathbf{x})$
is Pareto optimal for maximizing $\mathbf f(\mathbf x)=[-|\mu_{\hat y}(\mathbf x)|,\ \sigma_{\hat y}(\mathbf x)]^\top$.
\end{proposition}

\begin{proof}
Recall that REIF is a linear scalarization of the two objectives $f_\mu(\mathbf{x})$ and $f_\sigma(\mathbf{x})$ with weight $\xi$:
\begin{equation}
\mathrm{REIF}(\mathbf{x}) = \xi\, f_{\sigma}(\mathbf{x}) + f_{\mu}(\mathbf{x})
= \phi_a\!\left(f_\mu(\mathbf{x}),f_\sigma(\mathbf{x})\right),
\qquad
\phi(u_1,u_2)\coloneqq u_1 + \xi u_2.
\label{eq:reif_linear_scalarization_app}
\end{equation}
For $\xi>0$, $\phi_a$ is strictly increasing in each coordinate:
if $u_1>u_1'$ then $\phi_a(u_1,u_2)>\phi_a(u_1',u_2)$, and if $u_2>u_2'$ then
$\phi_a(u_1,u_2) > \phi_a(u_1,u_2')$.
Therefore, $\mathrm{REIF}(\mathbf{x})$ satisfies the monotonicity condition of
Proposition~\ref{prop:monotonicity_condition_app}. It follows immediately that any maximizer
$\mathbf{x}^{\rm REIF}$ is non-dominated, i.e., Pareto optimal.
\end{proof}

\paragraph{Note}
The condition $\xi>0$ is essential: if $\xi=0$, REIF reduces to pure exploitation ($f_\mu$ only), and if $\xi<0$ it
penalizes exploration, in which case coordinate-wise monotonicity with respect to $(f_\mu,f_\sigma)$ is violated.

\subsection{REIF2 is not guaranteed to yield a Pareto-optimal point}
\label{subsec:proof_REIF2_not_pareto}

\paragraph{Key observation}
Proposition~\ref{prop:monotonicity_condition_app} provides a sufficient condition for Pareto optimality:
the acquisition must be strictly monotone in \emph{each} coordinate of $(f_\mu,f_\sigma)$.
Multiplication by a non-constant factor $f_{\mathbf X}(\mathbf{x})$ generally destroys this property because
increasing $(f_\mu,f_\sigma)$ can be offset by a decrease in $f_{\mathbf X}$.

\begin{proposition}[REIF2 is not Pareto-guaranteed]
\label{prop:reif2_not_pareto}
If $f_{\mathbf X}(\mathbf{x})$ is not constant over $\mathcal{X}$, then maximizers of
$\mathrm{REIF2}(\mathbf{x})=\mathrm{REIF}(\mathbf{x})f_{\mathbf X}(\mathbf{x})$ are \emph{not} guaranteed to be Pareto optimal
with respect to maximizing $\mathbf f(\mathbf x)=[f_\mu(\mathbf x),\,f_\sigma(\mathbf x)]^\top$.
\end{proposition}

\begin{proof}
We provide a constructive counterexample. Consider a design space containing at least two points
$(\mathbf{x}_A,\mathbf{x}_B)\in\mathcal{X}$ such that, for some $\xi>0$:
\begin{equation}
f_\mu(\mathbf{x}_A) > f_\mu(\mathbf{x}_B),
\qquad
f_\sigma(\mathbf{x}_A) > f_\sigma(\mathbf{x}_B).
\label{eq:AB_domination_app}
\end{equation}
Then $\mathbf{x}_A \succ \mathbf{x}_B$ (i.e., $\mathbf{x}_A$ strictly dominates $\mathbf{x}_B$) in the bi-objective
maximization of $(f_\mu,f_\sigma)$. In particular, since $\mathrm{REIF}(\mathbf{x})=f_\mu(\mathbf{x})+\xi f_\sigma(\mathbf{x})$
is strictly increasing in each coordinate for $\xi>0$, \eqref{eq:AB_domination_app} implies:
\begin{equation}
\mathrm{REIF}(\mathbf{x}_A) > \mathrm{REIF}(\mathbf{x}_B).
\label{eq:REIF_strict_order_app}
\end{equation}
Assume the input density is non-constant and satisfies:
\begin{equation}
f_{\mathbf X}(\mathbf{x}_B) > f_{\mathbf X}(\mathbf{x}_A).
\label{eq:pdf_order_app}
\end{equation}
Because $f_{\mathbf X}(\mathbf{x}_B)/f_{\mathbf X}(\mathbf{x}_A)$ can be arbitrarily large in principle
(e.g., when $\mathbf{x}_B$ lies in a high-density region and $\mathbf{x}_A$ lies in a tail), we can select (or
encounter) a pair $(\mathbf{x}_A,\mathbf{x}_B)$ such that:
\begin{equation}
\mathrm{REIF}(\mathbf{x}_B)\, f_{\mathbf X}(\mathbf{x}_B)
\ >\
\mathrm{REIF}(\mathbf{x}_A)\, f_{\mathbf X}(\mathbf{x}_A).
\label{eq:product_flip_app}
\end{equation}
Indeed, \eqref{eq:product_flip_app} holds whenever:
\begin{equation}
\frac{f_{\mathbf X}(\mathbf{x}_B)}{f_{\mathbf X}(\mathbf{x}_A)}
\ >\
\frac{\mathrm{REIF}(\mathbf{x}_A)}{\mathrm{REIF}(\mathbf{x}_B)}.
\label{eq:ratio_condition_app}
\end{equation}
Under \eqref{eq:product_flip_app}, the REIF2 objective ranks $\mathbf{x}_B$ above $\mathbf{x}_A$ even though
$\mathbf{x}_A \succ \mathbf{x}_B$ in $(f_\mu,f_\sigma)$. Therefore, a maximizer $\mathbf{x}^{\rm REIF2}$ can be
dominated (e.g., $\mathbf{x}^{\rm REIF2}=\mathbf{x}_B$ while $\mathbf{x}_A$ dominates it), which proves that
REIF2 does not guarantee Pareto optimality.
\end{proof}

\subsection{Pareto-optimality of the Portfolio acquisition strategy}
\label{subsec:proof_portfolio_pareto}

\paragraph{Portfolio definition (strategy-level selection)}
Following~\cite{portfolio_hong2023}, consider a portfolio of $n$ scalar acquisition functions
\(
\mathcal{A}=\{a_1(\mathbf{x}),\dots,a_n(\mathbf{x})\}.
\)
At iteration $t$, each acquisition $a_i$ proposes its own candidate:
\begin{equation}
\mathbf{x}_i^*(t) = \arg\rm opt_{\,\mathbf{x}\in\mathcal{X}} a_i(\mathbf{x}),
\label{eq:portfolio_candidates_app}
\end{equation}
where $\rm opt$ is either $\arg\max$ or $\arg\min$ depending on the convention of $a_i$.
The portfolio mechanism then selects an index $I(t)\in\{1,\dots,n\}$ (e.g., via the reward/Softmax scheme
described in~\cite{portfolio_hong2023}) and outputs the portfolio sample:
\begin{equation}
\mathbf{x}^{\rm Port}(t)\ \coloneqq\ \mathbf{x}_{I(t)}^*(t).
\label{eq:portfolio_output_app}
\end{equation}

\begin{proposition}[Portfolio preserves Pareto optimality under Pareto-optimal base selection]
\label{prop:portfolio_pareto}
Assume that, at iteration $t$, the selected base strategy $a_{I(t)}$ returns a Pareto-optimal candidate:
\begin{equation}
\mathbf{x}_{I(t)}^*(t)\ \text{is Pareto optimal for maximizing }\mathbf f(\mathbf x).
\label{eq:selected_base_is_pareto_app}
\end{equation}
Then, the portfolio output $\mathbf{x}^{\rm Port}(t)$ is Pareto optimal for maximizing $\mathbf f(\mathbf x)$.
\end{proposition}

\begin{proof}
By definition of the portfolio output in \eqref{eq:portfolio_output_app},
\(
\mathbf{x}^{\rm Port}(t)=\mathbf{x}_{I(t)}^*(t).
\)
Under assumption \eqref{eq:selected_base_is_pareto_app}, $\mathbf{x}_{I(t)}^*(t)$ is non-dominated
(with respect to $\mathbf f$), hence so is $\mathbf{x}^{\rm Port}(t)$. Therefore, $\mathbf{x}^{\rm Port}(t)$
is Pareto optimal.
\end{proof}

\paragraph{Remark (why REIF2 breaks the guarantee)}
If the selected strategy can return dominated candidates (e.g., REIF2 as shown in
Section~\ref{subsec:proof_REIF2_not_pareto}), then the portfolio allocation cannot guarantee Pareto optimality, as
it can yield a dominated point whenever it selects such a strategy.

\newpage

\section{Additional results}
\label{sec:Appendix_support}

\begin{figure}[hb]
  \graphicspath{ {./figures/} }
  \centering
  \includegraphics{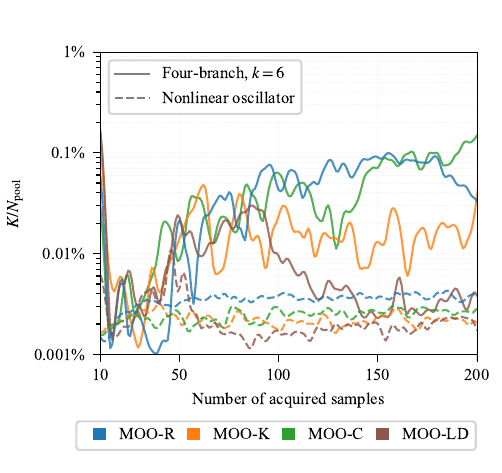}
    \caption{Pareto set size as a fraction of the candidate pool over active learning iterations. The number of samples on the Pareto front ($K$) relative to the candidate pool $(N_{\mathrm{pool}})$ is shown for the four-branch ($k=6$) and nonlinear oscillator limit state functions. Results correspond to those reported in Figure \ref{invertals_evolution_comparison}. Results are shown for the MOO-based acquisition strategies: MOO-R, MOO-LD, MOO-C, and MOO-K. All curves are smoothed with a Gaussian filter ($\text{std}=1.5$). The original candidate pool of $N_{\text{pool}}=10^6$ is consistently reduced to a much smaller Pareto set of size $K$.}
  \label{Pareto_size_evolution_moo}
\end{figure}

\begin{figure}
  \graphicspath{ {./figures/} }
  \centering
  \includegraphics{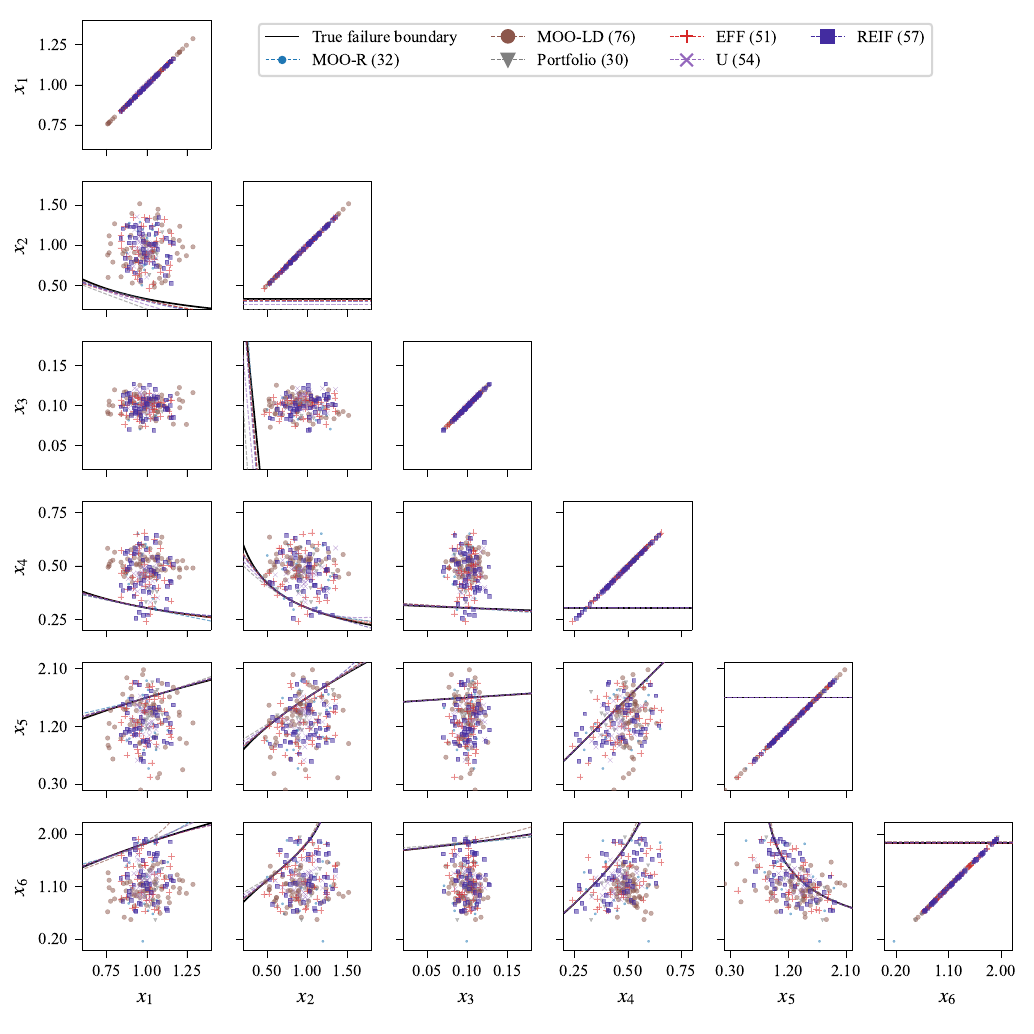}
    \caption{Acquired samples in the input space for the nonlinear oscillator limit-state function. Results are compared across the MOO-R, MOO-LD, Portfolio, EFF, U, and REIF strategies. Each lower‑triangle subplot shows the 2D projection of samples onto the ($x_i,x_j$) plane, with the failure boundary $g(\mathbf{x})=0$ drawn as a solid black contour. In contrast, the strategy-specific GP-predicted boundaries are shown as dashed lines, color-coded to each method. 
    For visualization purposes, active samples are represented up to the iteration where the relative error in the failure probability error remains below $\delta P_{F,\text{target}} = 5.0 \times 10^{-3}$ for three consecutive steps. The number of samples required by each strategy to reach the target is reported in the legend.}
  \label{activesamples_evolution_nonlinear_oscillator}
\end{figure}

\begin{figure}[t]
  \graphicspath{ {./figures/} }
  \centering
  \includegraphics{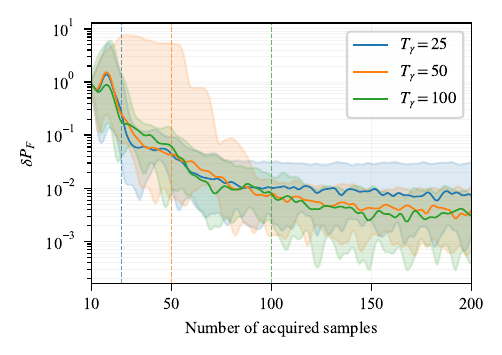}
    \caption{Parametric study of the MOO-LD decay period $T_\gamma$ on the four-branch ($k=6$) limit-state function. The plot shows the relative error $\delta P_F$ over 200 active learning iterations for $T_\gamma \in \{25, 50, 100\}$. Solid lines denote the median relative error over 10 independent random seeds, while shaded areas indicate the $2.5^{\text{th}}$ to $97.5^{\text{th}}$ percentile range. Vertical dashed lines mark the completion of the linear decay.}
  \label{moold_hyperp_fb6}
\end{figure}

\begin{figure}[t]
  \graphicspath{ {./figures/} }
  \centering
  \includegraphics{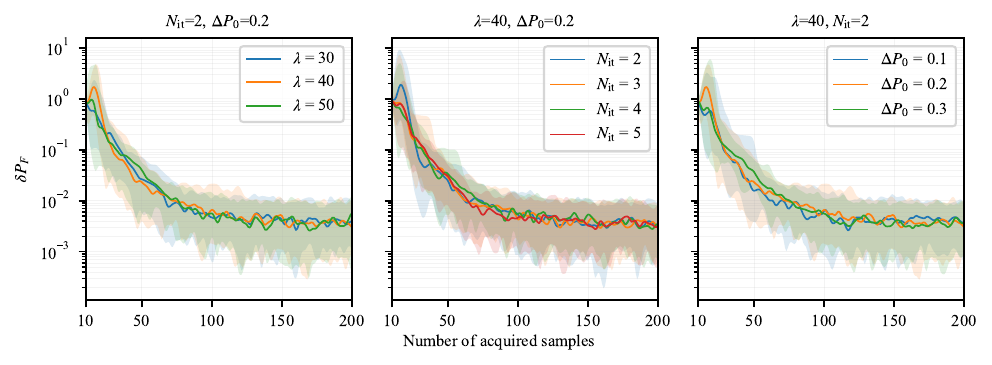}
    \caption{Parametric study of the MOO-R parameters on the four-branch ($k=6$) limit-state function. The plots show the relative error $\delta P_F$ over 200 active learning iterations for variations in: (left) sigmoid steepness $\lambda$, (center) sliding window length $N_{\mathrm{it}}$, and (right) convergence threshold $\Delta P_0$. Solid lines denote the median relative error over 10 independent random seeds, while shaded areas indicate the $2.5^{\text{th}}$ to $97.5^{\text{th}}$ percentile range. In each subplot, parameters not under study are fixed at their default values ($\lambda=40$, $N_{\mathrm{it}}=2$, $\Delta P_0=0.2$).}
  \label{moor_hyperp_fb6}
\end{figure}

This appendix presents additional results to support and clarify the findings reported in Sections \ref{sec:experiments} and \ref{sec:discussion}: 
\begin{itemize}
    \item Figure~\ref{Pareto_size_evolution_moo} presents the Pareto-set cardinality $K$ over active learning iterations for the tested MOO-based strategies, all executed with the same random seed.
    The cardinality, $K$, denotes the number of samples in the Pareto-optimal set, which is substantially smaller than the original candidate pool (i.e., $N_\text{pool}= 10^6$). 
    \item Figure~\ref{activesamples_evolution_nonlinear_oscillator} shows the acquired samples projected in the input space together with the failure boundary, revealing differences in spatial coverage and boundary refinement among MOO-R, MOO-LD, EFF, REIF, U, and Portfolio on the nonlinear oscillator limit-state function.
    \item Figure~\ref{moold_hyperp_fb6} presents the parametric study of the MOO-LD strategy on the four-branch $k=6$ limit-state function, focusing on the sensitivity to the decay period, $T_\gamma$ . 
    The results show how the timing of the transition from exploration to exploitation influences both the evolution and the final accuracy of the relative error $\delta P_F$.
    \item Figure~\ref{moor_hyperp_fb6} presents the parametric study of the MOO-R strategy on the four-branch $k=6$ limit-state function, analyzing the sensitivity to the sigmoid steepness $\lambda$, sliding window length $N_{\mathrm{it}}$, and threshold $\Delta P_0$. The results illustrate how the adaptive exploration–exploitation mechanism adjusts the acquisition strategy based on real-time reliability estimates.
\end{itemize}

\end{document}